\documentclass[aps,prd,twocolumn,showpacs,showkeys,superscriptaddress,nofootinbib, 10pt]{revtex4-2}

\usepackage[utf8]{inputenc} 
\usepackage{amsmath, amssymb} 
\usepackage{mathrsfs}
\usepackage{graphicx, subcaption} 
\usepackage{hyperref} 
\usepackage{float} 
\usepackage{color, caption} 
\usepackage{natbib} 
\usepackage{lineno}

\newcommand{\rp}{r_+}
\newcommand{\der}{\mathrm{d}}
\newcommand{\vel}{\vartheta}
\newcommand{\mnras}{Mon.\ Not.\ R.\ Astron.\ Soc.}
\newcommand{\apjs}{Astrophys.\ J.\ Suppl.\ Ser.}
\newcommand{\aap}{Astron.\ Astrophys.}
\begin{document}


\title{
On the spin dependence of the emergent gravity phenomena as
observed in axially symmetric black hole accretion with spatially
varying adiabatic index}

\author{Kalyanbrata Pal}
\email{kalyanbratapal@hri.res.in}
\affiliation{Harish-Chandra Research Institute (HRI),  Chhatnag Road, Jhunsi, Prayagraj (Allahabad), 211019, India}
\affiliation{Homi Bhabha National Institute (HBNI), Training School Complex, Anushakti Nagar, Mumbai,Maharashtra 400094, India}

\author{Souvik Ghose}
\email{dr.souvikghose@gmail.com}
\affiliation{HECRC, University of North Bengal, Raja Rammhunpur, West Bengal, 734013, India}
\affiliation{Harish-Chandra Research Institute (HRI),  Chhatnag Road, Jhunsi, Prayagraj (Allahabad), 211019, India}

\author{Ripon Sk}
\email{rsk31989@gmail.com}
\affiliation{West Bengal State University, Barasat, North 24 Parganas, India}

\author{Arpan Krishna Mitra}
\email{arpankmitra@gmail.com}
\affiliation{Aryabhatta Research Institute of Observational Sciences (ARIES), Nainital-263001, Uttarakhand, India}

\author{Tapas K. Das}
\email{das@hri.res.in}
\affiliation{Harish-Chandra Research Institute (HRI),  Chhatnag Road, Jhunsi, Prayagraj (Allahabad), 211019, India}
\affiliation{Homi Bhabha National Institute (HBNI), Training School Complex, Anushakti Nagar, Mumbai,Maharashtra 400094, India}

\date{\today}

\begin{abstract}

The present work addresses an axisymmetrically accreting black hole
system from three different perspectives. Firstly, it studies the
accretion process onto a rotating black hole under the influence of a
certain type of post-Newtonian pseudo-Kerr black hole potential from the astrophysical point of view, by considering the effect of the Kerr parameter and the geometric configuration of the flow on the flow dynamics. Secondly, by developing certain methodology which has been used to classify the critical points, it studies the dynamical systems aspects of large scale astrophysical flows under the influence of strong gravity, and last but not the least, the results presented in the current manuscript study the transonic accretion process from the standpoint of emergent gravity phenomena by investigating the nature of the acoustic space time embedded within the transonic accretion onto astrophysical black holes. The steady state equations governing low angular momentum axially symmetric accretion flow under the influence of pseudo-Kerr potential has been formulated, where the adiabatic index is a
function of radial distance and the flow contains multiple species. Transonic stationary integral accretion solutions are then constructed to show that such solutions may be multi-transonic and may accommodate a stationary shock. Different kinds of critical points encountered by the flow have been classified using the perturbative methods developed using the framework of dynamical systems phenomena. The linear stability analysis of the stationary solutions has been performed to check whether such solutions are stable under radial perturbation. Such linear perturbation of the transonic accretion leads to the formation of the acoustic geometry embedded inside the fluid where it contains acoustic black holes at the sonic points and acoustic white holes at the shock location. Corresponding acoustic horizons are then identified by constructing the causal structure of the acoustic metric by using the Carter Penrose diagram.
\end{abstract}

\maketitle


\section{Introduction}
\label{sec:intro}
Accretion flow onto astrophysical black holes  is usually transonic
\cite{1980ApJ...240..271L} where a subsonic flow starting from a large distance becomes supersonic after crossing a sonic point. For low angular momentum, axially symmetric advective accretion, such flow may have multiple sonic points, and may accommodate a stationary shock as well \cite{1981ApJ...246..314A, Fukue1983, Lu1985, Lu1986, Fukue1987, 1987MNRAS.227..975B, 1989ApJ...347..365C, cbook90, Nakayama1994, Yang1995, Chakrabarti1996, Pariev1996, Lu1997, Peitz1997, 2001ApJ...557..983D, Barai2004, Takahashi2007, Nagakura2008, Nagakura2009,Das2012,kumar2013effect,kc14,Tarafdar2015,Sukova2015,ck16,Le2016,Sukova2017,kc17,Palit2019,Palit2020b,scp20,Tarafdar2021}. Majority of the works dealing with the multi-transonic shocked accretion, however, study the matter flow described by a certain barotropic equation of state where the adiabatic index of the flow, $\Gamma$, remains constant throughout the flow. Since a number of radiative processes of different kinds take place during the accretion, momentum deposition by thermal photon on the infalling matter is expected and it is probably somewhat unlikely that $\Gamma$ will remain invariant throughout the flow from the outer boundary up to the black hole event horizon. It would thus be a relatively more realistic approach to model the flow for which the adiabatic index remains a function of the position, i.e., $\Gamma$ varies along the radial distance measured from the black hole event horizon.

Keeping that idea in mind, multi transonic shocked accretion onto astrophysical black holes has recently been studied \cite{kumar2013effect, kc14, kc17, sc19a, scp20} where the accretion flow is described by certain equation of state proposed in \cite{2006ApJS..166..410R,2009ApJ...694..492C} for which
$\Gamma$ is a function of local position, and such an equation of state
has been used to study the accretion composed of various species like
the electrons, the ions, and the positrons, respectively.   Such
studies, however, focus on stationary solutions of the steady state
accretion. Transient phenomena are, however, not quite rare in a
large-scale astrophysical setup. Hence, it is important to establish
that the information obtained from the stationary solutions is reliable
by ensuring that the steady states are stable for black hole accretion –
at least for astrophysically relevant time scales. Such a task may be
accomplished by perturbing the set of equations describing the steady
state accretion flow and by demonstrating that such perturbations do not
diverge with time.

Since a linear perturbation of any transonic flow may lead to the
emergence of the analogue space time embedded within that fluid (see,
e.g., \cite{Barcelo2011} for comprehensive discussions on the analogue gravity phenomena), analogue gravity
phenomena can be studied as a consequence of the linear perturbation
analysis of the stationary integral transonic accretion solutions. Such
attempts have recently been made where accreting black hole systems have
been considered as unique, natural models to study the emergent gravity
(hereafter we shall use the phrases analogue gravity and emergent
gravity synonymously) phenomena \cite{Shaikh_2017, PhysRevD.100.043024, PhysRevD.106.044062}. To accomplish such a task, the steady state equations governing black hole
accretion have been constructed and the corresponding stationary
transonic  integral solutions have been obtained for spherical, as well
as for the axial symmetric flow, respectively. The flow has been assumed
to be under the influence of the post-Newtonian pseudo Schwarzschild or
pseudo Kerr potentials, as well as fully general relativistic accretion
discs in the Schwazschild as well as in the Kerr metric have also been
studied. It has been observed that for low angular momentum axisymmetric
accretion, infalling material may exhibit multi-transonicity, and may
accommodate a stationary shock as well. A novel methodology has been
developed to classify the nature of the critical points corresponding to
the sonic points of the flow. The stationary solutions had been linearly
perturbed to obtain the curved acoustic metric embedded within the
accreting matter, which describes the propagation of the linear
perturbation, and such a metric had been found to possess
acoustic horizons - for multi-transonic accretion with a stationary shock, the sonic points have been identified with the acoustic black holes, and the shock location has been identified with the acoustic white holes.
Such identification had been accomplished by constructing the causal
structure by the use of the Carter Penrose diagrams. A suitable trial
solution of the massless scalar field like wave equation governing the
propagation of the linear perturbation has been used to establish that
the linear radial perturbation can not jeopardize the steady state, thus
the stationary solutions  remains valid even under the application of
such perturbation.

In the present work, we would like to investigate the analogue gravity
phenomena as observed in the axially symmetric accreting blackhole
systems, for accreting matter described by the equation of state as
proposed by \cite{2009ApJ...694..492C}. Accretion dynamics is assumed to be under the influence  of a particular post-Newtonian pseudo Kerr potential first proposed in \cite{1996ApJ...461..565A}. Recent use of powerful computational tools
and large scale numerical codes have paved the way to  directly handle the
time-dependent, fully general relativistic, magnetohydrodynamics (GRMHD)
astrophysical flows in the Kerr metric described (see,
e.g.,\cite{Dhruv_2025}, and \cite{Porth2019}),
as well as semi analytical stationary solutions for general relativistic
inviscid magnetohydrodynamic flow described by relativistic equation of
state as has been used in the present
work (\cite{Mitra_2024}).  However, a generalised study of the complete
general relativistic dissipative GRMHD accretion disk structure is
rather involved. Radiative processes associated with such
accretion is intractable, since the full general relativistic theory of
radiative transfer is yet to be fully
explored. Works related to the aforementioned topics  address the issue
in a
case by case basis \cite{Sharma_2025}, \cite{10.1093/mnras/stw1526},\cite{Schnittman_2013}, \cite{Hu2022}, \cite{10.1093/mnras/stv1046},\cite{2017IAUS..324..347P}, \cite{10.1111/j.1365-2966.2010.17502.x}
in general. The primary goal of working with accretion astrophysics is
to provide the observational signature of accreting black hole
candidates, for which the precise study of the spectral features of accreting black holes is necessary. These are still a very difficult
task to accomplish for GRMHD flow with proper formulation of GR
radiative processes in general, whereas such analysis is relatively more
tractable within a pseudo-Newtonian set-up. The practical advantages  of
the study of hydrodynamic accretion under the influence of pseudo-Kerr
potentials, thus, cannot be undermined.

A number of pseudo-Kerr black hole potentials have been proposed by
several authors \cite{1992MNRAS.256..300C, lovaas1998modified, 1999A&A...343..325S, Mukhopadhyay_2002, 10.1111/j.1365-2966.2006.10350.x, ghosh2007generalized, Ghosh_2014, 2014bhns.work..121K}. The essential purpose  behind the introduction of
a pseudo-Kerr black hole potential is to compromise between the
convenience of handling of a Newtonian description of gravity and the
space time structure described by the rather involved calculations
required to handle the general relativistic Kerr metric. That said, the
intrinsic simplicity of the algebraic presentation of a pseudo-potential
should be  of paramount importance for that potential  to qualify as the
useful one. Among all the pseudo-Kerr potentials available in the
literature, no particular pseudo-potential stands alone as a
significantly superior one to mimic all the properties of the full
general relativistic Kerr metric in connection to the description of the
dynamics of the multi-transonic flow. Thus for our purpose, the
potential with the simplest algebraic form will be of significant
interest. The pseudo-Kerr potential proposed by Artemova et al. (1996) \cite{1996ApJ...461..565A}
has a fairly
simple algebraic construct, and at the same time,  is competent enough
to mimic the Kerr metric (only along the equatorial plane, though) for
studying  multi-transonic accretion flow, hence we choose that potential
in the present work.  The generalized nature of the calculations
presented in this work ensures that  the procedure highlighted in the
current manuscript for handling the stationary flow as well the linear
stability analysis of such flow is  not  an artifact of a particular
type of black hole potential,
rather any pseudo-Kerr potential will be possible to accommodate to study
the flow properties.  It may be possible   that in future  a pseudo-Kerr
potential better than  all  existing potentials present in the
literature will be introduced, which may be the
best approximation for complete general relativistic investigation of
multi transonic shocked flow. In the present manuscript we  create a
generalized template  for a multi transonic shocked accretion disk for a
generalized form of $\Phi(r)$, hence
it will be able to readily accommodate any novel  $\Phi(r)$ without
having any significant change in the fundamental structure of the
formulation and solution scheme of the model, and one need not have to
worry about providing any new scheme exclusively valid only for the
upgraded new potential, if any.

We have performed our calculations for several models of the geometric
configuration of the accretion flow. Usually, multi-transonic shocked
accretion are studied for three different flow geometries, accretion
flow with constant flow thickness, quasi spherical conical flows, and
for flow in hydrostatic equilibrium along the vertical directions.
Details about such flow structures may be available at \cite{https://doi.org/10.1046/j.1365-8711.2001.04758.x, NAG2012285, Bilić_2014}. Use of a pseudo-Kerr potential along with these different geometric
structures of the axially symmetric flow configurations allows us to
study the stationary and the perturbative properties of the flow as a
function of black hole spins, as well as the spin dependence of the
acoustic geometry can be studied as functional properties of disc
geometries. At this point it is to be noted  that the  mathematical
expressions obtained for the flow thickness in aforementioned three flow
geometries are calculated by employing a set of idealized assumptions
only. In principle, a more realistic derivation of the flow thickness
may be worked out by employing the non-LTE radiative transfer (\cite{1998ApJ...505..558H, 2006ApJS..164..530D}) or by taking recourse to the
Grad–Shafranov equations for the MHD flow (\cite{1997PhyU...40..659B,Beskin_2005}),
see, e.g., \cite{2009mfca.book.....B} for details about the
computations of flow geometry for generalized dissipative
magnetohydrodynamic accretion flows. Such flows structures, however, can
not be used in our present work since the  additional complexities
(various types of magneto rotational instabilities, for example, see,
e.g., \cite{RevModPhys.70.1} for further
details) may lead to the complete breakdown of the Lorentz symmetries,
and the analytical modelling of the flow profile, as well as of the
sonic geometry embedded into it, will be intractable. Similarly,
introduction of the viscous dissipation will destroy the Lorentz
invariance (\cite{Barcelo2011}) as well, hence
we concentrate on low angular momentum advective flow for which the
inviscid assumption holds good. Aforementioned analysis of the stationary
solutions and their stability properties for spherically symmetric flow using the
post Newtonian set up (\cite{paul2025gravity})
as well as for complete general relativistic
Bondi \cite{1952MNRAS.112..195B} type
Michel flow (REF after the submission of Apasanka's paper). A preliminary version
with limited scope has also been communicated quite recently (\cite{pal2025effectspindynamicsmulticomponent})
where the stationary axisymmetric multi-transonic shocked accretion with variable $\Gamma$
has been studied for accretion flow in hydro static equilibrium along the vertical direction.
The present work, thus, addresses an accreting black hole system from
three different perspectives. Firstly, it studies the accretion process
from the astrophysical point of view, by considering the effect of the
Kerr parameter and the geometric configuration of the flow on the flow
dynamics, secondly, through the development of certain methodology which
has been used to classify the critical points, this work studies the
dynamical systems aspects of large scale astrophysical flows under the
influence of strong gravity, and last but not the least, the results
presented in the current manuscript study the transonic accretion
process from the standpoint of emergent gravity phenomena by
investigating the nature of the acoustic space time embedded within
the transonic accretion onto astrophysical black holes.
\section{Model Description}
\label{sec:modeldes}
In this section we describe the main building blocks of our system, namely the equation of state, back ground geometry of the flow, governing equation of the flow and flow geometry.
\subsection{Equation of state}
In this subsection we describe the equation of state (EoS) that relates thermodynamical variables of the accreting fluid under consideration. We take the accretion flow as a multispecies flow. The flow is composed of electrons ($\mathrm {e^-}$), positrons ($\mathrm{e^+}$) and protons ($\mathrm{p^+}$) in such a proportion that the fluid is overall charge neutral. Thus we could write:
\begin{equation}
    \label{eq:1}
    n_{\mathrm {e^-}}=n_{\mathrm {e^+}}+n_{\mathrm {p^+}}
\end{equation}
where $n_{\mathrm {e^-}}$, $n_{\mathrm {e^+}}$ and $n_{\mathrm {p^+}}$ are the number density of electron, positron and proton respectively.
Mass density ($\rho$) of the accreting matter is given by:
\begin{equation}
    \label{eq:2}
    \rho=\Sigma_{\mathrm{i}} n_{\mathrm{i}} m_{\mathrm{i}}=n_{\mathrm{e}^{-}} m_{\mathrm{e}^{-}}[2-\xi(1-1 / \eta)]=\rho_{\mathrm{e}^{-}}\tau
\end{equation}
where $\rho_{\rm e^-}=n_{\rm e^-} m_{\rm e^-}$, $\tau=[2-\xi(1-1 / \eta)]$, $\eta=m_{\mathrm{e}^{-}} / m_{\mathrm{p}^{+}}$ and $ \xi=n_{\mathrm{p}^{+}} / n_{\mathrm{e}^{-}}$. $m_{\mathrm{e}^{-}}$ is the mass of electron and $m_{\mathrm{p}^{+}}$ is the mass of proton. The quantity $\xi$ is the relative proportion of protons compared with electrons, known as the composition parameter. The isotropic thermal pressure ($p$) of the accreting fluid is expressed as:
\begin{equation}
    \label{eq:3}
    p=\Sigma_{\mathrm{i}} p_{\mathrm{i}}=2n_{\mathrm{e^-}}k_\mathrm{B}T=2 \rho_{\mathrm{e^-}}c^2 \Theta = \frac{2\rho c^2\Theta}{\tau}
\end{equation}
where $T$ is the temperature of the flow and $k_\mathrm{B}$ is the Boltzmann constant. The quantity $\Theta = k_\mathrm{B}T/(m_\mathrm{e^-}c^2)$, is known as the non-dimensional temperature. 

Equation of state for the multi-species accreting fluid is written as:
\begin{equation}
    \label{eq:4}
    \bar{e}=n_{\mathrm{e}^{-}} m_{\mathrm{e}^{-}} c^2 f=\rho_{\mathrm{e}^{-}} c^2 f=\frac{\rho c^2 f}{[2-\xi(1-1 / \eta)]} =   \frac{\rho  f}{\tau} [c = 1]
\end{equation}
where
$$f(\Theta) =(2-\xi)\left[1+\Theta\left(\frac{9 \Theta+3}{3 \Theta+2}\right)\right]+\xi\left[\frac{1}{\eta}+\Theta\left(\frac{9 \Theta+3 / \eta}{3 \Theta+2 / \eta}\right)\right]$$ 
and $\bar{e}$ is the energy density. Now we could write down the enthalpy ($h$) of the system as:
\begin{equation}
    \label{eq:5}
    h(\Theta) =\frac{(\bar{e}+p)}{\rho} =\frac{f c^2}{\tau}+\frac{2 c^2\Theta}{\tau} =\frac{f+2  \Theta}{\tau}
\end{equation}
The polytropic index ($N$) and the adiabatic index ($\Gamma$) are expressed as:
\begin{equation}
    \label{eq:6}
    N=\frac{1}{2} \frac{ \der f}{\der \Theta}
\end{equation}
and
\begin{equation}
    \label{eq:7}
    \Gamma = 1 + \frac{1}{N}
\end{equation}
In addition local sound speed ($c_\mathrm{s}$) is defined as:
\begin{equation}
    \label{eq:8}
    c_\mathrm{s}^{2} = \frac{2\Theta \Gamma}{\tau}
\end{equation}

\subsection{Background geometry}
The background geometry is expressed using a pseudo potential proposed by Artemova, Bjornsson and Novikov (hereafter ABN) \cite{1996ApJ...461..565A} to mimic Kerr space time. Free fall acceleration is given by: 
\begin{equation}
    \label{eq:9}
    F_\mathrm{ABN} = - \frac{1}{r^{2 -\beta}(r - \rp)^{\beta}}
\end{equation}
 Explicit form of terms used in the above expression:
\[
\begin{aligned}
r_{+} 
&= 1 + (1 - a^2)^{1/2},\\[6pt]    
Z_{1}
&= 1 + (1 - a^2)^{1/3}[(1 + a)^{1/3} + (1 -a)^{1/3}],\\[6pt]  
Z_{2} 
&= (3a^2 + Z_{1}^2)^{1/2},\\[6pt]   
r_\mathrm{in} 
&= 3 + Z_{2} - [(3 - Z_{1})(3 + Z_{1} + 2Z_{2})]^{1/2}, \\[6pt]
\beta 
&= \frac{r_\mathrm{in}}{r_{+}} - 1,  
\end{aligned}
\]
where $a$ is the black hole spin parameter (also known as Kerr parameter) and $r$ is the radial coordinate, measured along the equatorial plane. Here, $r_{+}$ and $r_\mathrm{in}$ denote the horizon of the rotating black hole and the innermost stable circular orbit (ISCO), respectively.

The pseudo-potential is obtained by integrating the above free-fall acceleration and can be written as:
\begin{equation}
    \label{eq:10}
    \Phi_\text{ABN}=\frac{1}{(1-\beta) \rp}\left(1-\frac{\rp}{r}\right)^{1-\beta}-\frac{1}{(1-\beta) \rp}.   
\end{equation}


\subsection{Governing equations of the flow dynamics}
We solve low angular momentum, inviscid, multi-species, axially symmetric accretion flow around a Kerr black hole for three types of flow geometry. Taking into account the symmetry of the problem, we adopt a cylindrical polar coordinate system ($r$, $\phi$, $z$). Rotation axis of the black hole is assumed to be in the $z$ direction. In addition to that we also consider that the flow is symmetric around the $z$ axis, $i.e.$, the flow variables are independent of the $\phi$ coordinate and the mid-plane of the flow is assumed to coincide with the equatorial plane ($z = 0$ plane). Flow of matter around the black hole is governed by two dynamical equations:

i)Mass conservation equation:
\begin{equation}
    \label{eq:11}
    \frac{\partial \Sigma}{\partial t} + \frac{1}{r}\frac{\partial (\Sigma \vel r)}{\partial r} = 0,
\end{equation}
where $\Sigma = 2\rho H$ is the vertically averaged surface density, $\rho$ is the density of the fluid and $H$ is the disk half height being measured from the flow mid plane. Depending on the flow geometry, $H$ is generally a function of $r$ and $\Theta$ $i.e.,$ $H \equiv H(r,\Theta)$. Here, $\vel$ is the inward radial flow velocity. 

ii) Euler equation:
\begin{equation}
    \label{eq:12}
    \frac{\partial \vel}{\partial \text{t}} + \vel \frac{\der \vel}{\der r} + \frac{1}{\rho}\frac{\der p}{\der r} - \frac{\lambda^2}{r^3} - F_\mathrm{ABN} = 0.
\end{equation}
Here $p$ is the gas pressure and the specific angular momentum $i.e.,$ angular momentum per unit mass is denoted by $\lambda$.
In addition to the above equations, we need the first law of thermodynamics:
\begin{equation}
    \label{eq:13}
    \frac{\der e}{\der r} - \frac{p}{\rho^2}\frac{\der \rho}{\der r} =0,
\end{equation}
where the expression of the quantity $e$ is: $e = \bar{e}/\rho = f/\tau$.

We assume that our accreting system is in steady state $i.e.,$ the thermodynamical and dynamical variables associated with the accreting system are independent of time ($t$). Now we could write the mass conservation equation (Eq. (\ref{eq:11})) in the following form:
\begin{equation}
    \label{eq:14}
     \frac{1}{r}\frac{\partial (\Sigma \vel r)}{\partial r} = 0,
\end{equation}
and the Euler equation (Eq. (\ref{eq:12}))becomes:
\begin{equation}
    \label{eq:15}
     \vel \frac{\der \vel}{\der r} + \frac{1}{\rho}\frac{\der p}{\der r} - \frac{\lambda^2}{r^3} - F_\mathrm{ABN} = 0.
\end{equation}

By solving the steady state mass conservation equation (Eq. (\ref{eq:14})) we construct the steady state mass accretion rate ($\dot{M}$) as follows:
\begin{equation}
    \label{eq:16}
    \dot M = 2\pi\Sigma \vel r ,  
\end{equation}
which is constant through out the accretion flow. By integrating the steady state Euler equation (Eq. \ref{eq:15}) we obtain the specific energy ($\mathcal{E}$) $i.e.,$ energy per unit mass as follows:
\begin{equation}
    \label{eq:17}
    \mathcal{E} = \frac{\vel^2}{2} + h + \frac{\lambda^2}{2r^2} + \Phi_\mathrm{ABN}.
\end{equation}
The specific energy ($\mathcal{E}$) is constant along the streamline of the accretion flow.
Entropy accretion rate (see \citep{kumar2013effect}) is given by:
\begin{equation}
    \label{eq:18}
    \mathcal{\dot M}= \text{H}\vel r \exp \left(k_3\right) \Theta^{3 / 2}(3 \Theta+2)^{k_1}(3 \Theta+2 / \eta)^{k_2}, 
\end{equation}
where $k_1=3(2-\xi) / 4$, $k_2=3 \xi / 4$ and $k_3=(f-\tau) /(2 \Theta)$.

Using Eqs. (\ref{eq:4}) and (\ref{eq:6}) we could write:
\begin{equation}
    \label{eq:19}
    \frac{\der e}{\der r} = \frac{2N}{\tau}\frac{\der \Theta}{\der r},
\end{equation}
and by using Eqs. (\ref{eq:3}) and (\ref{eq:16}), we get:
\begin{equation}
    \label{eq:20}
    \frac{p}{\rho^2}\frac{\der \rho}{\der r} = -\frac{2\Theta}{\tau}\bigg[\frac{1}{\vel}\frac{\der \vel}{\der r} + \frac{1}{r}+ \frac{1}{H}\frac{\partial H}{\partial r}+ \frac{1}{H}\frac{\partial H}{\partial \Theta}\frac{\der \Theta}{\der r}\bigg].
\end{equation}
Combining above two equations we obtain:
\begin{equation}
    \label{eq:21}
    \frac{\der \Theta}{\der r} = \Omega_{1} + \Omega_{2}\dfrac{\der \vel}{\der r},
\end{equation}
where
\[\Omega_{1} = -\frac{\Theta\bigg[\dfrac{1}{r}+\dfrac{1}{H}\dfrac{\partial H}{\partial r}\bigg]}{\bigg[N + \dfrac{\Theta}{H}\dfrac{\partial H}{\partial \Theta}\bigg]} \]
and
\[\Omega_{2} = - \frac{\Theta}{\vel\bigg[N + \dfrac{\Theta}{H}\dfrac{\partial H}{\partial \Theta}\bigg]}. \]
Eq. (\ref{eq:21}) describes the rate of change in temperature along the radial direction. By using Eqs. (\ref{eq:16}) and (\ref{eq:21}) we could recast Eq. (\ref{eq:15}) in the following way:
\begin{equation}
\label{eq:22}
\frac{\der \vel}{\der r}=\dfrac{\dfrac{c_\mathrm{s}^2\bigg(\dfrac{1}{r}+\dfrac{1}{H}\dfrac{\partial H}{\partial r}\bigg)}
{\left(1 + \dfrac{\Theta}{NH}\dfrac{\partial H}{\partial \Theta}\right)}+ \dfrac{\lambda^{2}}{r^3}- \dfrac{1}{r^{2-\beta}(r-\rp)^{\beta}}}{\left[\vel
- \dfrac{c_\mathrm{s}^2}{\vel\left(1 + \dfrac{\Theta}{NH}\dfrac{\partial H}{\partial \Theta}\right)}\right]}= \frac{\mathcal{N}}{\mathcal{D}} .
\end{equation}
Here, $\mathcal{N}$ and $\mathcal{D}$ stands for numerator and denominator respectively.
Above two equations (Eqs. (\ref{eq:21}) and (\ref{eq:22})) are coupled first order nonlinear differential equations in $\Theta$ and $\vel$. So, our goal is to solve these two equations in order to get the velocity and temperature profiles of the accreting flow. We represent the velocity profile in terms of phase portrait (Mach number ($M$) vs radial distance ($r$)) diagram.

To solve the above equations we adapt a technique from dynamical system analysis, known as critical point analysis method.
\subsection*{Critical Point Analysis}
In general accreting matter starts with subsonic velocity ($\vel<c_\mathrm{s}$) at the outer boundary and the velocity increases as the flow moves inward. Due to the inner boundary condition (see \citep{1980ApJ...240..271L}) the flow velocity exceeds the local sound speed at some radial distance and becomes supersonic ($\vel>c_\mathrm{s}$). The point at which the flow velocity is equal to the local sound speed, $i.e.,$ mach number $M$ =1, is known as the sonic point. If the accretion flow changes its state from subsonic to supersonic or from supersonic to subsonic, then the flow is a transonic flow. Therefore, flow that passes through more than one sonic point is known to be a multitransonic flow. Owing to the relative dominance between the flow velocity ($\vel$) and local sound speed ($c_\mathrm{s}$), the denominator ($\mathcal{D}$) of the velocity gradient equation (Eq. (\ref{eq:22})) may vanish at one or more locations. Since the flow of matter is assumed to be smooth, the numerator ($\mathcal{N}$) must also vanish at the radial locations where the denominator ($\mathcal{D}$) vanishes. These points are called critical points and denoted by $r_\mathrm{c}$, where `$\mathrm{c}$' stands for critical. Therefore, the critical point conditions are given by:
\begin{equation}
    \label{eq:23}
    \mathcal{N}_\mathrm{c}=\mathcal{D}_\mathrm{c} = 0.
\end{equation}
Using the critical point conditions we obtain the following relations:
\begin{equation}
    \label{eq:24}
    c_\mathrm{sc}^2 = \frac{\left(1 + \dfrac{\Theta_\mathrm{c}}{N_\mathrm{c}H_\mathrm{c}}\dfrac{\partial H}{\partial \Theta}\bigg\lvert_{\mathrm{c}}\right)\left( \dfrac{1}{r^{2-\beta}_\mathrm{c}(r_\mathrm{c}-\rp)^{\beta}} - \dfrac{\lambda^{2}}{r^3_\mathrm{c}}\right)}{\bigg(\dfrac{1}{r_\mathrm{c}}+\dfrac{1}{H_\mathrm{c}}\dfrac{\partial H}{\partial r}\bigg\lvert_{\mathrm{c}}\bigg)}
\end{equation}
and
\begin{equation}
    \label{eq:25}
    \vel_\mathrm{c}^2 = \frac{c_\mathrm{sc}^2}{\left(1 + \dfrac{\Theta_\mathrm{c}}{N_\mathrm{c}H_\mathrm{c}}\dfrac{\partial H}{\partial \Theta}\bigg\lvert_{\mathrm{c}}\right)},
\end{equation}
where each quantity is computed at the critical point.

To solve  Eqs. (\ref{eq:21}) and (\ref{eq:22}) numerically, we start the integration from the critical point ($r_\mathrm{c}$). In order to accomplish the task, we need the slope at the critical point, where $d\vel/dr \rightarrow 0/0$. So, we use the L'Hopital rule to calculate the slope as given below: 
\begin{equation}
    \label{eq:26}
    \frac{\der \vel}{\der r}\bigg\lvert_{\mathrm{c}} = \frac{\dfrac{\der \mathcal{N}}{\der r}\bigg\lvert_{\mathrm{c}}}{\dfrac{\der \mathcal{D}}{\der r}\bigg\lvert_{\mathrm{c}}}.
\end{equation}
In general we could write $\der \mathcal N/\der r \lvert _\mathrm{c}$ in the following way:
\begin{equation}
    \label{eq:27}
    \frac{\der\mathcal{N}}{\der r}\bigg\lvert _\mathrm{c} = \mathcal{N}_{1\mathrm{c}} + \mathcal{N}_{2\mathrm{c}}\frac{\der \vel}{\der r}\bigg\lvert _\mathrm{c},
\end{equation}
where
\[
\begin{aligned}
 \mathcal{N}_{1\mathrm{c}} 
 &= \frac{\partial \mathcal{N}}{\partial r}\bigg\lvert _\mathrm{c} + \Omega_{1\mathrm{c}}\frac{\partial \mathcal{N}}{\partial \Theta}\bigg\lvert _\mathrm{c},\\[6pt]   
 \mathcal{N}_{2\mathrm{c}} 
 &= \Omega_{2\mathrm{c}}\frac{\partial \mathcal{N}}{\partial \Theta}\bigg\lvert _\mathrm{c}. 
\end{aligned}
\]
On the other hand $\der \mathcal D/\der r\lvert _\mathrm{c}$ can be written as:
\begin{equation}
    \label{eq:28}
    \frac{\der \mathcal{D}}{\der r}\bigg\lvert _\mathrm{c} = \mathcal{D}_{1\mathrm{c}} + \mathcal{D}_{2\mathrm{c}}\frac{\der \vel}{\der r}\bigg\lvert _\mathrm{c},
\end{equation}
where
\[
\begin{aligned}
 \mathcal{D}_{1\mathrm{c}} 
 &= \Omega_{1\mathrm{c}} \frac{\partial \mathcal{D}}{\partial \Theta}\bigg\lvert _\mathrm{c},\\[6pt] 
 \mathcal{D}_{2\mathrm{c}} 
 &=  \frac{\partial \mathcal{D}}{\partial \vel}\bigg\lvert _\mathrm{c} + \Omega_{2\mathrm{c}}\frac{\partial \mathcal{D}}{\partial \Theta}\bigg\lvert _\mathrm{c}. 
\end{aligned}
\]
Using Eqs. (\ref{eq:27}) and (\ref{eq:28}) we could rewrite Eq. (\ref{eq:26}) as:
\begin{equation}
    \label{eq:29}
    \mathcal{A}_\mathrm{c}\bigg(\frac{\der \vel}{\der r}\bigg\lvert _{\rm c}\bigg)^2+ \mathcal{B}_\mathrm{c}\bigg(\frac{\der \vel}{\der r}\bigg\lvert _{\rm c}\bigg) + \mathcal{C}_\mathrm{c} = 0,
\end{equation}
where
\[
\begin{aligned}
\mathcal{A}_\mathrm{c} 
&= \mathcal{D}_{2\mathrm{c}},\\[6pt] 
\mathcal{B}_\mathrm{c}
&= \mathcal{D}_{1\mathrm{c}} -\mathcal{N}_{2\mathrm{c}},\\[6pt] 
\mathcal{C}_\mathrm{c} 
&= -\mathcal{N}_{1\mathrm{c}}. 
\end{aligned}
\]
Now, Eq. (\ref{eq:29}) has two roots given by:
\begin{equation}
    \label{eq:30}
   \frac{\der \vel}{\der r}\bigg\lvert_{\mathrm{c}}= \frac{-\mathcal{B}_\mathrm{c} \overset{+}{-} \sqrt{\mathcal{B}^2_\mathrm{c} -4\mathcal{A}_\mathrm{c}\mathcal{C}_\mathrm{c}}}{2\mathcal{A}_\mathrm{c}}.
\end{equation}
Negative value of root corresponds to negative slope at the critical point, which represent accretion type solution. Depending on the boundary conditions, the accretion flow may contain one critical point or three critical points which are physically pertinent .
In next section we elaborate this explicitly. But before that we analyze the accretion flow dynamics of matter for conical flow model ($CF$ model)  for demonstration. Exact formulation and related calculations for vertical equilibrium model ($VE$ model) and constant height model ($CH$ model) are given in the Appendix~\ref{app:eqns}.
\subsection*{Conical flow ($CF$) model}
For conical flow model the disk height is given by:
\begin{equation}
    \label{eq:31}
    H_{CF}(r,\Theta) = \alpha r
\end{equation}
where $\alpha$ is some constant. For this type of flow geometry:
\[\left(\frac{\der \Theta}{\der r}\right)_{CF} = (\Omega_{1})_{CF} + (\Omega_2)_{CF}\left(\dfrac{\der \vel}{\der r}\right)_{CF},\]
where
\[
\begin{aligned}
(\Omega_{1})_{CF} 
&= -\frac{2\Theta}{Nr},\\[6pt] 
(\Omega_{2})_{CF} 
&= -\frac{\Theta}{N\vel}.
\end{aligned}
\]
Therefore, we could write the rate of change in temperature (Eq.(\ref{eq:21})) in the radial direction as follows:
\begin{equation}
    \label{eq:32}
    \left(\frac{\der \Theta}{\der r}\right)_{CF} = -\frac{\Theta}{N}\left(\frac{2}{r} + \frac{1}{\vel}\dfrac{\der \vel}{\der r} \right).
\end{equation}
Radial velocity gradient equation (Eq. (\ref{eq:22})) can be written as:
\begin{equation}
    \label{eq:33}
    \left(\frac{\der \vel}{\der r}\right)_{CF} = \frac{\dfrac{2c_\mathrm{s}^2}{r} + \dfrac{\lambda^{2}}{r^3}- \dfrac{1}{r^{2-\beta}(r-\rp)^{\beta}}}{\vel - \dfrac{c_\mathrm{s}^2}{\vel}} = \frac{\mathcal{N}_{CF}}{\mathcal{D}_{CF}}.
\end{equation}
Using critical point condition (Eq. (\ref{eq:23})), which takes the following form:
\begin{equation}
    \label{eq:34}
    (\mathcal{N}_{CF})_\mathrm{c} = (\mathcal{D}_{CF})_\mathrm{c} = 0,
\end{equation}
we could write the local sound speed and flow velocity at the critical point as:
\begin{equation}
    \label{eq:35}
    (c_\mathrm{sc}^2)_{CF} = \frac{r_\mathrm{c}}{2}\left( \dfrac{1}{r^{2-\beta}_\mathrm{c}(r_\mathrm{c}-\rp)^{\beta}} - \dfrac{\lambda^{2}}{r^3_\mathrm{c}}\right),
\end{equation}
and
\begin{equation}
    \label{eq:36}
    (\vel_\mathrm{c}^2)_{CF} = (c_\mathrm{sc}^2)_{CF}.
\end{equation}
For conical flow model we could write:
\begin{equation}
\label{eq:37}
\begin{aligned}
\frac{\partial \mathcal{N}_{CF}}{\partial r}
&= -\frac{2c_\mathrm{s}^2}{r^2} - \frac{3\lambda^2}{r^4} +F'_{\mathrm{ABN}},\\[6pt]
\frac{\partial \mathcal{N}_{CF}}{\partial \Theta} 
&= \frac{4}{\tau r}\bigg[\Gamma + \Theta \frac{\der \Gamma}{\der \Theta}\bigg],\\[6pt]
\frac{\partial \mathcal{D}_{CF}}{\partial \Theta} 
&= -\frac{2}{\tau \vel}\bigg[\Gamma + \Theta \frac{\der \Gamma}{\der \Theta}\bigg],\\[6pt]
\frac{\partial \mathcal{D}_{CF}}{\partial \vel} 
&= 1 + \frac{c_\mathrm{s}^2}{\vel^2},
\end{aligned}
\end{equation}
where
\[
\begin{aligned}
 F'_{\mathrm{ABN}} = \frac{\partial F_{\mathrm{ABN}}}{\partial r} = \frac{2r - \rp(2 - \beta)}{r^{3-\beta}(r - \rp)^{\beta+1}}.   
\end{aligned}
\]
Now using Eqs. (\ref{eq:35}), (\ref{eq:36}) and (\ref{eq:37}) we can write:
\begin{equation}
\label{eq:38}
 \begin{aligned}
(\mathcal{A}_\mathrm{c})_{CF} 
&= 2 + \frac{1}{N_\mathrm{c}\Gamma_\mathrm{c}}\left(\Gamma_\mathrm{c}+\Theta_\mathrm{c}\frac{\der \Gamma}{\der \Theta}\bigg\lvert _{\rm c}\right),\\[6pt] 
(\mathcal{B}_\mathrm{c})_{CF}
&= \frac{8\Theta_\mathrm{c}}{\vel_\mathrm{c}r_\mathrm{c}N_\mathrm{c}\tau}\left(\Gamma_\mathrm{c}+\Theta_\mathrm{c}\frac{\der \Gamma}{\der \Theta}\bigg\lvert _{\rm c}\right),\\[6pt] 
(\mathcal{C}_\mathrm{c})_{CF} 
&= \frac{2c_\mathrm{sc}^2}{r_\mathrm{c}^2} + \frac{3\lambda^2}{r_\mathrm{c}^4} -F'_{\mathrm{ABN}}\lvert _{\rm c}\\
&+\frac{8\Theta_\mathrm{c}}{r^2_\mathrm{c}N_\mathrm{c}\tau}\left(\Gamma_\mathrm{c}+\Theta_\mathrm{c}\frac{\der \Gamma}{\der \Theta}\bigg\lvert _{\rm c}\right). 
\end{aligned}
\end{equation}
where
\[
\begin{aligned}
    \Theta_{\mathrm{c}} = \frac{\tau r_\mathrm{c}}{4\Gamma_{\mathrm{c}}}\left( \dfrac{1}{r^{2-\beta}_\mathrm{c}(r_\mathrm{c}-\rp)^{\beta}} - \dfrac{\lambda^{2}}{r^3_\mathrm{c}}\right).  
\end{aligned}
\]
Finally, using Eq. (\ref{eq:38}), we can write the slope at the critical point for conical flow ($CF$) model in the following way:
\begin{equation}
    \label{eq:39}
   \left(\frac{\der \vel}{\der r}\right)_{CF}\bigg\lvert_{\mathrm{c}}= \frac{-(\mathcal{B}_\mathrm{c})_{CF} \overset{+}{-} \sqrt{(\mathcal{B}_\mathrm{c})^2_{CF} -4(\mathcal{A}_\mathrm{c})_{CF}(\mathcal{C}_\mathrm{c})_{CF}}}{2(\mathcal{A}_\mathrm{c})_{CF}}.
\end{equation}


\section{Multicritical Parameter Space}
\label{sec:mcpspace}
In section (\ref{sec:modeldes}), we have mentioned that the accretion flow may contain one or three critical points depending on the initial conditions. To solve the global accretion flow we need to specify a set of initial parameters: [$\mathcal{E}$, $\lambda$, $a$, $\xi$]. These initial parameters dictate whether the flow will contain one or three critical points, which are physically relevant in this context. For global accretion solution with three critical points, critical point adjacent to the black hole horizon is called inner critical point ($r_\mathrm{in}$), that formed far away from the black hole horizon is called the outer critical point ($r_\mathrm{out}$) and the one formed in between the inner and outer critical points, is referred to as the middle critical point ($r_\mathrm{mid}$).

\begin{figure*}
    \centering
    \includegraphics[width=1\linewidth]{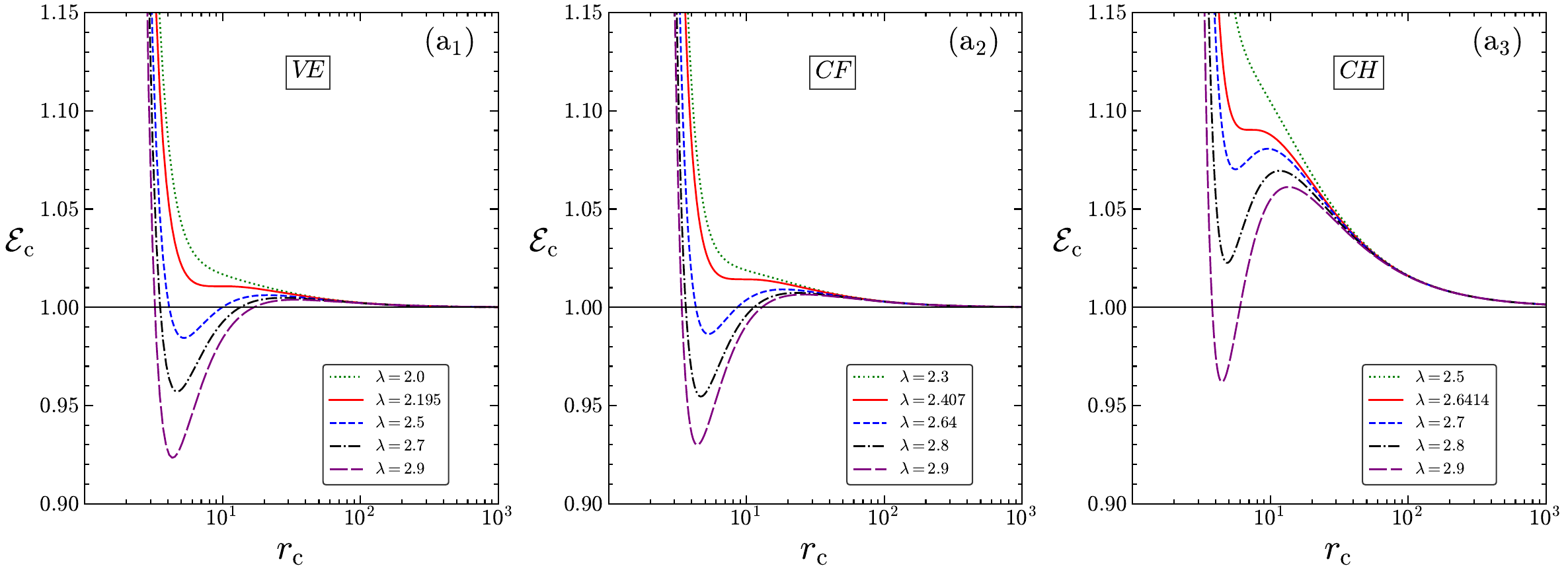}
    \caption{The critical specific energy ($\mathcal{E}_\text{c}$) is plotted against the location of critical points ($r_\text{c}$) for (a$_1$) $a$ = 0.0, (a$_2$) $a$ = 0.5 and (a$_3$) $a$ = 0.95, considering various values of specific angular momentum ($\lambda$). Specific angular momentum corresponding to each curve is as follows: (a$_1$) $\lambda$ = 2.4 (dotted), $\lambda$ = 2.76 (solid), $\lambda$ = 3.0 (dashed), $\lambda$ = 3.3 (dashed dotted) and $\lambda$ = 3.6 (long dashed), (a$_2$) $\lambda$ = 2.0 (dotted), $\lambda$ = 2.195 (solid), $\lambda$ = 2.5 (dashed), $\lambda$ = 2.7 (dashed dotted) and $\lambda$ = 2.9 (long dashed), (a$_3$) $\lambda$ = 1.0 (dotted), $\lambda$ = 1.261 (solid), $\lambda$ = 1.4 (dashed), $\lambda$ = 1.5 (dashed dotted) and $\lambda$ = 1.6 (long dashed).
    The number of intersection points between the horizontal line plotted at $\mathcal{E}_\mathrm{c} = 1.0001$ in each diagram and the curves represents the number of critical points for that particular flow. Curves corresponding to limiting value of specific angular momentum ($\lambda$) are shown as solid lines. In all cases, the composition parameter is set to $\xi = 1.0$.  }
    \label{fig:E_vs_r}
\end{figure*}

Therefore, our goal is to find the number of critical points associated with the accretion flow and to do so we plot how the specific energy of the flow at the critical point ($\mathcal{E}_\mathrm{c}$) varies with the location of critical point ($r_\mathrm{c}$) as well as the energy momentum parameter space ($\mathcal{E}-\lambda$) for different flow geometry. In Fig.~\ref{fig:E_vs_r}(a$_1$) we have plotted $\mathcal{E}_\mathrm{c}$ vs $r_\mathrm{c}$ for vertical equilibrium ($VE$) model with specific angular momentum value: $\lambda$ =2.0, 2.195, 2.5, 2.7 and 2.9. Fig.~\ref{fig:E_vs_r}(a$_2$) shows the variation of $\mathcal{E}_\mathrm{c}$ with $r_\mathrm{c}$ for conical flow ($CF$) model with specific angular momentum value: $\lambda$ =2.3, 2.407, 2.64, 2.8 and 2.9. Whereas Fig.~\ref{fig:E_vs_r}(a$_3$) describes variation of $\mathcal{E}_\mathrm{c}$ with $r_\mathrm{c}$ for constant height ($CH$) model with specific angular momentum value: $\lambda$ =2.5, 2.6414, 2.7, 2.8 and 2.9. In all the above cases we have fixed the black hole spin parameter value at $a$= 0.5 and the composition parameter at $\xi$ = 1.0, i.e. the flow is a pure electron-proton ($e^- - p^+$) flow. To get an idea about the number and location of critical points we draw a horizontal line for a fixed critical specific energy ($\mathcal{E}_\mathrm{c}$) value, which intersects the curve corresponding to fixed specific angular momentum ($\lambda$) at various locations. The  intersecting points represent the location of critical points. In our case for demonstration we have drawn the line for $\mathcal{E}_\mathrm{c}$=1.0001, in Fig.~\ref{fig:E_vs_r}. It is interesting to note that there exists a limiting specific angular momentum value below which the flow contains only a single critical point irrespective of the specific energy of the flow. For each panel in Fig.~\ref{fig:E_vs_r} we depict the curve with solid red line corresponding to this limiting angular momentum value.

\begin{figure*}
    \centering
    \includegraphics[width=1\linewidth]{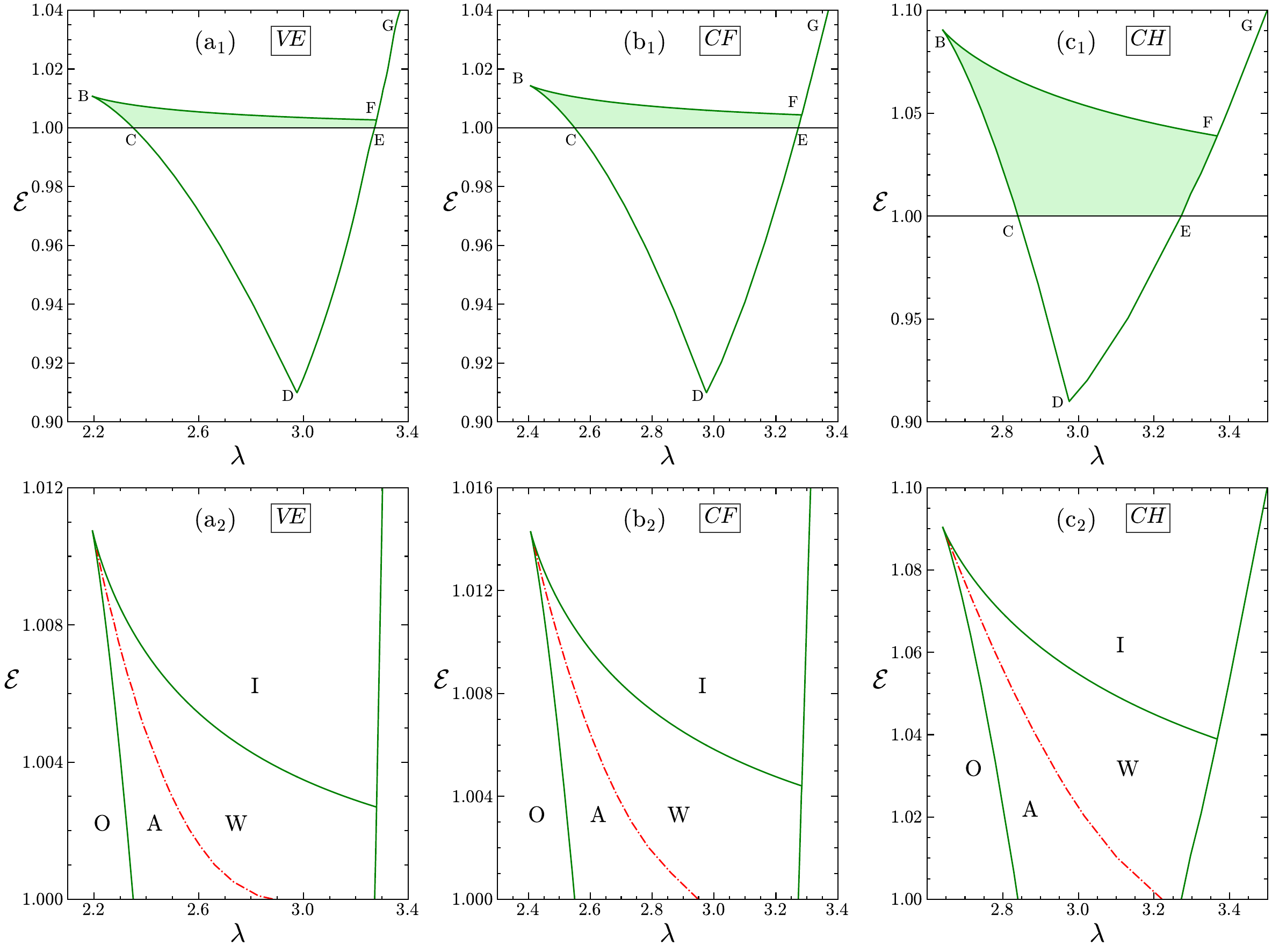}
    \caption{The specific energy--angular momentum ($\mathcal{E}$--$\lambda$) parameter space for the (a$_1$) vertical equilibrium (VE) model, (b$_1$) conical flow (CF) model, and (c$_1$) constant height (CH) model (upper panels), along with the enlarged multi-critical sub-regions (lower panels a$_2$, b$_2$, c$_2$). In the upper panels the green-shaded region BCEFB contains three critical points and admits multi-transonic solutions; the region CDEC contains two critical points but does not support global accretion; the region bounded by DC and DG is either unphysical or contains no critical solution; the remaining region with $\mathcal{E}>1.0$ supports a single critical point. In the lower panels the multi-critical region is further divided into four sub-regions: \textbf{O} (single outer critical point), \textbf{I} (single inner critical point), \textbf{A} (three critical points, $\dot{\mathcal{M}}(r_\mathrm{in}) > \dot{\mathcal{M}}(r_\mathrm{out})$, shock possible for accretion-type solution), and \textbf{W} (three critical points, $\dot{\mathcal{M}}(r_\mathrm{in}) < \dot{\mathcal{M}}(r_\mathrm{out})$, shock possible for wind-type solution). The dashed red line separates the \textbf{NS} (no-shock) and \textbf{S} (shock) sub-regions within \textbf{A}. All panels are computed for spin parameter $a=0.5$ and composition parameter $\xi=1.0$.}
    \label{fig:mcp3model}
\end{figure*}

From the above discussion and from Fig.~\ref{fig:E_vs_r} we get an elementary idea about the critical points. But it is very useful to plot the specific energy-angular momentum ($\mathcal{E} - \lambda$) parameter space for describing the nature of the accretion flow precisely. In the upper panel of Fig.~\ref{fig:mcp3model} the specific energy-momentum ($\mathcal{E}-\lambda$) parameter space is shown for the (a$_1$) vertical equilibrium (VE) model, (a$_2$) conical flow (CF) model and (a$_3$) constant height (CH) model. We can divide  each diagram of the upper panel in four regions: 
\begin{enumerate}
\renewcommand{\labelenumi}{(\Roman{enumi})}
\item Region marked by BCEFB and shaded in green. This region contains three critical points and the flow can be multitransonic in this region.
\item Region bounded by CDEC  contains two critical points. But global accretion flow, which connects the flow from large distance to the black hole horizon is not possible in this zone.
\item Region marked by left part of the curve DC and right part of the curve DG. Either there is no critical solution or the flow is not physical in this region.
\item Remaining part of the diagram i.e. left part of curve EG not shaded in green and for which $\mathcal{E}>1.0$. Here the flow contains single critical point.
\end{enumerate}
See \cite{Fukue1987} for a detailed discussion of the above regions. In the lower panel of Fig.~\ref{fig:mcp3model} we have explicitly plotted the upper region of the ($\mathcal{E}-\lambda$) parameter space for each flow geometry. Each diagram is plotted for  [$a$, $\xi$] = [0.5, 1.0]. Based on the flow topology, the upper region is further divided into four distinct part, labeled as $\boldsymbol{\textbf{O}}$, $\boldsymbol{\textbf{A}}$, $\boldsymbol{\textbf{W}}$, and $\boldsymbol{\textbf{I}}$. Region marked by $\boldsymbol{\textbf{O}}$ or $\boldsymbol{\textbf{I}}$  corresponds to flow with a single critical point. Particularly, the $\boldsymbol{\textbf{O}}$ region exhibits flows with an \textbf{O}uter type critical point ($r_\text{out}$) whereas the $\boldsymbol{\textbf{I}}$ region contains  flows with an \textbf{I}nner critical point ($r_\text{in}$).  
On the other hand, parameter values within the $\boldsymbol{\textbf{A}}$ or $\boldsymbol{\textbf{W}}$ regions produce flows with three critical points: two saddle-type critical points and one center-type critical point. It could be shown that the saddle-type critical points are located close to the black hole horizon (inner critical point, $r_\text{in}$) and far from the horizon (outer critical point, $r_\text{out}$). Between the above two critical points lies the center-type critical point, known as the middle critical point ($r_\text{mid}$). In the $\boldsymbol{\textbf{A}}$ region entropy satisfies:~$\mathcal{\dot{M}}(r_\text{in}) > \mathcal{\dot{M}}(r_\text{out})$, while in $\boldsymbol{\textbf{W}}$:~ $\mathcal{\dot{M}}(r_\text{in}) < \mathcal{\dot{M}}(r_\text{out})$. As a consequence only the $\boldsymbol{\textbf{A}}$ccretion type flow solution that passes through the outer critical point may experience the shock and becomes a multi-transonic flow in the $\boldsymbol{\textbf{A}}$ region. In case of $\boldsymbol{\textbf{W}}$ region only the $\boldsymbol{\textbf{W}}$ind type flow solution may experience the shock and become a multi-transonic flow. Whether the accretion solution will be multi-transonic or not depends on the satisfaction of the Rankine-Hugonoit shock conditions, which we discuss next. The region denoted by $\boldsymbol{\textbf{A}}$ and $\boldsymbol{\textbf{W}}$ as a whole known as the multi-critical parameter space, which is basically a sub-region of the full ($\mathcal{E}- \lambda$) parameter space.

\section{Shocked Flow}
\label{sec:shockflow}
Accretion flow that accommodates three critical points (parameter values chosen from the $\boldsymbol{\textbf{A}}$ region) may encounter shock and become multi-transonic flow. Flow starting subsonically at the outer boundary becomes supersonic crossing the outer sonic point. If the shock conditions are met the flow encounter shock and becomes subsonic again. However, due to  the inner boundary conditions at the horizon, the flow must pass through the inner sonic point and it  becomes supersonic again when it enters the black hole horizon. The flow should obey the Rankine-Hugonoit shock conditions at the shock transition, given as:
\begin{equation}
\label{eq:shock1}
    \dot M_{+} = \dot M_{-} ,    
\end{equation}
\begin{equation}
\label{eq:shock2}
    W_{+} + \Sigma_{+}\vel_{+}^2 = W_{-} + \Sigma_{-}\vel_{-}^2 ,
\end{equation}
and
\begin{equation}
\label{eq:shock3}
    \mathcal{E}_{+} = \mathcal{E}_{-}.
\end{equation}
In the above equations $W = 2Hp$ represents the vertically integrated thermal pressure. Pre-shocked quantities are labeled with subscripts $(-)$ and $(+)$ denotes post-shocked quantities. The above shock equations give  us information about the quantities that are conserved throughout the supersonic and subsonic flow of accreting matter. Now we have to calculate the location of shock ($r_{\mathrm{sh}}$). By defining a shock invariant quantity (I) using the above shock conditions, whose value is same only at the shock location for subsonic and supersonic branches, we can compute the shock location. The shock invariant quantity is given as (see Appendix D of \cite{pal2025effectspindynamicsmulticomponent})
\begin{equation}
\label{eq:shock4}
    I = \frac{[\Theta + \frac{\tau}{2}\vel^2]}{\vel}.
\end{equation}
Now from below condition
\begin{equation}
     \frac{[\Theta_{+} + \frac{\tau}{2}\vel^2_{+}]}{\vel_{+}} = \frac{[\Theta_{-} + \frac{\tau}{2}\vel^2_{-}]}{\vel_{-}},
\end{equation}
we can calculate shock location.

\subsection*{Phase diagram and Shock space}
\begin{figure*}[ht]
    \centering
    \includegraphics[width=1\linewidth]{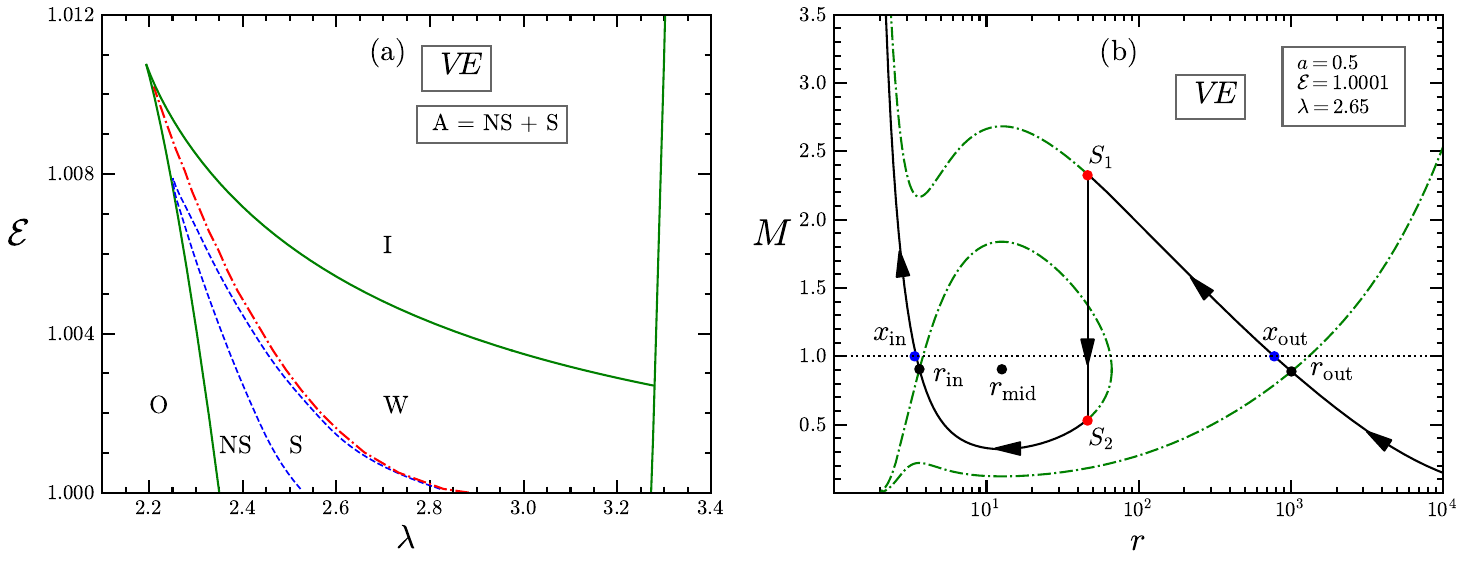}
    \caption{(a) The $\mathcal{E}$--$\lambda$ parameter space for the vertical equilibrium (VE) model showing the shock parameter space. The multi-critical region \textbf{A} is divided into the \textbf{S} (shock) and \textbf{NS} (no-shock) sub-regions, separated by the dashed red curve. (b) Phase portrait (Mach number $M = \vel/c_\mathrm{s}$ versus radial distance $r$) for a representative multi-transonic shocked accretion flow in the VE model with $a=0.5$, $\mathcal{E}=1.0001$, and $\lambda=2.65$. The global shocked solution (solid black curve with arrows) passes through the outer sonic point $x_\mathrm{out}$ and the inner sonic point $x_\mathrm{in}$ (solid blue circles), and undergoes a discontinuous shock transition $\overline{S_1 S_2}$ between the two branches. Open circles mark the locations of the outer ($r_\mathrm{out}$), middle ($r_\mathrm{mid}$), and inner ($r_\mathrm{in}$) critical points. The composition parameter is $\xi=1.0$.}
    \label{fig:VEphasediagram}
\end{figure*}

In previous section we have describe multi-critical parameter space and denoted a region by $\boldsymbol{\textbf{A}}$, which contains three critical points. Now in this region there is a possibility that the accretion flow may pass through both the outer critical point and inner critical point before reaching the black hole horizon, if it encounters shock transition. Fig.~\ref{fig:VEphasediagram}(a) shows multicritical parameter space with shock space for vertical equilibrium flow (VE) model. Here the $\boldsymbol{\textbf{A}}$ region (see Fig.~\ref{fig:mcp3model}) is subdivided into two sub-regions: $\boldsymbol{\textbf{NS}}$ (\textbf{N}o \textbf{S}hock region) and $\boldsymbol{\textbf{S}}$ (\textbf{S}hock region). This $\boldsymbol{\textbf{S}}$ region is known as the shock parameter space. Therefore, if we pick the parameter values from this shock space region the accretion flow will be multi-critical as well as multi-transonic flow. In contrast the  $\boldsymbol{\textbf{NS}}$ region represents multi-critical but mono-transonic accretion flow.

\begin{figure*}[ht]
    \centering
    \includegraphics[width=1\linewidth]{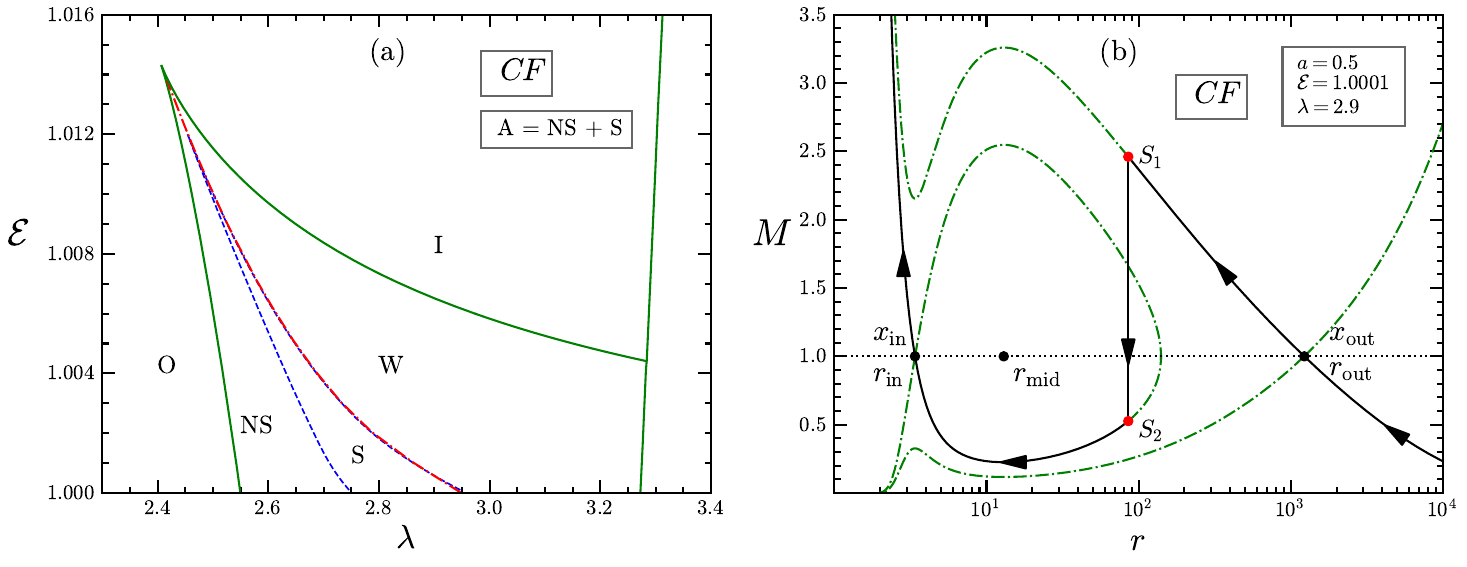}
    \caption{Same as Fig.~\ref{fig:VEphasediagram} but for the conical flow (CF) model. The shock parameter space (a) and the phase portrait (b) are shown for $a=0.5$, $\mathcal{E}=1.0001$, and $\lambda=2.9$. In the CF model the sonic points and critical points coincide, i.e.\ $x_\mathrm{in}=r_\mathrm{in}$ and $x_\mathrm{out}=r_\mathrm{out}$. All other labels are as defined in Fig.~\ref{fig:VEphasediagram}.}
    \label{fig:CFphasediagram}
\end{figure*}

\begin{figure*}[ht]
    \centering
    \includegraphics[width=1\linewidth]{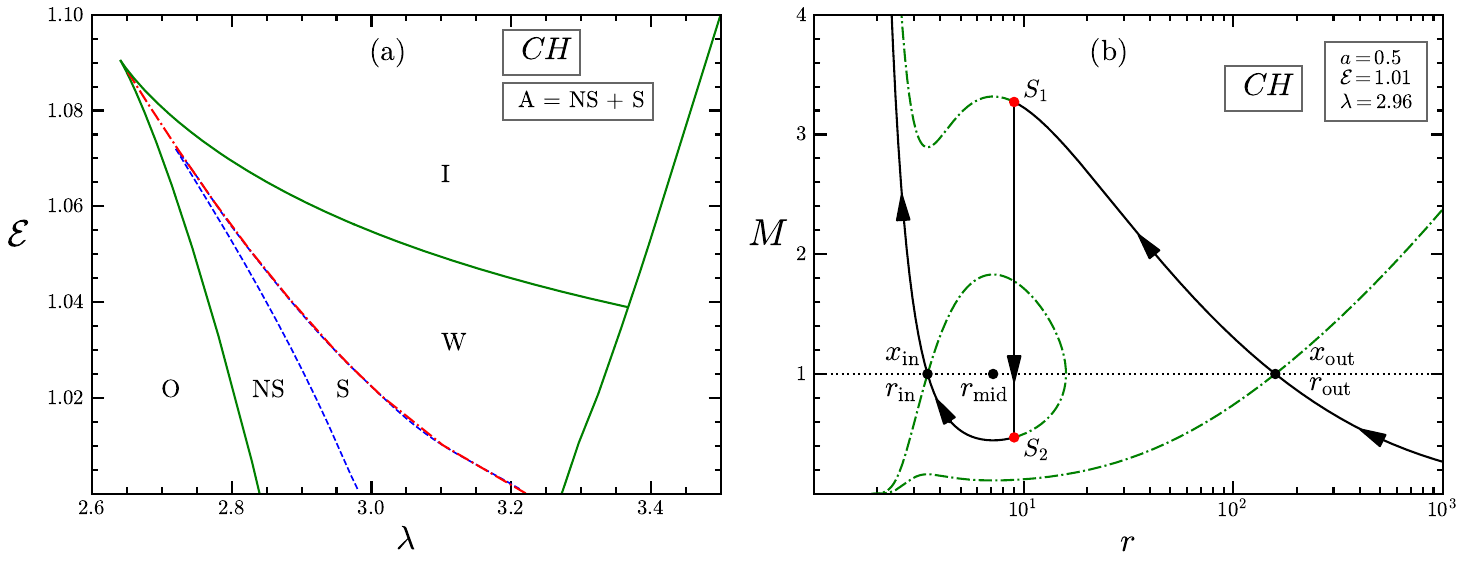}
    \caption{Same as Fig.~\ref{fig:VEphasediagram} but for the constant height (CH) model, with $a=0.5$, $\mathcal{E}=1.01$, and $\lambda=2.96$. As in the CF model, the sonic points and critical points coincide. All other labels are as defined in Fig.~\ref{fig:VEphasediagram}.}
    \label{fig:CHphasediagram}
\end{figure*}

\begin{figure}
    \centering
    \includegraphics[width=1\linewidth]{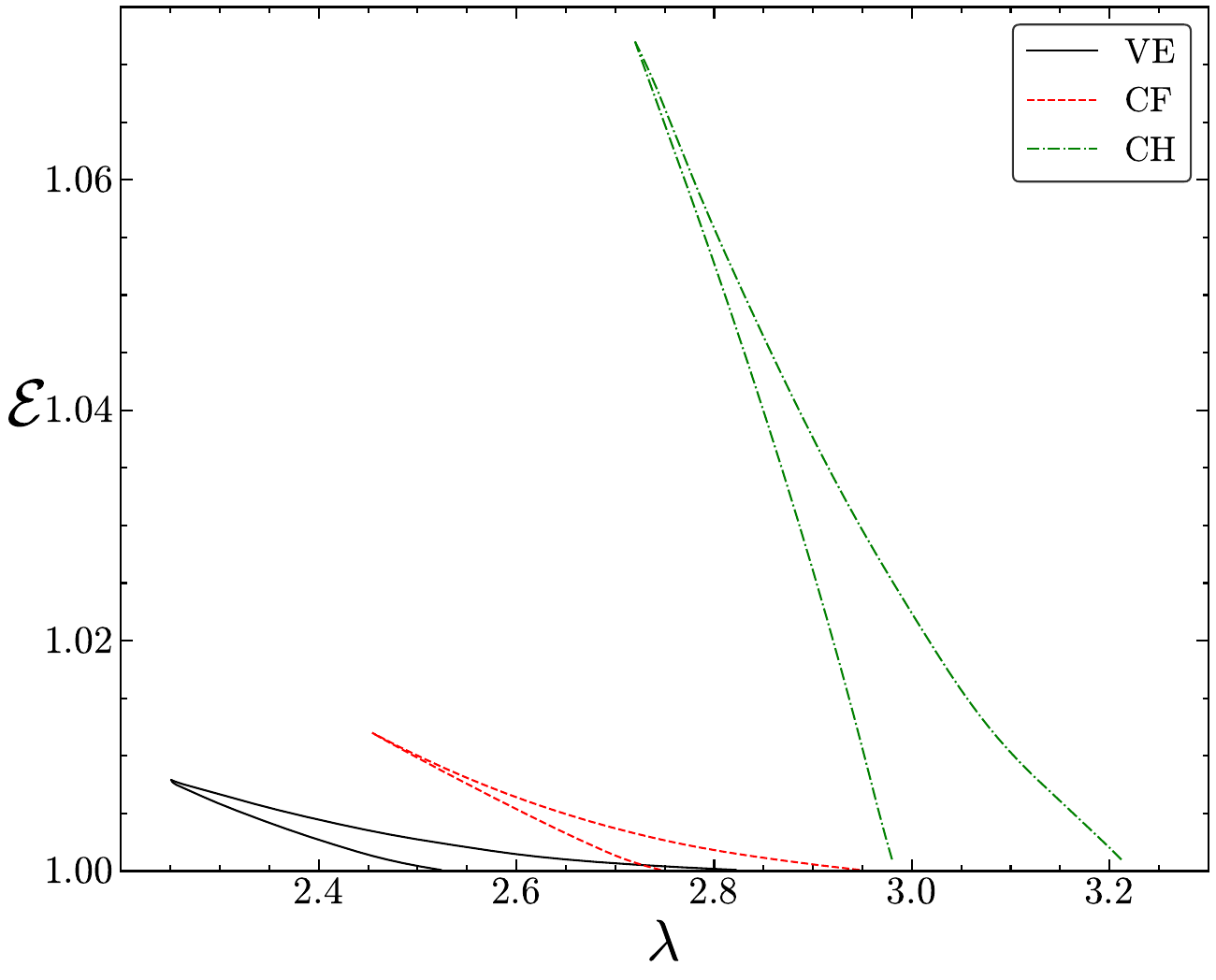}
    \caption{Comparison of the shock parameter spaces for the three disc geometry models: vertical equilibrium (VE, solid black boundary), conical flow (CF, dashed red boundary), and constant height (CH, dash-dotted green boundary). The region enclosed by each curve represents the locus of $(\lambda, \mathcal{E})$ values for which the accretion flow develops a standing shock. All curves are computed for spin parameter $a=0.5$ and composition parameter $\xi=1.0$.}
    \label{fig:Shockspace3models}
\end{figure}

In Fig.~\ref{fig:VEphasediagram}(b) we have shown the phase diagram for multi-transonic shocked accretion flow when the flow geometry is in hydrostatic equilibrium in the vertical direction. Phase diagram, which is the Mach number ($ M = \frac{\vel}{c_\mathrm{s}}$) vs radial distance (r) plot, is very useful in describing this type of accretion flow. In this diagram the solid black line with arrow is the multi-transonic shocked global accretion flow. Here, we have shown the location of critical points with solid black circle, where $r_\mathrm{out}$, $r_\mathrm{mid}$ and $r_\mathrm{in}$ are the outer, middle and inner critical point respectively. Solid blue circles in the diagram represent the location of sonic points, $x_\mathrm{out}$ being the outer sonic point and $x_\mathrm{in}$ being the inner sonic point. The discontinuous jump, indicated by the vertical line $\overline{S_1S_2}$ is the shock jump. Therefore, it could be seen very easily from the diagram that the accretion flow passes through more than one critical and sonic point, as well as it encounters shock. So, this flow is a multi-critical, multi-transonic, shocked accretion flow.
Fig.~\ref{fig:CFphasediagram} describe the shock space and phase diagram for conical flow (CF) model and Fig.~\ref{fig:CHphasediagram} describe the similar things for constant height (CH) model. One thing to note here is that the locations of sonic points ($x_\mathrm{in}$ and $x_\mathrm{out}$) and critical points ($r_\mathrm{in}$ and $r_\mathrm{out}$) are different in the case of VE model. In comparison, the location of the sonic points ($x_\mathrm{in}$ and $x_\mathrm{in}$) and critical points ($r_\mathrm{in}$ and $r_\mathrm{out}$) are the same in case for the CF and CH model.
Next, in Fig.~\ref{fig:Shockspace3models} we have compared the shock parameter space explicitly for the three flow geometry. The region bounded by the solid black line represents the shock space for the VE model, region bounded by the dashed red line shows shock space for CF model and region enclosed by the dash dotted green line depicts the shock space for CH model. For all the flow models shown in Fig.~\ref{fig:Shockspace3models}, the chosen value of spin parameter is $a = 0.5$ and composition parameter is $\xi = 1.0$.


\subsection*{Effect of black hole spin}
 In case of accretion  of low angular momentum, multi-transonic flow around a rotating black hole, formation of shock is essential. In one hand, shock formation has immense effects on the dynamical and thermodynamical variables related to the flow. These variables could change drastically when it encounters shock. Various interesting and important phenomena related to accretion physics, e.g. jet formation, quasi-periodic oscillation, could be explained by invoking the shock formation. On the other hand, various parameters related to the accretion flow or related to space-time geometry have stupendous effects on the formation of shock or physical quantities related to shock. Here we investigate the effect of black hole spin ($a$) on shock location and other quantities related to shock in details.

\begin{figure*}
    \centering
    \includegraphics[width=1\linewidth]{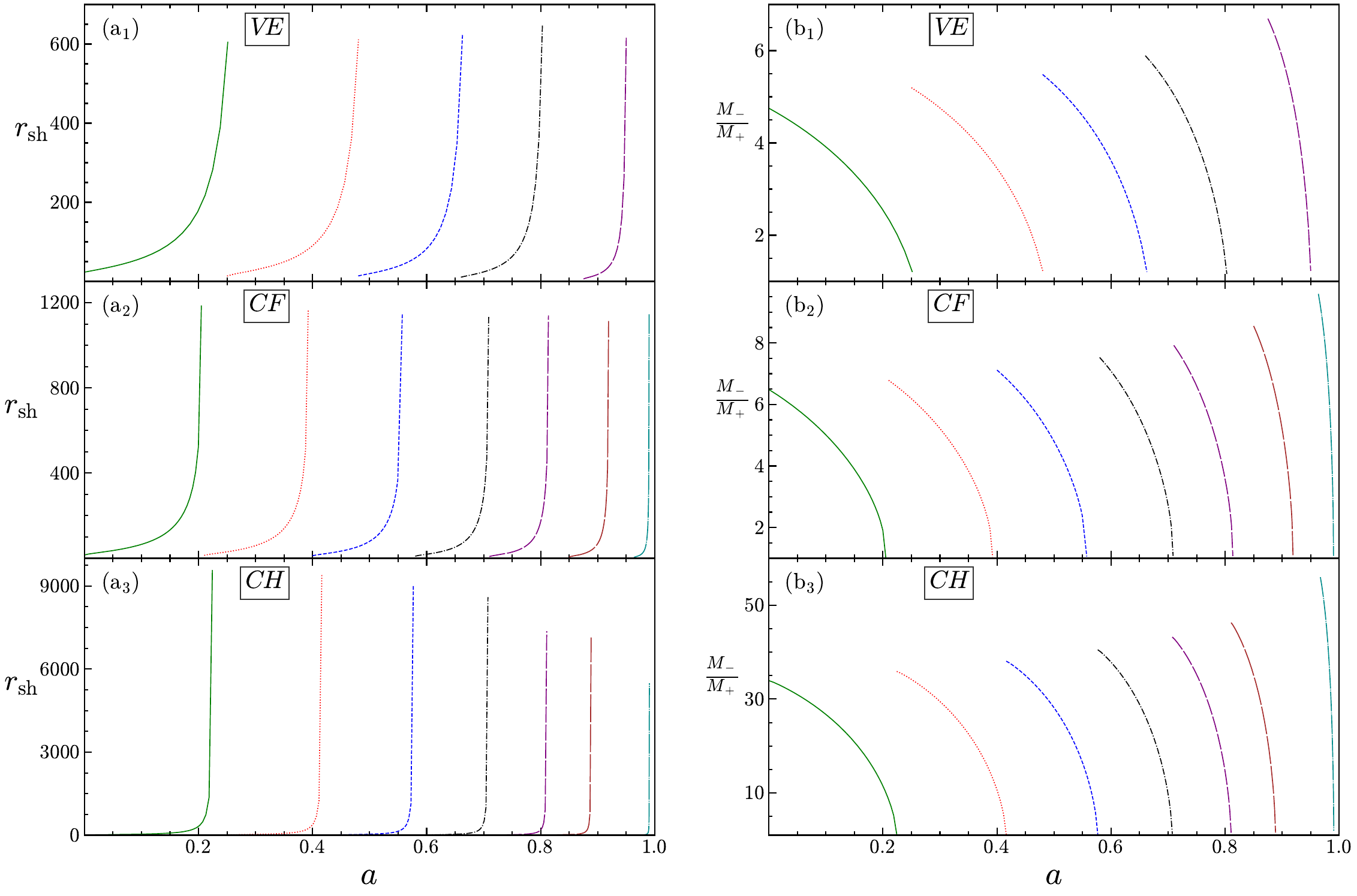}
    \caption{Left panels (a$_1$--a$_3$): Variation of the shock location $r_\mathrm{sh}$ with the black hole spin parameter $a$ for the (a$_1$) VE, (a$_2$) CF, and (a$_3$) CH disc models. Right panels (b$_1$--b$_3$): Corresponding variation of the shock strength, measured as the ratio of pre-shock to post-shock Mach number ($M_-/M_+$), with $a$ for the same three models. In each panel, different curves correspond to different values of specific angular momentum $\lambda$ (increasing from left to right as listed in the text). All panels are computed with $\mathcal{E}=1.0001$ and $\xi=1.0$. For a fixed $\lambda$, increasing black hole spin shifts the shock to larger radii due to enhanced frame-dragging (centrifugal support), while the shock strength decreases monotonically with $a$.}
    \label{fig:shockLocation}
\end{figure*}

 In Fig.~\ref{fig:shockLocation}, the left panel describes the variation of shock location ($r_\mathrm{sh}$) with spin parameter ($a$) for (a$_1$) VE model, (a$_2$) CF model and (a$_3$) CH model. In right panel we have plotted the change of shock strength, defined as the ratio of pre-shock to post-shock Mach number ($M_-/M_+$) with spin parameter ($a$) for (b$_1$) VE model, (b$_2$) CF model and (b$_3$) CH model. In each diagram the specific energy value is fixed at, $\mathcal{E} = 1.0001$ and composition parameter value, $\xi = 1.0$. For vertical equilibrium model (VE), the value of specific angular momentum for each curve from left to right is: $\lambda$ = 3.15, 2.85, 2.56, 2.275 and 1.8 respectively. In case of conical flow model (CF), specific angular momentum value for each curve from left to right is: $\lambda$ = 3.3979, 3.1479, 2.8946, 2.6165, 2.3766, 2.0437 and 1.5945 respectively. Finally for constant height model (CH), specific angular momentum value for each curve from left to right is: $\lambda$ = 3.68019, 3.39404, 3.11792, 2.85169, 2.5959, 2.3478 and 1.74 respectively. As could be seen from the diagram, for a fixed specific angular momentum if we increase the spin value, the shock forms at larger distance from the horizon. This happens because the increases in spin value, increases the outward centrifugal force at each radial distance and as a result flow halts due to shock at larger distance. For all the models the shock strength decreases with the increase in spin value and is inversely proportional to the shock location ($r_\mathrm{sh}$).

\begin{figure*}
    \centering
    \includegraphics[width=1\linewidth]{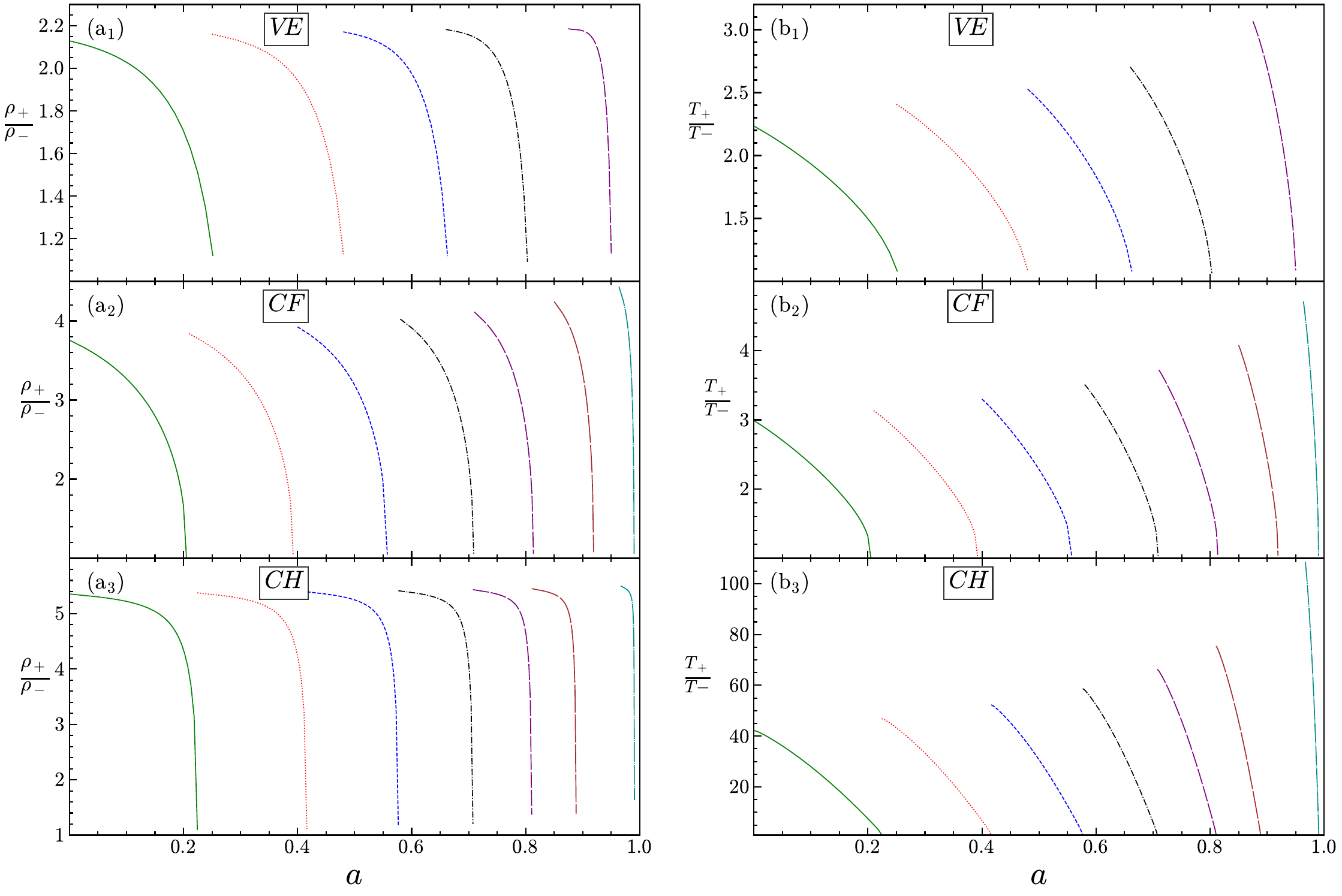}
    \caption{Left panels (a$_1$--a$_3$): Variation of the post-shock to pre-shock density (compression) ratio $\rho_+/\rho_-$ with the black hole spin parameter $a$ for the (a$_1$) VE, (a$_2$) CF, and (a$_3$) CH disc models. Right panels (b$_1$--b$_3$): Corresponding variation of the post-shock to pre-shock temperature ratio $T_+/T_-$ with $a$ for the same three models. Different curves within each panel correspond to different values of specific angular momentum $\lambda$ (same set as in Fig.~\ref{fig:shockLocation}). All panels are computed with $\mathcal{E}=1.0001$ and $\xi=1.0$. Both the compression ratio and the temperature ratio decrease with increasing spin for all three disc geometries, consistent with the outward shift and weakening of the shock at higher values of $a$.}
    \label{fig:CompressionRatio}
\end{figure*}

 In Fig.~\ref{fig:CompressionRatio}, the left panel depicts the compression ratio, defined as the ratio of the post-shock to pre-shock flow density ($\rho_+/\rho_-$) with spin value ($a$) for (a$_1$) VE model, (a$_2$) CF model and (a$_3$) CH model. The right panel shows the variation of post-shock to pre-shock temperature ratio ($T_+/T_-$) with spin parameter for (b$_1$) VE model, (b$_2$) CF model and (b$_3$) CH model. In each diagram the specific energy value is fixed at, $\mathcal{E} = 1.0001$ and composition parameter is $\xi = 1.0$. For all the flow models the value of the specific angular momentum is same as before. The compression ratio and the temperature ratio for all the models varies inversely with the spin value.


\section{Nature of The Critical Points}
\label{sec:nature}
It is evident from the phase portraits that when the parameters of the accretion flow allows for multiple (three) critical points no flow line  actually passes through the critical point at the center. This is the classic case of dynamical system analogues where two `saddle' type critical points are separated by a `centre' type critical point. It is possible to analytically explore the nature of the sonic points by a treatment analogous to the one used in dynamical systems \cite{}. In order to do so, let us consider the Eq. (\ref{eq:22}) and write it as:
\begin{equation}
    \label{eq:dyn}
    \frac{\der \vel}{\der r} = \frac{\mathcal{N}(\Theta, r)}{\mathcal{D}(\Theta, \vel)},
\end{equation}
and linearize both the numerator and the denominator around the critical point.
\begin{align}
    \mathcal{N}(\Theta, r) & \sim  \left. \mathcal{N} \right|_c + \left.\frac{\partial \mathcal{N}}{\partial \Theta} \right |_c \der\Theta + \left.\frac{\partial \mathcal{N}}{\partial r} \right |_c \der r \\ \notag
    \mathcal{D}(\Theta, r) & \sim  \left. \mathcal{D} \right|_c + \left.\frac{\partial \mathcal{D}}{\partial \Theta} \right |_c \der\Theta + \left.\frac{\partial \mathcal{D}}{\partial r} \right |_c \der r.
\end{align}
However, $\mathcal{N}\rvert _\mathrm{c} = \mathcal{D}\rvert _\mathrm{c} = 0$, and $\der \Theta$ can be connected to $\der\vel$ and $\der r$ through eq. (\ref{eq:21}). Consequently, one can write:
\begin{equation}
    \label{eq:dyn2}
    \frac{\der\vel}{\der r}=\frac{\left.\frac{\partial \mathcal{N}}{\partial \vel} \right |_c d\vel + \left.\frac{\partial \mathcal{N}}{\partial r} \right |_c dr}{\left.\frac{\partial \mathcal{D}}{\partial \vel} \right |_c d\vel + \left.\frac{\partial \mathcal{D}}{\partial r} \right |_c dr}.
\end{equation}
Equation~(\ref{eq:dyn2}) may now be rearranged to obtain a closed set of
linearized equations describing the evolution of perturbations in the
$(r,\vel)$ phase space. Writing
\begin{align}
\left.\frac{\partial \mathcal{N}}{\partial r}\right|_c &\equiv \lambda_{11},
&
\left.\frac{\partial \mathcal{N}}{\partial \vel}\right|_c &\equiv \lambda_{12}, \notag\\
\left.\frac{\partial \mathcal{D}}{\partial r}\right|_c &\equiv \lambda_{21},
&
\left.\frac{\partial \mathcal{D}}{\partial \vel}\right|_c &\equiv \lambda_{22},
\end{align}
Eq.~(\ref{eq:dyn2}) can be written as
\begin{equation}
\left( \lambda_{22}\, d\vel + \lambda_{21}\, dr \right)\frac{d\vel}{dr}
=
\lambda_{12}\, d\vel + \lambda_{11}\, dr .
\label{eq:dyn3}
\end{equation}

Introducing an auxiliary parameter $\tau$ to recast the system into an
autonomous form, we write
\begin{align}
\label{eq:auto1}
\frac{d\vel}{d\tau} & = \lambda_{12}\, \delta\vel + \lambda_{11}\, \delta r,\\
\label{eq:auto2}
\frac{dr}{d\tau} & = \lambda_{22}\, \delta\vel + \lambda_{21}\, \delta r.
\end{align}

Equations~(\ref{eq:auto1})–(\ref{eq:auto2}) may be expressed in matrix form as
\begin{equation}
\frac{d}{d\tau}
\begin{pmatrix}
\delta\vel \\ \delta r
\end{pmatrix}
=
\mathbf{J}
\begin{pmatrix}
\delta \vel \\ \delta r
\end{pmatrix},
\label{eq:matrix}
\end{equation}
where the Jacobian matrix evaluated at the critical point is
\begin{equation}
\mathbf{J} =
\begin{pmatrix}
\lambda_{12} & \lambda_{11} \\
\lambda_{22} & \lambda_{21}
\end{pmatrix}.
\label{eq:jacobian}
\end{equation}

The corresponding eigenvalue equation is
\begin{equation}
\mu^2 - (\mathrm{Tr}\,\mathbf{J})\,\mu + \det\mathbf{J} = 0 ,
\label{eq:eigen}
\end{equation}
with
\begin{align}
\mathrm{Tr}\,\mathbf{J} &= \lambda_{21} + \lambda_{12},
\label{eq:trace}\\
\det\mathbf{J} &= \lambda_{21}\lambda_{12}
- \lambda_{22}\lambda_{11}.
\label{eq:det}
\end{align}

The nature of the critical point is determined entirely by the signs of
$\det\mathbf{J}$ and $\mathrm{Tr}\,\mathbf{J}$, along with the discriminant
\begin{equation}
\Delta = (\mathrm{Tr}\,\mathbf{J})^2 - 4\,\det\mathbf{J}.
\label{eq:disc}
\end{equation}

In particular, $\det\mathbf{J}<0$ implies real eigenvalues of opposite signs,
corresponding to a saddle-type critical point, which alone permits physically
acceptable transonic solutions. When $\det\mathbf{J}>0$ and
$\mathrm{Tr}\,\mathbf{J}=0$, the eigenvalues are purely imaginary, yielding a
center-type critical point around which closed orbits exist in phase space. Jus to demonstrate, let us plot the three cases for the Conical Disk. Similar conditions hold for the other disk models.
\begin{figure}[H]
    \centering
    \includegraphics[width=.4\columnwidth]{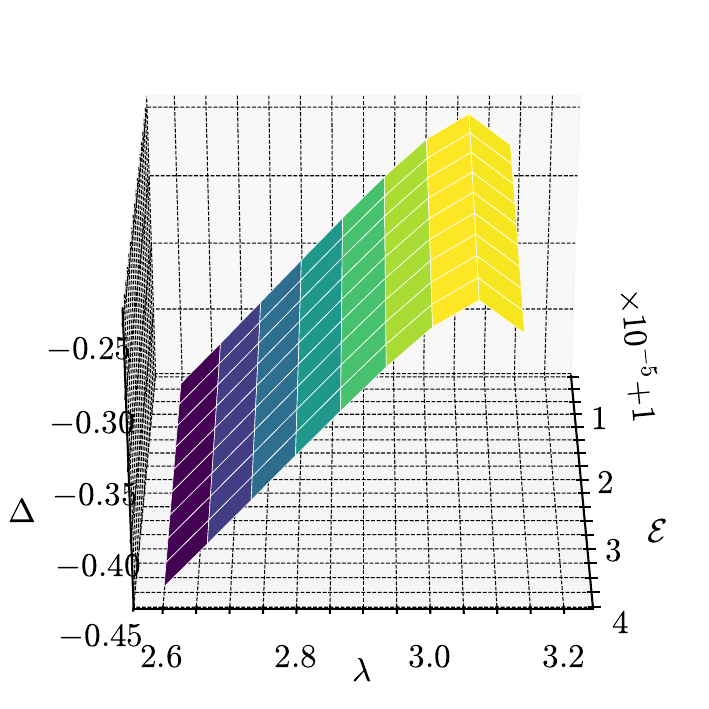}\\ \vspace{-0cm}
    \includegraphics[width=.4\columnwidth]{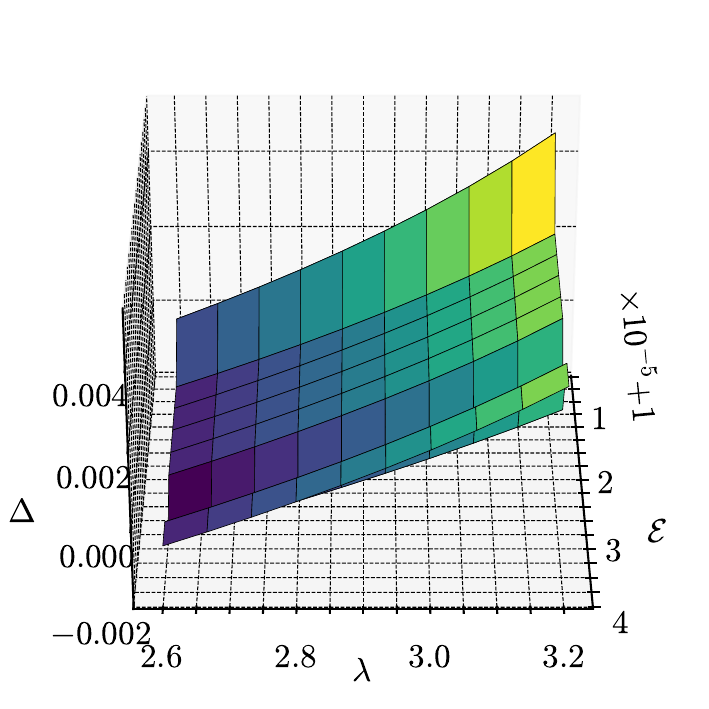}\\\vspace{-0cm}
    \includegraphics[width=.4\columnwidth]{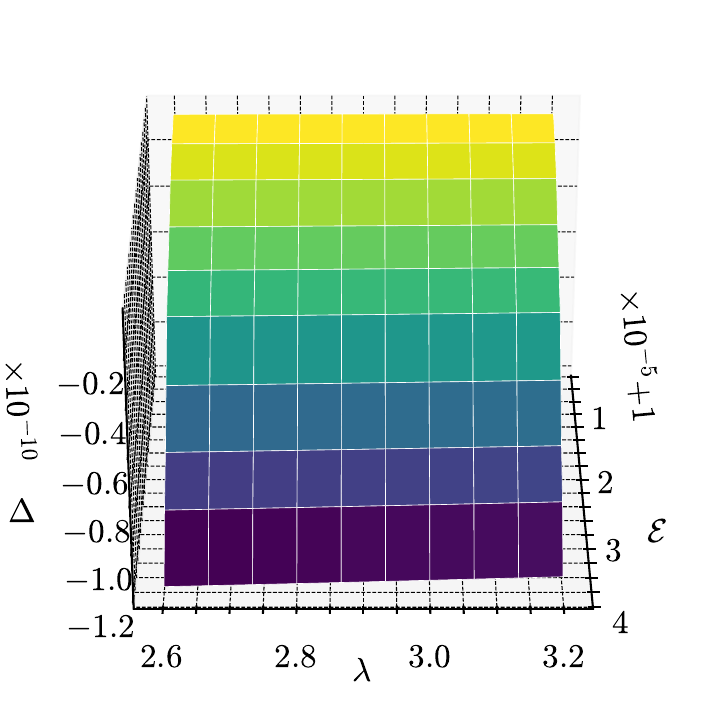}
    \caption{
    (\textbf{a}) $\det\mathbf{J}<0$ for the inner sonic point (saddle).
    (\textbf{b}) $\det\mathbf{J}>0 $for the sonic point in the middle (centre).
    (\textbf{c}) $\det\mathbf{J}<0$ for the outer sonic point (saddle).
    }
    \label{fig:three_panel_vertical}
\end{figure}


\section{Stability of the Stationary Solution}
To study the stability of the steady-state solutions, we examine the time evolution of linear perturbations imposed on the equations governing the fluid flow. We consider the stationary solutions stable if the perturbations remain bounded and do not diverge over time.

We start with the continuity and the Euler equation for an irrotational fluid. The general form of these two equations for the system is, 

\begin{equation}
\partial_t \rho + \nabla \cdot (\rho \mathbf{u}) = 0
\label{connuu}
\end{equation}
and
\begin{equation}
\frac{\partial \mathbf{u}}{\partial t} + \nabla \left( \frac{u^2}{2} \right) + \nabla h + \nabla \phi = 0,
\label{eluu}
\end{equation}
respectively.

Here $\phi$ is the external potential (e.g., gravitational potential), and $h$ is the specific enthalpy.

 \subsection{Linear Perturbation Analysis}
For a steady-state background flow, the fluid variables can be decomposed into a stationary background flow, added with a small time-dependent perturbation part.

Further analysis proceeds by combining the linearized continuity and  Euler equations to obtain a second-order wave equation for the perturbation variable, which determines the flow's stability properties.

Since the flow is irrotational, we can define a velocity potential $\psi$ such that
\begin{equation}
\mathbf{u} = - \nabla \psi.
\label{gru}
\end{equation}

Substituting this into the Euler equation and integrating, we obtain
\begin{equation}
- \frac{\partial \psi}{\partial t} + \frac{1}{2} (\nabla \psi)^2 + h + \phi = 0,
\label{bern}
\end{equation}
which is a form of Bernoulli’s equation.

\noindent
Note that any arbitrary function of time arising from the integration can be absorbed into the definition of $\psi$, since the velocity potential is not unique.

\medskip

We consider the exact stationary background solution of the fluid equations:
\[
\rho_0(\mathbf{r}), \quad \mathbf{u}_0(\mathbf{r}), \quad \psi_0(\mathbf{r}).
\]

We introduce small time-dependent perturbations for all the fluid variables around this background as:
\begin{align}
\label{yiyut}
\rho(\mathbf{r}, t) &= \rho_0(\mathbf{r}) + \epsilon \, \tilde{\rho}(\mathbf{r}, t), \\
p(\mathbf{r}, t) &= p_0(\mathbf{r}) + \epsilon \, \tilde{p}(\mathbf{r}, t), \\
\psi(\mathbf{r}, t) &= \psi_0(\mathbf{r}) + \epsilon \, \tilde{\psi}(\mathbf{r}, t), \\
\mathbf{u}(\mathbf{r}, t) &= \mathbf{u}_0(\mathbf{r}) + \epsilon \, \tilde{\mathbf{u}}(\mathbf{r}, t),
\end{align}
where $\epsilon \ll 1$ is a small parameter.

\medskip

Substituting these perturbed quantities into the continuity and Euler equations and retaining only terms linear in $\epsilon$, we obtain the linearised equations governing the evolution of perturbations.

\noindent
where the overhead tilde denotes the perturbed quantity and
$0 < \epsilon \ll 1$. Quantities with subscript $0$ correspond
to stationary (background) values.

Substituting the perturbed variables into the equations of motion 
[Eqs. (43) and (45)], and retaining terms only up to first order in 
$\epsilon$, we obtain the linearised equations governing the perturbations.

\medskip

From the continuity equation, the linearised form becomes
\begin{equation}
\frac{\partial \tilde{\rho}}{\partial t}
+ \nabla \cdot (\rho_0 \tilde{\mathbf{u}} + \tilde{\rho} \mathbf{u}_0) = 0.
\label{hiui}
\end{equation}

Since $\mathbf{u} = -\nabla \psi$, the perturbation in velocity is
\begin{equation}
\tilde{\mathbf{u}} = - \nabla \tilde{\psi}.
\label{uguu}
\end{equation}

Thus, the continuity equation reduces to
\begin{equation}
\frac{\partial \tilde{\rho}}{\partial t}
- \nabla \cdot (\rho_0 \nabla \tilde{\psi})
+ \nabla \cdot (\tilde{\rho} \mathbf{u}_0) = 0.
\label{lin_cont}
\end{equation}

\medskip

From the Bernoulli equation \eqref{bern}, linearization gives
\begin{equation}
- \frac{\partial \tilde{\psi}}{\partial t}
+ \mathbf{u}_0 \cdot \nabla \tilde{\psi}
+ \frac{c_s^2}{\rho_0} \tilde{\rho} = 0,
\label{lin_bern}
\end{equation}
where we have used
\begin{equation}
\tilde{h} = \frac{c_s^2}{\rho_0} \tilde{\rho}, 
\quad
c_s^2 = \left( \frac{\partial p}{\partial \rho} \right)_0.
\end{equation}
These linearised equations (\ref{lin_cont}, \ref{lin_bern}), when combined, yield a second-order wave equation for the perturbation in the velocity potential $\tilde{\psi}$, which forms the basis of the analogue gravity description.

\medskip

Eliminating $\tilde{\rho}$ between Eqs. \eqref{lin_cont} and \eqref{lin_bern}, we obtain the cherished differential equation governing the propagation 
of $\tilde{\psi}(\mathbf{r}, t)$:


\begin{equation}
\begin{split}
&\partial_t \left( \frac{\rho_0}{c_s^2}
\left[ \partial_t \tilde{\psi} + \mathbf{u}_0 \cdot \nabla \tilde{\psi} \right] \right)
\\
&\quad - \nabla \cdot \left( \rho_0 \nabla \tilde{\psi}
- \frac{\rho_0}{c_s^2} \mathbf{u}_0
\left[ \partial_t \tilde{\psi} + \mathbf{u}_0 \cdot \nabla \tilde{\psi} \right] \right)
= 0.
\end{split}
\label{uyjuu}
\end{equation}

\medskip

This equation can be written in a compact covariant form as
\begin{equation}
\partial_\mu \left( f^{\mu\nu} \partial_\nu \tilde{\psi} \right) = 0,
\label{covu}
\end{equation}
where $f^{\mu\nu}$ depends on the background flow variables.

\begin{equation}
f^{\mu\nu}= \frac{\rho_0}{c_{s0}^2}
\begin{pmatrix}
1 & u_0 ^1 & u_0^2 & u_0^3 \\
u_0^1 & u_0^1u_0^1 - c_{s0}^2 & u_0^1 u_0^2 & u_0^1 u_0^3 \\
u_0^2 & u_0^2 u_0^1 & u_0^2u_0^2 - c_{s0}^2 & u_0^2 u_0^3 \\
u_0^3 & u_0^3 u_0^1 & u_0^3 u_0^2 & u_0^3u_0^3 - c_{s0}^2
\end{pmatrix}
    \label{metu}
\end{equation} 

here $(u_0^1, u_0^2, u_0^3)$ are the  three components of the background 
velocity $\mathbf{u}_0$.
\medskip

\subsection{Analogue Gravity Metric}
The general propagation equation for a massless scalar field  is given by:
\begin{equation}
    \label{meu}
\partial_\mu \left( \sqrt{-g}\, g^{\mu\nu} \partial_\nu \phi \right) = 0
\end{equation}
Comparing \eqref{covu} with \eqref{meu} we get,

\begin{equation}
f^{\mu\nu} = \sqrt{-g}\, g^{\mu\nu}
\label{eq:fmunu_relation}
\end{equation}
with $g= det(g_{\mu\nu})$. A simple algebraic manipulation leads us to the relation
\begin{equation}
\det\left(f^{\mu\nu}\right) = g
\label{eq:det_relation}
\end{equation}
 Exploiting \eqref{metu} and \eqref{eq:det_relation}, we found the metric structure that governs the propagation of the linear perturbation of the fluid variable (velocity potential in this case).  
 
\begin{equation}
\label{tutu}
g_{\mu\nu} =-
\frac{\rho_0 }{c_{s0}}
\begin{pmatrix}
-(c_{s0}^2 - u_0^2) & -u_0^1 & -u_0^2 & -u_0^3 \\
- u_0^1 & 1 & 0 & 0 \\
- u_0^2 & 0 & 1 & 0 \\
- u_0^3 & 0 & 0 & 1
\end{pmatrix}
\end{equation}
\label{ssec:analogue}
The corresponding line element is given by
\begin{equation}
ds^2 =-
\frac{\rho_0 }{c_{s0}}
\left[
- c_{s0}^2 (dx^0)^2
+ \delta_{ij}(dx^i - u^i dx^0)(dx^j - u^j dx^0)
\right]
\label{rutuu}
\end{equation}

The resulting line element bears a close resemblance to the Schwarzschild metric expressed in Painlevé–Gullstrand coordinates and is characterised by a non-zero curvature. However, it is important to emphasize that \eqref{covu} governs only the acoustic perturbations. Consequently, this emergent curved spacetime is perceived exclusively by the perturbations, while the underlying background spacetime remains fundamentally Newtonian.

The appearance of such an effective geometry—commonly referred to as the acoustic metric—within an inviscid, irrotational, classical fluid was first recognised in the pioneering work of \cite{Unruh1981}, and subsequently formalised in a broader framework by \cite{Visser1998}. Systems described within this paradigm are collectively known as analogue gravity models (see \cite{Barcelo2011} for a comprehensive review).

Astrophysical accretion flows provide a particularly intriguing realization of analog gravity (see \cite{Das2004}), as they can simultaneously accommodate both a gravitational event horizon and an acoustic horizon, especially when the accretor is a black hole. Once such an analogy is established, one can meaningfully apply techniques developed in general relativity to analyze the causal structure of the emergent spacetime.

While the analogy between the sound horizon and a black hole event horizon has been noted earlier, it is of particular interest to demonstrate that the sound horizon indeed corresponds to a null surface in the causal structure of the acoustic spacetime. Given a steady-state flow solution—supplemented, for instance, by a phase portrait—the associated acoustic metric can be constructed at every spatial point, thereby enabling a detailed causal analysis.
\subsection*{Perturbation in Axisymmetric Flow}

\subsection*{Governing Equations}
When we consider the dominance of radial velocity $v_{r}$ in the disc accretion around the BH and neglect the other velocity components, we find that the mass accretion rate $f =\rho vrH$ is conserved for the steady state.  The governing fluid equations then take the form :

\textbf{Continuity Equation:}

\begin{equation}
\frac{\partial \rho}{\partial t}
+ \frac{1}{r}\frac{\partial}{\partial r}(\rho u r) = 0
\label{conu}
\end{equation}

\textbf{Euler Equation:}
\begin{equation}
\frac{\partial u}{\partial t}
+ u\frac{\partial u}{\partial r}
+ \frac{\partial \phi}{\partial r}
+ \frac{1}{\rho}\frac{\partial h}{\partial r}
- \frac{\lambda^2}{r^3} = 0
\label{euu}
\end{equation}

\textbf{Mass Accretion Rate:}
\begin{equation}
f = \rho u r H
\label{massu}
\end{equation}
$H =H(r)$ is the local disc height.

Now, we will apply a similar perturbation scheme for this particular case. We here choose to perturb the aforementioned conserved quantity of the motion, namely, mass accretion rate $f$.
Perturbations of the flow variables other than density $\rho$ and velocity $\bar{v}$ are listed below

\begin{align}
f(r,t) &= f_0(r) + \epsilon f'(r,t) \\
H(r) &= H_0(r) \quad \text{(assumed unperturbed)}
\label{heiu}
\end{align}

We now expand $f$ order by order

\begin{align}
f &= (\rho_0 + \epsilon \rho')(u_0 + \epsilon u') r H_0 \\
&= \rho_0 u_0 r H_0
+ \epsilon (\rho' u_0 + \rho_0 u') r H_0
\end{align}

\textbf{Zeroth Order:}
\begin{equation}
f_0 = \rho_0 u_0 r H_0
\label{zeru}
\end{equation}

\textbf{First Order:}
\begin{equation}
f' = (\rho' u_0 + \rho_0 u') r H_0
\label{firsu}
\end{equation}
  Combining \eqref{zeru} and \eqref{firsu} we achive the relation
\begin{equation}
\frac{f'}{f_0}
= \frac{\rho'}{\rho_0} + \frac{u'}{u_0}
\label{u6u}
\end{equation}
Following the same path as discussed in the earlier section, after a long and complex algebra, we achieve the second-order differential equation governing the propagation of the linear order perturbation of the mass accretion rate $\tilde{f}$

The final Wave Equation looks like:

\begin{align}
\nonumber
\frac{\partial}{\partial t}
\left(\frac{u_0}{f_0}\frac{\partial f'}{\partial t}\right)
+ \frac{\partial}{\partial t}
\left(\frac{u_0^2}{f_0}\frac{\partial f'}{\partial r}\right)\\
+ \frac{\partial}{\partial r}
\left(\frac{u_0^2}{f_0}\frac{\partial f'}{\partial t}\right)
+ \frac{\partial}{\partial r}
\left(\frac{u_0}{f_0}(u_0^2 - c_{s0}^2)
\frac{\partial f'}{\partial r}\right)
= 0
\label{equu}
\end{align}

\subsection*{Acoustic Metric Form}
We represent \eqref{equu} in the compact form 
\begin{equation}
\partial_\mu \left( \Sigma^{\mu\nu} \partial_\nu f' \right) = 0.
\label{comu}
\end{equation}
with
\begin{equation}
\Sigma^{\mu\nu} =
\frac{u_0}{f_0}
\begin{pmatrix}
1 & u_0 \\
u_0 & u_0^2 - c_{s0}^2
\end{pmatrix}
\label{dumu}
\end{equation}
The explicit $4\text{X}4$ form 
\begin{equation}
\Sigma^{\mu\nu} =
\frac{u_0}{f_0}
\begin{pmatrix}
1 & u_0 ^1 & u_0^2 & u_0^3 \\
u_0^1 & u_0^1u_0^1 - c_{s0}^2 & u_0^1 u_0^2 & u_0^1 u_0^3 \\
u_0^2 & u_0^2 u_0^1 & u_0^2u_0^2 - c_{s0}^2 & u_0^2 u_0^3 \\
u_0^3 & u_0^3 u_0^1 & u_0^3 u_0^2 & u_0^3u_0^3 - c_{s0}^2
\end{pmatrix}
\label{permatu}
\end{equation}

Following the same scheme developed earlier, we outline the structure of the metric and the line element governing the linear perturbation of the mass accretion rate in this conical disc geometry. 

\begin{equation}
g_{\mu\nu} =-
\frac{u_0 c_{s0}}{f_0}
\begin{pmatrix}
-(c_{s0}^2 - u_0^2) & -u_0^j \\
- u_0^i & \delta_{ij}
\end{pmatrix}
\label{eq:gmunu_final}
\end{equation}

\begin{equation}
g_{\mu\nu} =-
\frac{u_0 c_{s0}}{f_0}
\begin{pmatrix}
-(c_{s0}^2 - u_0^2) & -u_0^1 & -u_0^2 & -u_0^3 \\
- u_0^1 & 1 & 0 & 0 \\
- u_0^2 & 0 & 1 & 0 \\
- u_0^3 & 0 & 0 & 1
\end{pmatrix}
\label{gmuu}
\end{equation}
\label{ssec:analogue}
\begin{equation}
ds^2 =-
\frac{u_0 c_{s0}}{f_0}
\left[
- c_{s0}^2 (dx^0)^2
+ \delta_{ij}(dx^i - u_0^i dx^0)(dx^j - u_0^j dx^0)
\right]
\label{linu}
\end{equation}
\section{Standing  and traveling Wave}

We start with  the perturbation equation \eqref{comu} for this conical disc accretion,

Assuming harmonic time dependence of the trial solution ,
\begin{equation}
f'(r,t) = f_\omega(r) e^{-i\omega t},
\end{equation}
we obtain the radial wave equation
\begin{equation}
-\omega^2 f_\omega - 2i\omega \partial_r(u f_\omega)
+ \frac{1}{u}\partial_r \left[ u(u^2 - c_s^2)\partial_r f_\omega \right] = 0.
\label{wave_eq_radial}
\end{equation}

\subsection{Standing Wave Analysis}

For an accreting system with a physical boundary, we impose
Dirichlet boundary conditions:
\begin{equation}
f_\omega(r_1) = f_\omega(r_2) = 0.
\end{equation}

Multiplying Eq.~(\ref{wave_eq_radial}) by $u f_\omega$ and integrating
over the domain $[r_1, r_2]$, we obtain
\begin{equation}
\label{ytu}
\int_{r_1}^{r_2} \left[
-u\omega^2 f_\omega^2
-2i\omega u f_\omega \partial_r(u f_\omega)
+ f_\omega \partial_r\left(u(u^2-c_s^2)\partial_r f_\omega\right)
\right] dr = 0.
\end{equation}

Using integration by parts and the boundary conditions, the surface terms vanish.
The second term reduces to a total derivative:
\begin{equation}
u f_\omega \partial_r(u f_\omega)
= \frac{1}{2} \partial_r (u^2 f_\omega^2),
\end{equation}
which integrates to zero.

Thus we obtain
\begin{equation}
\label{yu}
-\omega^2 \int u f_\omega^2 dr
- \int u(u^2 - c_s^2)(\partial_r f_\omega)^2 dr = 0.
\end{equation}

Solving for $\omega^2$:
\begin{equation}
\label{labu}
\omega^2 =
- \frac{\int u(u^2 - c_s^2)(\partial_r f_\omega)^2 dr}
{\int u f_\omega^2 dr}
\end{equation}

For subsonic flow ($u^2 < c_s^2$), the numerator is negative,
ensuring $\omega^2 > 0$, and hence $\omega$ is real.
Therefore, the solution is oscillatory and stable.

Note that at the sonic point ($u = c_s$), the coefficient of the highest derivative
vanishes, rendering the equation singular. Hence, global standing wave solutions
cannot exist across transonic flows.

\label{ssec:standing}
\subsection{Traveling Wave}
\label{ssec:travelling}

\subsection*{Traveling Wave Analysis (WKB Approximation)}

We consider a trial solution of the form
\begin{equation}
f'(r,t) = f_\omega(r), e^{-i\omega t}.
\end{equation}

Substituting into Eq.~(\ref{equu}), we obtain the following differential equation for the amplitude $( f_\omega(r) )$:
\begin{align}
(u_0^2 - c_{s0}^2) \frac{d^2 f_\omega}{dr^2}
+\left[
3u_0 \frac{du_0}{dr} -\frac{1}{u_0}\frac{d}{dr}(u_0 c_{s0}^2)
-2i\omega u_0\right] \frac{d f_\omega}{dr}\\-
\left[
2i\omega \frac{du_0}{dr}
\omega^2
  \right] f_\omega = 0.
  \label{wkb_eq}
  \end{align}

Following the standard Wentzel–Kramers–Brillouin (WKB) approximation, we assume a solution of the form
\begin{equation}
f_\omega(r) =
\exp\left[
\sum_{n=-1}^{\infty} \frac{k_n(r)}{\omega^n}
\right].
\label{wkb_ansatz}
\end{equation}

Substituting Eq.~(\ref{wkb_ansatz}) into Eq.~(\ref{wkb_eq}) and collecting coefficients of different powers of $(\omega)$, we obtain a hierarchy of equations.

\subsubsection*{Order $(\omega^2)$}
\begin{equation}
(u_0^2 - c_{s0}^2)
\left(\frac{dk_{-1}}{dr}\right)^2
- 2i u_0 \frac{dk_{-1}}{dr}
- 1 = 0.
  \end{equation}

Solving,
\begin{equation}
\frac{dk_{-1}}{dr} = \frac{i}{u_0 \mp c_{s0}},
\end{equation}
\begin{equation}
k_{-1} = i \int \frac{dr}{u_0 \mp c_{s0}}.
\end{equation}

\subsubsection*{Order $(\omega^1)$}
\begin{align}
\nonumber
(u_0^2 - c_{s0}^2)
\left(
\frac{d^2 k_{-1}}{dr^2} + 2 \frac{dk_{-1}}{dr}\frac{dk_0}{dr}
  \right)\\
  \nonumber
+\left[
3u_0 \frac{du_0}{dr} -\frac{1}{u_0}\frac{d}{dr}(u_0 c_{s0}^2)
  \right] \frac{dk_{-1}}{dr}\\
  \
- 2i \left(
  u_0 \frac{dk_0}{dr}
+ \frac{du_0}{dr}
  \right) = 0.
  \end{align}

Solving for $(k_0)$, we obtain
\begin{equation}
k_0 = -\frac{1}{2} \ln\left(\frac{u_0 c_{s0}}{f_0}\right).
\end{equation}

\subsubsection*{Order $(\omega^0)$}

\begin{align}
\nonumber
(u_0^2 - c_{s0}^2)
\left(
\frac{d^2 k_0}{dr^2} +2 \frac{dk_{-1}}{dr}\frac{dk_1}{dr} +\left(\frac{dk_0}{dr}\right)^2
  \right)\\
+\left[
3u_0 \frac{du_0}{dr}-
 \frac{1}{u_0}\frac{d}{dr}(u_0 c_{s0}^2)
  \right] \frac{dk_0}{dr}
-2i u_0 \frac{dk_1}{dr} = 0.
  \end{align}

Using the previously obtained expressions for $(k_{-1})$ and $(k_0)$, we obtain the following expression for $(k_1)$:

\begin{align}
\frac{dk_1}{dr}
=-\frac{1}{2i(\pm c_{s0})}(\mathcal{A}+\mathcal{B})
\end{align}
Integrating, we obtain
\begin{align}
\label{iu}
\nonumber
k_1(r)
=
\int \frac{i}{2(\pm c_{s0})}
(\mathcal{A}+\mathcal{B}) dr.
  \end{align}
Where,
$\mathcal{A}=
(u_0^2 - c_{s0}^2)
\left(
\frac{d^2 k_0}{dr^2}
 +\left(\frac{dk_0}{dr}\right)^2
  \right)
\text{ and } \mathcal{B}=\left(
3u_0 \frac{du_0}{dr}
-\frac{1}{u_0}\frac{d}{dr}(u_0 c_{s0}^2)
  \right)\frac{dk_0}{dr}
  .$
\subsubsection*{Convergence of the Series}

For sufficiently large $(\omega)$, the hierarchy satisfies
\begin{equation}
\omega |k_{-1}| \ll |k_0| \ll \frac{1}{\omega} |k_1|,
\end{equation}
ensuring that the series remains convergent. Hence, the travelling wave solution remains well-behaved and finite.

\subsubsection*{Near-Horizon Behaviour}

Near the sonic horizon $r = r_c$, where $u_{0} = c_{s0}$,
\begin{equation}
u_{0} - c_{s0} \approx (u_{0}' - c_{s0}')(r - r_c),
\end{equation}
leading to
\begin{equation}
\int \frac{dr}{u_{0} - c_{s0}} \sim \ln|r - r_c|.
\end{equation}

This logarithmic divergence is analogous to the behaviour near a black hole
event horizon, leading to Hawking-like radiation in the analogue system.

\section{Causal Structure}
\label{sec:cpd}
The acoustic metric derived in Sec.~\ref{ssec:analogue} governs the
propagation of linear perturbations through the background accretion
flow, in precise analogy with the propagation of a massless scalar
field on a curved spacetime. In this section we exploit that analogy
to construct the causal structure of the emergent acoustic spacetime
for the multi-transonic shocked flow, and represent it through a
Carter--Penrose diagram.

\subsection*{Identification of Acoustic Horizons}

The causal character of the acoustic spacetime is entirely controlled
by the sign of $(c_{s0}^2 - u_0^2)$.  Projecting the line element
\eqref{linu} onto the $(t, r)$ plane yields the effective
$(1+1)$-dimensional metric
\begin{equation}
  \label{eq:ds2_2d}
  ds^2_{(1+1)} = \frac{u_0 c_{s0}}{f_0}
  \Bigl[
    (c_{s0}^2 - u_0^2)\,dt^2
    - 2u_0\,dt\,dr
    - dr^2
  \Bigr],
\end{equation}
which admits the null-ray factorisation
\begin{equation}
  \label{eq:null_factor}
  ds^2_{(1+1)} = -\frac{u_0 c_{s0}}{f_0}
  \bigl[dr - (u_0 + c_{s0})\,dt\bigr]
  \bigl[dr - (u_0 - c_{s0})\,dt\bigr].
\end{equation}
The two families of acoustic null geodesics therefore propagate with
coordinate velocities
\begin{equation}
  \label{eq:null_vel}
  \frac{dr}{dt} = u_0 + c_{s0}
  \quad \text{and} \quad
  \frac{dr}{dt} = u_0 - c_{s0}.
\end{equation}
In the subsonic region ($u_0 < c_{s0}$), both $u_0 + c_{s0} > 0$ and
$u_0 - c_{s0} < 0$, so the two families propagate in opposite
directions and acoustic information can escape to large $r$. In the
supersonic region ($u_0 > c_{s0}$), both velocities are positive
(inward, since the accretion flow is directed toward the black hole),
and no acoustic signal can propagate outward. The locus $u_0 = c_{s0}$
— i.e., the sonic surface — therefore constitutes an acoustic horizon,
in direct analogy with a black hole event horizon.

For the multi-transonic shocked accretion considered here, the flow
passes through the outer sonic point $x_\mathrm{out}$ (from subsonic
to supersonic), then encounters the shock at $r_\mathrm{sh}$ (abrupt
deceleration back to subsonic), and finally crosses the inner sonic
point $x_\mathrm{in}$ (again from subsonic to supersonic before
reaching the black hole). This topology generates three distinct
acoustic horizons:
\begin{enumerate}
  \item The \emph{outer acoustic black hole horizon}
        $\mathfrak{B}_{BH}^\mathrm{out}$ at $r = x_\mathrm{out}$,
        where the outgoing family of null rays is momentarily frozen.
  \item The \emph{acoustic white hole horizon}
        $\mathfrak{B}_{WH}$ at $r = r_\mathrm{sh}$, across which
        acoustic perturbations can only emerge (from the compressed
        post-shock region) but cannot enter.
  \item The \emph{inner acoustic black hole horizon}
        $\mathfrak{B}_{BH}^\mathrm{in}$ at $r = x_\mathrm{in}$,
        which irreversibly captures all inward-propagating perturbations.
\end{enumerate}

\subsection*{Tortoise Coordinate and Null Coordinates}

To resolve the coordinate singularity at each acoustic horizon and
construct the global extension, we introduce the acoustic tortoise
coordinate $r^*$ defined by
\begin{equation}
  \label{eq:tortoise}
  \frac{dr^*}{dr} = \frac{1}{c_{s0}^2 - u_0^2}.
\end{equation}
The associated outgoing ($U$) and ingoing ($V$) null coordinates are
\begin{equation}
  \label{eq:null_uv}
  U = t - r^*, \qquad V = t + r^*,
\end{equation}
in terms of which the reduced $(1+1)$-dimensional metric takes the
conformally flat form
\begin{equation}
  \label{eq:metric_uv}
  ds^2_{(1+1)} = -\frac{u_0 c_{s0}}{f_0}(c_{s0}^2 - u_0^2)\,dU\,dV.
\end{equation}
Near each acoustic horizon $r = r_c$ (where $u_0 = c_{s0}$), the
approximation $u_0 - c_{s0} \approx \sigma_c\,(r - r_c)$ with
$\sigma_c \equiv \left.\bigl(du_0/dr - dc_{s0}/dr\bigr)\right|_{r_c}$
gives
\begin{equation}
  \label{eq:tort_log}
  r^* \approx \frac{1}{\sigma_c}\ln|r - r_c|,
\end{equation}
reproducing the logarithmic divergence already encountered in the
WKB analysis of Sec.~\ref{ssec:travelling}, and directly analogous
to the Regge--Wheeler tortoise coordinate of the Schwarzschild
geometry near the event horizon.

\subsection*{Kruskal Extension}

To obtain a globally regular coordinate system across each acoustic
horizon, we employ a Kruskal-like extension.  For a given acoustic
horizon at $r_c$ with surface gravity $\kappa_c$ (to be computed in
Sec.~\ref{ssec:surfgrav}), we define
\begin{equation}
  \label{eq:kruskal}
  \widetilde{U}_c = -e^{-\kappa_c U}, \qquad
  \widetilde{V}_c =  e^{+\kappa_c V}.
\end{equation}
In the Kruskal coordinates $(\widetilde{U}_c, \widetilde{V}_c)$, the
conformal factor $(c_{s0}^2 - u_0^2)$ in Eq.~\eqref{eq:metric_uv}
remains finite and non-zero at $r = r_c$; the horizon corresponds to
the locus $\widetilde{U}_c = 0$ (future horizon) or
$\widetilde{V}_c = 0$ (past horizon), and the metric extends smoothly
across both.  An acoustic black hole horizon admits only the future
branch ($\widetilde{U}_c = 0$), whereas an acoustic white hole horizon
admits only the past branch ($\widetilde{V}_c = 0$).

\subsection*{Carter--Penrose Diagram}

\begin{figure}[H]
  \centering
  \includegraphics[width=0.5\textwidth]{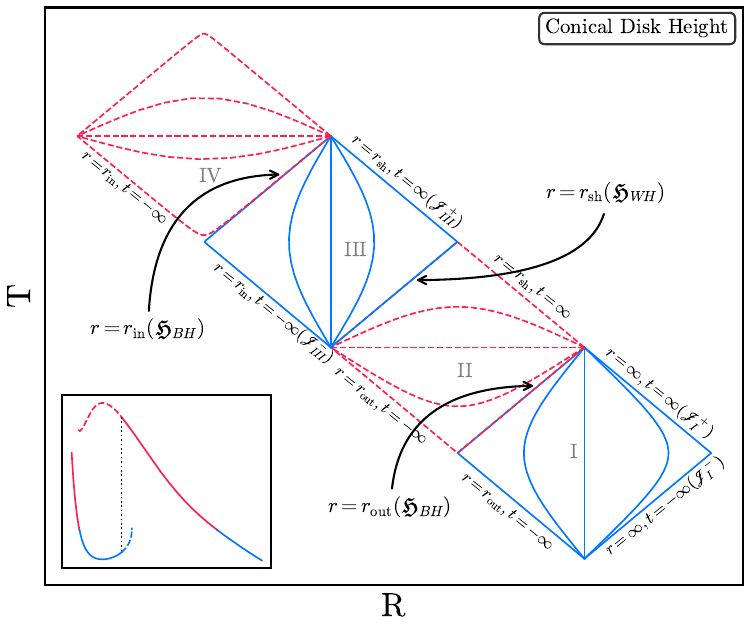}
  \caption{Carter--Penrose diagram for the multi-transonic shocked accretion flow in the conical disc geometry. The causal structure is governed by two acoustic black hole horizons at the outer sonic point $r_\mathrm{out}$ ($\mathfrak{B}_{BH}^\mathrm{out}$) and the inner sonic point $r_\mathrm{in}$ ($\mathfrak{B}_{BH}^\mathrm{in}$), and one acoustic white hole horizon at the shock location $r_\mathrm{sh}$ ($\mathfrak{B}_{WH}$). These three horizons divide the conformal diagram into four causally disconnected regions: Region~I (outer subsonic, $r > x_\mathrm{out}$), which is asymptotically flat and communicates with future null infinity $\mathscr{I}^+$; Region~II (outer supersonic, $r_\mathrm{sh} < r < x_\mathrm{out}$), bounded by $\mathfrak{B}_{BH}^\mathrm{out}$ and $\mathfrak{B}_{WH}$; Region~III (post-shock subsonic, $x_\mathrm{in} < r < r_\mathrm{sh}$), sourced by the acoustic white hole; and Region~IV (inner supersonic, $r < x_\mathrm{in}$), from which no acoustic signal escapes. The topology is formally identical to the maximally extended Schwarzschild geometry supplemented with a white hole sector.}
  \label{fig:conical_pcd}
\end{figure}

The full Carter--Penrose diagram is obtained through the further
compactification
\begin{equation}
  \label{eq:penrose}
  \begin{split}
    T &= \arctan\widetilde{U}_c + \arctan\widetilde{V}_c,\\
    R &= \arctan\widetilde{V}_c - \arctan\widetilde{U}_c,
  \end{split}
\end{equation}
which maps the infinite acoustic spacetime onto a finite diamond with
$|T \pm R| \leq \pi$.  The resulting diagram for the conical flow
model is displayed in Fig.~\ref{fig:conical_pcd}.

The presence of two acoustic black hole horizons ($x_\mathrm{out}$ and
$x_\mathrm{in}$) and one acoustic white hole horizon ($r_\mathrm{sh}$)
divides the causal diagram into four causally disconnected regions:
\begin{itemize}
  \item \textbf{Region~I} (outer subsonic, $r > x_\mathrm{out}$):
        The asymptotically flat exterior.  Acoustic signals generated
        here can reach future null infinity $\mathscr{I}^+$ and are
        received by a distant observer.
  \item \textbf{Region~II} (outer supersonic,
        $r_\mathrm{sh} < r < x_\mathrm{out}$): The pre-shock
        supersonic region.  Bounded above by the acoustic black hole
        horizon $\mathfrak{B}_{BH}^\mathrm{out}$ and below by the
        white hole horizon $\mathfrak{B}_{WH}$, no signal from
        Region~II can escape to $\mathscr{I}^+$.
  \item \textbf{Region~III} (post-shock subsonic,
        $x_\mathrm{in} < r < r_\mathrm{sh}$):  The compressed
        subsonic layer between the shock and the inner sonic point.
        Perturbations here originate on the acoustic white hole
        $\mathfrak{B}_{WH}$ and can propagate both inward and outward
        within this region.
  \item \textbf{Region~IV} (inner supersonic, $r < x_\mathrm{in}$):
        The innermost supersonic region, separated from Region~III by
        $\mathfrak{B}_{BH}^\mathrm{in}$.  All acoustic perturbations
        are irreversibly advected toward the physical black hole
        horizon; no signal can escape.
\end{itemize}

This four-region causal topology is formally identical to the maximal
Kruskal extension of the Schwarzschild geometry with an additional
white hole sector: the outer sonic point $x_\mathrm{out}$ plays the
role of the future event horizon, the shock $r_\mathrm{sh}$ plays the
role of the past (white hole) horizon, and the inner sonic point
$x_\mathrm{in}$ provides a second, nested acoustic black hole horizon.
The acoustic white hole at $r_\mathrm{sh}$ is a direct physical
consequence of the Rankine--Hugoniot conditions: the compression and
heating of the post-shock gas effectively converts kinetic energy of
the supersonic flow into acoustic radiation that propagates outward
across $r_\mathrm{sh}$, in precise analogy with the particle creation
attributed to white hole evaporation in the gravitational setting.

\subsection{Surface Gravity}
\label{ssec:surfgrav}
The surface gravity associated with the acoustic horizon can be expressed in terms of the normal derivative of the quantity $(c^2 - v^2)$ as
\begin{equation}
\kappa_H =
\left|
\frac{1}{2} \frac{\partial}{\partial n} (c^2 - v^2)
\right|_H,
\label{eq:kappa_def1}
\end{equation}
where the derivative is taken along the direction normal to the acoustic horizon, and all quantities are evaluated at the horizon.

Noting that at the acoustic horizon $v = c$, we can rewrite Eq.~(\ref{eq:kappa_def1}) as
\begin{equation}
\kappa_H =
c_H
\left|
\frac{\partial}{\partial n} (c - v)
\right|_H.
\label{eq:kappa_def2}
\end{equation}

This expression differs from the original result obtained by Unruh~\cite{Unruh1981}, where the speed of sound was implicitly assumed to be position-independent. Under that approximation, $\partial c / \partial n = 0$, and the surface gravity reduces to a form proportional only to the gradient of the flow velocity. While this assumption is appropriate for simple laboratory fluids such as water, it is not valid for general astrophysical flows where the sound speed varies spatially.

Therefore, Eq.~(\ref{eq:kappa_def2}) provides the more general expression for the acoustic surface gravity, incorporating both the velocity and sound speed gradients.

\subsection*{Acoustic Surface Gravity in the Present Model}

Since we are dealing with a stationary acoustic metric, there is no explicit time dependence. Therefore, we can choose the timelike Killing vector as
\begin{equation}
\psi^\mu = \delta^\mu_t,
\label{eq:killing_vector}
\end{equation}
where $\delta^\mu_t$ is the Kronecker delta.

The expression for the acoustic surface gravity has been derived in~\cite{Bilić_2014} (see Sec.~VI therein) and is given by
\begin{equation}
\kappa =
\left.
\frac{\sqrt{\psi^\mu \psi_\mu}}{\sqrt{-g_{rr}}}
\frac{1}{1 - c_s^2}
\left(
\frac{du}{dr} - \frac{dc_s}{dr}
\right)
\right|_{r_c},
\label{eq:kappa_general}
\end{equation}
where all quantities are evaluated at the acoustic horizon $r_c$.

Now, using the background metric properties, we have
\begin{equation}
\sqrt{\psi^\mu \psi_\mu} = \sqrt{g_{tt}} = \sqrt{1 + 2\Phi} \approx \sqrt{1 + \Phi},
\label{eq:killing_norm}
\end{equation}
and
\begin{equation}
g_{rr} = -1.
\label{eq:grr}
\end{equation}

Substituting Eqs.~(\ref{eq:killing_norm}) and (\ref{eq:grr}) into Eq.~(\ref{eq:kappa_general}), we obtain the final expression for the acoustic surface gravity at the horizon $r_c$ (see also Sec .~3 of~\cite{Barcelo2011}):
\begin{equation}
\kappa =
\left.
\frac{\sqrt{1 + \Phi}}{1 - c_s^2}
\left(
\frac{du}{dr} - \frac{dc_s}{dr}
\right)
\right|_{r_c}.
\label{eq:kappa_final}
\end{equation}

\section{Conclusion}
\label{sec:conclusion}

In this work we have carried out a comprehensive study of low angular momentum, inviscid, multi-species axisymmetric accretion onto a rotating black hole, modelled through the Artemova--Bj\"{o}rnsson--Novikov pseudo-Kerr potential, for three distinct disc geometries: the vertical equilibrium (VE), conical flow (CF), and constant height (CH) models. The accretion flow is described by the relativistic equation of state of Chattopadhyay \& Ryu~\cite{2009ApJ...694..492C}, in which the adiabatic index $\Gamma$ varies with local temperature, rendering the flow inherently non-barotropic. Our principal findings are summarized as follows.

\begin{enumerate}

\item \textit{Multi-transonic critical point structure.}
Steady-state axisymmetric accretion with the spatially varying equation of state admits up to three critical points for flow parameters drawn from appropriate regions of the $\mathcal{E}$--$\lambda$ parameter space. Using the dynamical systems formalism developed here, we have confirmed that the inner and outer critical points are saddle-type ($\det\mathbf{J}<0$), which alone support physically acceptable transonic transitions, while the intermediate critical point is of center type ($\det\mathbf{J}>0$, $\mathrm{Tr}\,\mathbf{J}=0$). These results hold consistently across all three disc geometries.

\item \textit{Stationary shocks and spin dependence.}
Within the multi-critical ($\mathbf{A}$) region of the parameter space, a subset --- the shock parameter space $\mathbf{S}$ --- supports global accretion solutions that pass through both the outer and inner sonic points and undergo a standing Rankine--Hugoniot shock. For all three disc models, increasing the black hole spin parameter $a$ causes the shock to form at progressively larger radii, a direct consequence of the enhanced frame-dragging centrifugal barrier. Correspondingly, the shock strength $M_-/M_+$, the density compression ratio $\rho_+/\rho_-$, and the temperature jump $T_+/T_-$ all decrease monotonically with spin. The size of the shock parameter space and the quantitative values of the shock-related quantities differ among the three disc geometries, reflecting the role of the geometric factor $H(r,\Theta)$ in shaping the flow dynamics.

\item \textit{Linear stability.}
The stationary transonic solutions have been shown to be linearly stable under radial perturbations. A WKB analysis of the traveling-wave solutions confirms that the perturbation series converges for sufficiently large frequencies and that the perturbation amplitude remains finite everywhere in the flow domain. Near each acoustic horizon the perturbation displays the expected logarithmic divergence in the tortoise coordinate, in precise analogy with the behavior near a black hole event horizon in general relativity, thereby establishing the physical consistency of the acoustic analogy.

\item \textit{Emergent acoustic spacetime.}
Linear perturbation of the mass accretion rate yields a second-order wave equation that can be recast in the covariant form $\partial_\mu(\Sigma^{\mu\nu}\partial_\nu f')=0$, from which a curved acoustic metric $g_{\mu\nu}$ is extracted. For multi-transonic accretion with a standing shock, this acoustic metric possesses three horizons: two acoustic black hole horizons at the outer and inner sonic points ($x_\mathrm{out}$ and $x_\mathrm{in}$), and one acoustic white hole horizon at the shock location ($r_\mathrm{sh}$). The surface gravity $\kappa$ at each horizon has been computed using the general expression that accounts for the spatial variation of the local sound speed, generalizing the original Unruh formula.

\item \textit{Causal structure.}
The Carter--Penrose diagram for the conical flow model reveals a four-region causal topology that is formally identical to the maximally extended Schwarzschild geometry with an additional white hole sector. The outer sonic point acts as a future acoustic event horizon, the shock plays the role of the past (white hole) horizon, and the inner sonic point provides a second, nested acoustic black hole horizon. The acoustic white hole at the shock is a direct physical consequence of the Rankine--Hugoniot compression and heating, in analogy with white-hole evaporation in the gravitational setting.

\end{enumerate}

Taken together, these results establish that rotating black hole accretion with a spatially varying adiabatic index provides a rich, physically motivated arena for analogue gravity. The spin of the central black hole modulates both the hydrodynamic structure (shock location, strength, and thermodynamic jumps) and the acoustic causal geometry (surface gravities and the topology of the Penrose diagram) in a geometry-dependent manner. This spin dependence opens a potential observational window: variations in quasi-periodic oscillations and spectral features of accreting black holes may, in principle, carry imprints of the acoustic surface gravity and its spin dependence as derived here. Future work will extend this analysis to fully general relativistic disc configurations and will investigate the quantum aspects of the acoustic Hawking temperature associated with the emergent horizons.

\section*{Acknowledgment}
SG would like to thank HRI for supporting academic visits after completion of the post-doctoral tenure. AKM and RS would like to thank HRI for supporting academic visits.

%

\appendix
\section{}
\label{app:eqns}

Here we describe accretion flow dynamics for matter in hydrostatic equilibrium in the vertical direction ($VE$ model) and for matter flow with constant height ($CH$ model).

\subsection*{Vertical Equilibrium ($VE$) model}
Disk height for $VE$ model is given by
\begin{equation}
    \label{eq:app1}
    H_{VE}(r,\Theta) = c_\mathrm{s}\sqrt{\frac{ r}{\Gamma |F_\mathrm{ABN}|}} =\sqrt{\frac{2\Theta r^{3 - \beta}(r - \rp)^{\beta}}{\tau}},
\end{equation}
where $|F_\mathrm{ABN}|$ means the mod value of $F_\mathrm{ABN}$. Eq. (\ref{eq:21}) in this type of flow geometry takes the form
\[\left(\frac{\der \Theta}{\der r}\right)_{VE} = (\Omega_{1})_{VE} + (\Omega_2)_{VE}\left(\dfrac{\der \vel}{\der r}\right)_{VE},\]
where
\[
\begin{aligned}
(\Omega_{1})_{VE} 
&= -\frac{\Theta}{(2N + 1)} \bigg[ \frac{(5 - \beta)}{r} + \frac{\beta}{(r -\rp)} \bigg],\\[6pt] 
(\Omega_{2})_{VE} 
&= - \frac{2\Theta}{(2N + 1)\vel}.
\end{aligned}
\]
Therefore, explicit form of Eq. (\ref{eq:21}) is
\begin{equation}
    \label{eq:app2}
    \left(\frac{\der \Theta}{\der r}\right)_{VE} = -\frac{\Theta}{(2N + 1)}\left[\frac{(5-\beta)}{r}+\frac{\beta}{(r - \rp)}+\frac{2}{\vel}\frac{\der \vel}{\der r}\right].
\end{equation}
Radial velocity gradient equation (Eq. (\ref{eq:22})) takes the form
\begin{equation}
\begin{aligned}
\label{eq:app3}
\left(\frac{\der \vel}{\der r}\right)_{VE} = \frac{\dfrac{\Big[5r + \rp(\beta - 5)\Big]}{(\Gamma + 1)r(r-\rp)}c_\mathrm{s}^{2} + \dfrac{\lambda^{2}}{r^3} + F_\mathrm{ABN}} {\bigg[\vel - \dfrac{\mathrm{c_\mathrm{s}^2}}{\vel}\dfrac{2}{(\Gamma + 1)}\bigg]} = \frac{\mathcal{N}_{VE}}{\mathcal{D}_{VE}}.
\end{aligned}
\end{equation}
Using critical point condition
\[
\begin{aligned}
(\mathcal{N}_{VE})_\mathrm{c} = (\mathcal{D}_{VE})_\mathrm{c} = 0,
\end{aligned}
\]
we could write the critical sound speed and critical flow velocity as
\begin{equation}
    \label{eq:app4}
    (c_\mathrm{sc}^2)_{VE} = \frac{(\Gamma_\mathrm{c} + 1)r_\mathrm{c}(r_\mathrm{c}-\rp)}{\big(5r_\mathrm{c} + \rp(\beta - 5)\big)}\left( \dfrac{1}{r^{2-\beta}_\mathrm{c}(r_\mathrm{c}-\rp)^{\beta}} - \dfrac{\lambda^{2}}{r^3_\mathrm{c}}\right),
\end{equation}
and
\begin{equation}
    \label{eq:app5}
    (\vel_\mathrm{c}^2)_{CF} = \frac{2}{\Gamma_\mathrm{c} + 1}(c_\mathrm{sc}^2)_{CF}.
\end{equation}
For $VE$ model we can write
\begin{equation}
\label{eq:app6}
\begin{aligned}
\frac{\partial \mathcal{N}_{VE}}{\partial r}
&= -\frac{c_\mathrm{s}^2}{(\Gamma + 1)}\bigg[  \frac{(5 - \beta)}{r^2} + \frac{\beta}{(r -\rp)^2}\bigg] - \frac{3\lambda^2}{r^4} +F'_{\mathrm{ABN}},\\[6pt]
\frac{\partial \mathcal{N}_{VE}}{\partial \Theta} 
&= \frac{2}{\tau(\Gamma + 1)}\bigg[ \frac{(5 - \beta)}{r} + \frac{\beta}{(r - \rp)}        \bigg]\bigg[ \Gamma + \frac{\Theta}{(\Gamma + 1)}\frac{\der \Gamma}{\der \Theta}   \bigg] ,\\[6pt]
\frac{\partial \mathcal{D}_{VE}}{\partial \Theta} 
&= -\frac{4}{\tau(\Gamma + 1)\vel}\bigg[ \Gamma + \frac{\Theta}{(\Gamma + 1)}\frac{\der \Gamma}{\der \Theta}  \bigg],\\[6pt]
\frac{\partial \mathcal{D}_{VE}}{\partial \vel} 
&= 1 + \frac{2c_\mathrm{s}^2}{(\Gamma + 1)\vel^2}.
\end{aligned}
\end{equation}
Now using Eqs. (\ref{eq:app4}), (\ref{eq:app5}) and (\ref{eq:app6}) we can write:
\begin{equation}
\label{eq:app7}
 \begin{aligned}
(\mathcal{A}_\mathrm{c})_{VE} 
&= 2 + \frac{2}{(2N_\mathrm{c}+1)\Gamma_\mathrm{c}}\left(\Gamma_\mathrm{c}+\frac{\Theta_\mathrm{c}}{(\Gamma_\mathrm{c}+1)}\frac{\der \Gamma}{\der \Theta}\bigg\lvert _{\rm c}\right),\\[6pt] 
(\mathcal{B}_\mathrm{c})_{VE}
&= \frac{8\Theta_\mathrm{c}\big(5r_\mathrm{c}+\rp (\beta-5)\big)}{\tau\vel_\mathrm{c}N_\mathrm{c}(\Gamma_\mathrm{c}+1)^2 r_\mathrm{c}(r_\mathrm{c}-\rp)}\left(\Gamma_\mathrm{c}+\frac{\Theta_\mathrm{c}}{(\Gamma_\mathrm{c}+1)}\frac{\der \Gamma}{\der \Theta}\bigg\lvert _{\rm c}\right),\\[6pt] 
(\mathcal{C}_\mathrm{c})_{VE} 
&= \frac{c_\mathrm{sc}^2}{(\Gamma_\mathrm{c}+1)}\left[\frac{5-\beta}{r_\mathrm{c}^2}+\frac{\beta}{(r_\mathrm{c}-\rp)^2}\right] + \frac{3\lambda^2}{r_\mathrm{c}^4} -F'_{\mathrm{ABN}}\lvert _{\rm c}\\
&+\frac{2\Theta_\mathrm{c}}{(2N_\mathrm{c}+1)(\Gamma_\mathrm{c}+1)\tau}\left[\frac{5-\beta}{r_\mathrm{c}}+\frac{\beta}{(r_\mathrm{c}-\rp)}\right]^2\left(\Gamma_\mathrm{c}+\frac{\Theta_\mathrm{c}}{(\Gamma_\mathrm{c}+1)}\frac{\der \Gamma}{\der \Theta}\bigg\lvert _{\rm c}\right), 
\end{aligned}
\end{equation}
where
\[
\begin{aligned}
    \Theta_{\mathrm{c}} = \frac{\tau(\Gamma_\mathrm{c} + 1)r_\mathrm{c}(r_\mathrm{c}-\rp)}{2\Gamma_\mathrm{c}\big(5r_\mathrm{c} + \rp(\beta - 5)\big)}\left( \dfrac{1}{r^{2-\beta}_\mathrm{c}(r_\mathrm{c}-\rp)^{\beta}} - \dfrac{\lambda^{2}}{r^3_\mathrm{c}}\right).  
\end{aligned}
\]
Now, using Eq. (\ref{eq:app7}), the slope at the critical point for vertical equilibrium ($VE$) model can be expressed in the following way
\begin{equation}
    \label{eq:app8}
   \left(\frac{\der \vel}{\der r}\right)_{VE}\bigg\lvert_{\mathrm{c}}= \frac{-(\mathcal{B}_\mathrm{c})_{VE} \overset{+}{-} \sqrt{(\mathcal{B}_\mathrm{c})^2_{VE} -4(\mathcal{A}_\mathrm{c})_{VE}(\mathcal{C}_\mathrm{c})_{VE}}}{2(\mathcal{A}_\mathrm{c})_{VE}}.
\end{equation}

\subsection*{Constant Height ($CH$) model}
For constant height disk geometry, the disk height is given by
\begin{equation}
    \label{eq:app9}
    H_{CH}(r,\Theta) = \alpha 
\end{equation}
where $\alpha$ is some constant. For this type of flow geometry we can write Eq. (\ref{eq:21}) in the following way
\[\left(\frac{\der \Theta}{\der r}\right)_{CH} = (\Omega_{1})_{CH} + (\Omega_2)_{CH}\left(\dfrac{\der \vel}{\der r}\right)_{CH},\]
where
\[
\begin{aligned}
(\Omega_{1})_{CH} 
&= -\frac{\Theta}{Nr},\\[6pt] 
(\Omega_{2})_{CH} 
&= -\frac{\Theta}{N\vel}.
\end{aligned}
\]
Therefore, explicit form of the rate of change in temperature in the radial direction, as given by Eq.(\ref{eq:21}), can be written as
\begin{equation}
    \label{eq:app10}
    \left(\frac{\der \Theta}{\der r}\right)_{CH} = -\frac{\Theta}{N}\left(\frac{1}{r} + \frac{1}{\vel}\dfrac{\der \vel}{\der r} \right).
\end{equation}
We can write radial velocity gradient equation (Eq. (\ref{eq:22})) as
\begin{equation}
    \label{eq:app11}
    \left(\frac{\der \vel}{\der r}\right)_{CH} = \frac{\dfrac{c_\mathrm{s}^2}{r} + \dfrac{\lambda^{2}}{r^3}- \dfrac{1}{r^{2-\beta}(r-\rp)^{\beta}}}{\vel - \dfrac{c_\mathrm{s}^2}{\vel}} = \frac{\mathcal{N}_{CH}}{\mathcal{D}_{CH}}.
\end{equation}
Using critical point condition (Eq. (\ref{eq:23})), which takes the following form
\[
\begin{aligned}
(\mathcal{N}_{CH})_\mathrm{c} = (\mathcal{D}_{CH})_\mathrm{c} = 0,
\end{aligned}
\]
we could write the local sound speed and matter flow velocity at the critical point as
\begin{equation}
    \label{eq:app12}
    (c_\mathrm{sc}^2)_{CH} = r_\mathrm{c}\left( \dfrac{1}{r^{2-\beta}_\mathrm{c}(r_\mathrm{c}-\rp)^{\beta}} - \dfrac{\lambda^{2}}{r^3_\mathrm{c}}\right),
\end{equation}
and
\begin{equation}
    \label{eq:app13}
    (\vel_\mathrm{c}^2)_{CH} = (c_\mathrm{sc}^2)_{CH}.
\end{equation}
For constant height model we can write
\begin{equation}
\label{eq:app14}
\begin{aligned}
\frac{\partial \mathcal{N}_{CH}}{\partial r}
&= -\frac{c_\mathrm{s}^2}{r^2} - \frac{3\lambda^2}{r^4} +F'_{\mathrm{ABN}},\\[6pt]
\frac{\partial \mathcal{N}_{CH}}{\partial \Theta} 
&= \frac{2}{\tau r}\bigg[\Gamma + \Theta \frac{\der \Gamma}{\der \Theta}\bigg],\\[6pt]
\frac{\partial \mathcal{D}_{CH}}{\partial \Theta} 
&= -\frac{2}{\tau \vel}\bigg[\Gamma + \Theta \frac{\der \Gamma}{\der \Theta}\bigg],\\[6pt]
\frac{\partial \mathcal{D}_{CH}}{\partial \vel} 
&= 1 + \frac{c_\mathrm{s}^2}{\vel^2}.
\end{aligned}
\end{equation}
Now using Eqs. (\ref{eq:app12}), (\ref{eq:app13}) and (\ref{eq:app14}) we can write:
\begin{equation}
\label{eq:app15}
 \begin{aligned}
(\mathcal{A}_\mathrm{c})_{CH} 
&= 2 + \frac{1}{N_\mathrm{c}\Gamma_\mathrm{c}}\left(\Gamma_\mathrm{c}+\Theta_\mathrm{c}\frac{\der \Gamma}{\der \Theta}\bigg\lvert _{\rm c}\right),\\[6pt] 
(\mathcal{B}_\mathrm{c})_{CH}
&= \frac{4\Theta_\mathrm{c}}{\vel_\mathrm{c}r_\mathrm{c}N_\mathrm{c}\tau}\left(\Gamma_\mathrm{c}+\Theta_\mathrm{c}\frac{\der \Gamma}{\der \Theta}\bigg\lvert _{\rm c}\right),\\[6pt] 
(\mathcal{C}_\mathrm{c})_{CH} 
&= \frac{c_\mathrm{sc}^2}{r_\mathrm{c}^2} + \frac{3\lambda^2}{r_\mathrm{c}^4} -F'_{\mathrm{ABN}}\lvert _{\rm c}\\
&+\frac{2\Theta_\mathrm{c}}{r^2_\mathrm{c}N_\mathrm{c}\tau}\left(\Gamma_\mathrm{c}+\Theta_\mathrm{c}\frac{\der \Gamma}{\der \Theta}\bigg\lvert _{\rm c}\right). 
\end{aligned}
\end{equation}
where
\[
\begin{aligned}
    \Theta_{\mathrm{c}} = \frac{\tau r_\mathrm{c}}{2\Gamma_{\mathrm{c}}}\left( \dfrac{1}{r^{2-\beta}_\mathrm{c}(r_\mathrm{c}-\rp)^{\beta}} - \dfrac{\lambda^{2}}{r^3_\mathrm{c}}\right).  
\end{aligned}
\]
Finally, using Eq. (\ref{eq:app15}), the slope at the critical point for the constant height ($CH$) model is given by
\begin{equation}
    \label{eq:app16}
   \left(\frac{\der \vel}{\der r}\right)_{CH}\bigg\lvert_{\mathrm{c}}= \frac{-(\mathcal{B}_\mathrm{c})_{CH} \overset{+}{-} \sqrt{(\mathcal{B}_\mathrm{c})^2_{CH} -4(\mathcal{A}_\mathrm{c})_{CH}(\mathcal{C}_\mathrm{c})_{CH}}}{2(\mathcal{A}_\mathrm{c})_{CH}}.
\end{equation}



\begin{thebibliography}{74}%
\makeatletter
\providecommand \@ifxundefined [1]{%
 \@ifx{#1\undefined}
}%
\providecommand \@ifnum [1]{%
 \ifnum #1\expandafter \@firstoftwo
 \else \expandafter \@secondoftwo
 \fi
}%
\providecommand \@ifx [1]{%
 \ifx #1\expandafter \@firstoftwo
 \else \expandafter \@secondoftwo
 \fi
}%
\providecommand \natexlab [1]{#1}%
\providecommand \enquote  [1]{``#1''}%
\providecommand \bibnamefont  [1]{#1}%
\providecommand \bibfnamefont [1]{#1}%
\providecommand \citenamefont [1]{#1}%
\providecommand \href@noop [0]{\@secondoftwo}%
\providecommand \href [0]{\begingroup \@sanitize@url \@href}%
\providecommand \@href[1]{\@@startlink{#1}\@@href}%
\providecommand \@@href[1]{\endgroup#1\@@endlink}%
\providecommand \@sanitize@url [0]{\catcode `\\12\catcode `\$12\catcode
  `\&12\catcode `\#12\catcode `\^12\catcode `\_12\catcode `\%12\relax}%
\providecommand \@@startlink[1]{}%
\providecommand \@@endlink[0]{}%
\providecommand \url  [0]{\begingroup\@sanitize@url \@url }%
\providecommand \@url [1]{\endgroup\@href {#1}{\urlprefix }}%
\providecommand \urlprefix  [0]{URL }%
\providecommand \Eprint [0]{\href }%
\providecommand \doibase [0]{https://doi.org/}%
\providecommand \selectlanguage [0]{\@gobble}%
\providecommand \bibinfo  [0]{\@secondoftwo}%
\providecommand \bibfield  [0]{\@secondoftwo}%
\providecommand \translation [1]{[#1]}%
\providecommand \BibitemOpen [0]{}%
\providecommand \bibitemStop [0]{}%
\providecommand \bibitemNoStop [0]{.\EOS\space}%
\providecommand \EOS [0]{\spacefactor3000\relax}%
\providecommand \BibitemShut  [1]{\csname bibitem#1\endcsname}%
\let\auto@bib@innerbib\@empty
\bibitem [{\citenamefont {{Liang}}\ and\ \citenamefont
  {{Thompson}}(1980)}]{1980ApJ...240..271L}%
  \BibitemOpen
  \bibfield  {author} {\bibinfo {author} {\bibfnamefont {E.~P.~T.}\
  \bibnamefont {{Liang}}}\ and\ \bibinfo {author} {\bibfnamefont {K.~A.}\
  \bibnamefont {{Thompson}}},\ }\href {https://doi.org/10.1086/158231}
  {\bibfield  {journal} {\bibinfo  {journal} {\apj}\ }\textbf {\bibinfo
  {volume} {240}},\ \bibinfo {pages} {271} (\bibinfo {year}
  {1980})}\BibitemShut {NoStop}%
\bibitem [{\citenamefont {{Abramowicz}}\ and\ \citenamefont
  {{Zurek}}(1981)}]{1981ApJ...246..314A}%
  \BibitemOpen
  \bibfield  {author} {\bibinfo {author} {\bibfnamefont {M.~A.}\ \bibnamefont
  {{Abramowicz}}}\ and\ \bibinfo {author} {\bibfnamefont {W.~H.}\ \bibnamefont
  {{Zurek}}},\ }\href {https://doi.org/10.1086/158924} {\bibfield  {journal}
  {\bibinfo  {journal} {\apj}\ }\textbf {\bibinfo {volume} {246}},\ \bibinfo
  {pages} {314} (\bibinfo {year} {1981})}\BibitemShut {NoStop}%
\bibitem [{\citenamefont {Fukue}(1983)}]{Fukue1983}%
  \BibitemOpen
  \bibfield  {author} {\bibinfo {author} {\bibfnamefont {J.}~\bibnamefont
  {Fukue}},\ }\href@noop {} {\bibfield  {journal} {\bibinfo  {journal} {Publ.
  Astron. Soc. Jpn.}\ }\textbf {\bibinfo {volume} {35}},\ \bibinfo {pages}
  {355} (\bibinfo {year} {1983})}\BibitemShut {NoStop}%
\bibitem [{\citenamefont {Lu}(1985)}]{Lu1985}%
  \BibitemOpen
  \bibfield  {author} {\bibinfo {author} {\bibfnamefont {J.~F.}\ \bibnamefont
  {Lu}},\ }\href@noop {} {\bibfield  {journal} {\bibinfo  {journal} {Astron.
  Astrophys.}\ }\textbf {\bibinfo {volume} {148}},\ \bibinfo {pages} {176}
  (\bibinfo {year} {1985})}\BibitemShut {NoStop}%
\bibitem [{\citenamefont {Lu}(1986)}]{Lu1986}%
  \BibitemOpen
  \bibfield  {author} {\bibinfo {author} {\bibfnamefont {J.~F.}\ \bibnamefont
  {Lu}},\ }\href@noop {} {\bibfield  {journal} {\bibinfo  {journal} {Gen.
  Relativ. Gravit.}\ }\textbf {\bibinfo {volume} {18}},\ \bibinfo {pages} {45}
  (\bibinfo {year} {1986})}\BibitemShut {NoStop}%
\bibitem [{\citenamefont {Fukue}(1987)}]{Fukue1987}%
  \BibitemOpen
  \bibfield  {author} {\bibinfo {author} {\bibfnamefont {J.}~\bibnamefont
  {Fukue}},\ }\href@noop {} {\bibfield  {journal} {\bibinfo  {journal} {Publ.
  Astron. Soc. Jpn.}\ }\textbf {\bibinfo {volume} {39}},\ \bibinfo {pages}
  {309} (\bibinfo {year} {1987})}\BibitemShut {NoStop}%
\bibitem [{\citenamefont {{Blaes}}(1987)}]{1987MNRAS.227..975B}%
  \BibitemOpen
  \bibfield  {author} {\bibinfo {author} {\bibfnamefont {O.~M.}\ \bibnamefont
  {{Blaes}}},\ }\href {https://doi.org/10.1093/mnras/227.4.975} {\bibfield
  {journal} {\bibinfo  {journal} {\mnras}\ }\textbf {\bibinfo {volume} {227}},\
  \bibinfo {pages} {975} (\bibinfo {year} {1987})}\BibitemShut {NoStop}%
\bibitem [{\citenamefont {{Chakrabarti}}(1989)}]{1989ApJ...347..365C}%
  \BibitemOpen
  \bibfield  {author} {\bibinfo {author} {\bibfnamefont {S.~K.}\ \bibnamefont
  {{Chakrabarti}}},\ }\href {https://doi.org/10.1086/168125} {\bibfield
  {journal} {\bibinfo  {journal} {\apj}\ }\textbf {\bibinfo {volume} {347}},\
  \bibinfo {pages} {365} (\bibinfo {year} {1989})}\BibitemShut {NoStop}%
\bibitem [{\citenamefont {{Chakrabarti}}(1990)}]{cbook90}%
  \BibitemOpen
  \bibfield  {author} {\bibinfo {author} {\bibfnamefont {S.~K.}\ \bibnamefont
  {{Chakrabarti}}},\ }\href {https://doi.org/10.1142/1091} {\emph {\bibinfo
  {title} {{Theory of Transonic Astrophysical Flows}}}}\ (\bibinfo {year}
  {1990})\BibitemShut {NoStop}%
\bibitem [{\citenamefont {Nakayama}(1994)}]{Nakayama1994}%
  \BibitemOpen
  \bibfield  {author} {\bibinfo {author} {\bibfnamefont {K.}~\bibnamefont
  {Nakayama}},\ }\href@noop {} {\bibfield  {journal} {\bibinfo  {journal} {Mon.
  Not. R. Astron. Soc.}\ }\textbf {\bibinfo {volume} {270}},\ \bibinfo {pages}
  {871} (\bibinfo {year} {1994})}\BibitemShut {NoStop}%
\bibitem [{\citenamefont {Yang}\ and\ \citenamefont
  {Kafatos}(1995)}]{Yang1995}%
  \BibitemOpen
  \bibfield  {author} {\bibinfo {author} {\bibfnamefont {R.}~\bibnamefont
  {Yang}}\ and\ \bibinfo {author} {\bibfnamefont {M.}~\bibnamefont {Kafatos}},\
  }\href@noop {} {\bibfield  {journal} {\bibinfo  {journal} {Astron.
  Astrophys.}\ }\textbf {\bibinfo {volume} {295}},\ \bibinfo {pages} {238}
  (\bibinfo {year} {1995})}\BibitemShut {NoStop}%
\bibitem [{\citenamefont {Chakrabarti}(1996)}]{Chakrabarti1996}%
  \BibitemOpen
  \bibfield  {author} {\bibinfo {author} {\bibfnamefont {S.~K.}\ \bibnamefont
  {Chakrabarti}},\ }\href@noop {} {\bibfield  {journal} {\bibinfo  {journal}
  {Mon. Not. R. Astron. Soc.}\ }\textbf {\bibinfo {volume} {283}},\ \bibinfo
  {pages} {325} (\bibinfo {year} {1996})}\BibitemShut {NoStop}%
\bibitem [{\citenamefont {Pariev}(1996)}]{Pariev1996}%
  \BibitemOpen
  \bibfield  {author} {\bibinfo {author} {\bibfnamefont {V.~I.}\ \bibnamefont
  {Pariev}},\ }\href@noop {} {\bibfield  {journal} {\bibinfo  {journal} {Mon.
  Not. R. Astron. Soc.}\ }\textbf {\bibinfo {volume} {283}},\ \bibinfo {pages}
  {1264} (\bibinfo {year} {1996})}\BibitemShut {NoStop}%
\bibitem [{\citenamefont {J.~F.~Lu}\ and\ \citenamefont
  {Young}(1997)}]{Lu1997}%
  \BibitemOpen
  \bibfield  {author} {\bibinfo {author} {\bibfnamefont {F.~Y.}\ \bibnamefont
  {J.~F.~Lu}, \bibfnamefont {K.~N.~Yu}}\ and\ \bibinfo {author} {\bibfnamefont
  {E.~C.~M.}\ \bibnamefont {Young}},\ }\href@noop {} {\bibfield  {journal}
  {\bibinfo  {journal} {Astron. Astrophys.}\ }\textbf {\bibinfo {volume}
  {321}},\ \bibinfo {pages} {665} (\bibinfo {year} {1997})}\BibitemShut
  {NoStop}%
\bibitem [{\citenamefont {Peitz}\ and\ \citenamefont {Appl}(1997)}]{Peitz1997}%
  \BibitemOpen
  \bibfield  {author} {\bibinfo {author} {\bibfnamefont {J.}~\bibnamefont
  {Peitz}}\ and\ \bibinfo {author} {\bibfnamefont {S.}~\bibnamefont {Appl}},\
  }\href@noop {} {\bibfield  {journal} {\bibinfo  {journal} {Mon. Not. R.
  Astron. Soc.}\ }\textbf {\bibinfo {volume} {286}},\ \bibinfo {pages} {681}
  (\bibinfo {year} {1997})}\BibitemShut {NoStop}%
\bibitem [{\citenamefont {{Das}}\ \emph {et~al.}(2001)\citenamefont {{Das}},
  \citenamefont {{Chattopadhyay}},\ and\ \citenamefont
  {{Chakrabarti}}}]{2001ApJ...557..983D}%
  \BibitemOpen
  \bibfield  {author} {\bibinfo {author} {\bibfnamefont {S.}~\bibnamefont
  {{Das}}}, \bibinfo {author} {\bibfnamefont {I.}~\bibnamefont
  {{Chattopadhyay}}},\ and\ \bibinfo {author} {\bibfnamefont {S.~K.}\
  \bibnamefont {{Chakrabarti}}},\ }\href {https://doi.org/10.1086/321692}
  {\bibfield  {journal} {\bibinfo  {journal} {\apj}\ }\textbf {\bibinfo
  {volume} {557}},\ \bibinfo {pages} {983} (\bibinfo {year} {2001})},\ \Eprint
  {https://arxiv.org/abs/astro-ph/0107046} {arXiv:astro-ph/0107046 [astro-ph]}
  \BibitemShut {NoStop}%
\bibitem [{\citenamefont {P.~Barai}\ and\ \citenamefont
  {Wiita}(2004)}]{Barai2004}%
  \BibitemOpen
  \bibfield  {author} {\bibinfo {author} {\bibfnamefont {T.~K.~D.}\
  \bibnamefont {P.~Barai}}\ and\ \bibinfo {author} {\bibfnamefont {P.~J.}\
  \bibnamefont {Wiita}},\ }\href@noop {} {\bibfield  {journal} {\bibinfo
  {journal} {Astrophys. J. Lett.}\ }\textbf {\bibinfo {volume} {613}},\
  \bibinfo {pages} {L49} (\bibinfo {year} {2004})}\BibitemShut {NoStop}%
\bibitem [{\citenamefont {Takahashi}(2007)}]{Takahashi2007}%
  \BibitemOpen
  \bibfield  {author} {\bibinfo {author} {\bibfnamefont {R.}~\bibnamefont
  {Takahashi}},\ }\href@noop {} {\bibfield  {journal} {\bibinfo  {journal}
  {Mon. Not. R. Astron. Soc.}\ }\textbf {\bibinfo {volume} {382}},\ \bibinfo
  {pages} {567} (\bibinfo {year} {2007})}\BibitemShut {NoStop}%
\bibitem [{\citenamefont {Nagakura}\ and\ \citenamefont
  {Yamada}(2008)}]{Nagakura2008}%
  \BibitemOpen
  \bibfield  {author} {\bibinfo {author} {\bibfnamefont {H.}~\bibnamefont
  {Nagakura}}\ and\ \bibinfo {author} {\bibfnamefont {S.}~\bibnamefont
  {Yamada}},\ }\href@noop {} {\bibfield  {journal} {\bibinfo  {journal}
  {Astrophys. J.}\ }\textbf {\bibinfo {volume} {689}},\ \bibinfo {pages} {391}
  (\bibinfo {year} {2008})}\BibitemShut {NoStop}%
\bibitem [{\citenamefont {Nagakura}\ and\ \citenamefont
  {Yamada}(2009)}]{Nagakura2009}%
  \BibitemOpen
  \bibfield  {author} {\bibinfo {author} {\bibfnamefont {H.}~\bibnamefont
  {Nagakura}}\ and\ \bibinfo {author} {\bibfnamefont {S.}~\bibnamefont
  {Yamada}},\ }\href@noop {} {\bibfield  {journal} {\bibinfo  {journal}
  {Astrophys. J.}\ }\textbf {\bibinfo {volume} {696}},\ \bibinfo {pages} {2026}
  (\bibinfo {year} {2009})}\BibitemShut {NoStop}%
\bibitem [{\citenamefont {Das}\ and\ \citenamefont {Czerny}(2012)}]{Das2012}%
  \BibitemOpen
  \bibfield  {author} {\bibinfo {author} {\bibfnamefont {T.~K.}\ \bibnamefont
  {Das}}\ and\ \bibinfo {author} {\bibfnamefont {B.}~\bibnamefont {Czerny}},\
  }\href@noop {} {\bibfield  {journal} {\bibinfo  {journal} {New Astron.}\
  }\textbf {\bibinfo {volume} {17}},\ \bibinfo {pages} {254} (\bibinfo {year}
  {2012})}\BibitemShut {NoStop}%
\bibitem [{\citenamefont {Kumar}\ \emph {et~al.}(2013)\citenamefont {Kumar},
  \citenamefont {Singh}, \citenamefont {Chattopadhyay},\ and\ \citenamefont
  {Chakrabarti}}]{kumar2013effect}%
  \BibitemOpen
  \bibfield  {author} {\bibinfo {author} {\bibfnamefont {R.}~\bibnamefont
  {Kumar}}, \bibinfo {author} {\bibfnamefont {C.~B.}\ \bibnamefont {Singh}},
  \bibinfo {author} {\bibfnamefont {I.}~\bibnamefont {Chattopadhyay}},\ and\
  \bibinfo {author} {\bibfnamefont {S.~K.}\ \bibnamefont {Chakrabarti}},\
  }\href@noop {} {\bibfield  {journal} {\bibinfo  {journal} {Monthly Notices of
  the Royal Astronomical Society}\ }\textbf {\bibinfo {volume} {436}},\
  \bibinfo {pages} {2864} (\bibinfo {year} {2013})}\BibitemShut {NoStop}%
\bibitem [{\citenamefont {{Kumar}}\ and\ \citenamefont
  {{Chattopadhyay}}(2014)}]{kc14}%
  \BibitemOpen
  \bibfield  {author} {\bibinfo {author} {\bibfnamefont {R.}~\bibnamefont
  {{Kumar}}}\ and\ \bibinfo {author} {\bibfnamefont {I.}~\bibnamefont
  {{Chattopadhyay}}},\ }\href {https://doi.org/10.1093/mnras/stu1389}
  {\bibfield  {journal} {\bibinfo  {journal} {\mnras}\ }\textbf {\bibinfo
  {volume} {443}},\ \bibinfo {pages} {3444} (\bibinfo {year} {2014})},\ \Eprint
  {https://arxiv.org/abs/1407.2130} {arXiv:1407.2130 [astro-ph.HE]}
  \BibitemShut {NoStop}%
\bibitem [{\citenamefont {Tarafdar}\ and\ \citenamefont
  {Das}(2015)}]{Tarafdar2015}%
  \BibitemOpen
  \bibfield  {author} {\bibinfo {author} {\bibfnamefont {P.}~\bibnamefont
  {Tarafdar}}\ and\ \bibinfo {author} {\bibfnamefont {T.~K.}\ \bibnamefont
  {Das}},\ }\href@noop {} {\bibfield  {journal} {\bibinfo  {journal} {Int. J.
  Mod. Phys. D}\ }\textbf {\bibinfo {volume} {24}},\ \bibinfo {pages} {1550096}
  (\bibinfo {year} {2015})}\BibitemShut {NoStop}%
\bibitem [{\citenamefont {Suková}\ and\ \citenamefont
  {Janiuk}(2015)}]{Sukova2015}%
  \BibitemOpen
  \bibfield  {author} {\bibinfo {author} {\bibfnamefont {P.}~\bibnamefont
  {Suková}}\ and\ \bibinfo {author} {\bibfnamefont {A.}~\bibnamefont
  {Janiuk}},\ }\href@noop {} {\bibfield  {journal} {\bibinfo  {journal} {J.
  Phys. Conf. Ser.}\ }\textbf {\bibinfo {volume} {600}},\ \bibinfo {pages}
  {012012} (\bibinfo {year} {2015})}\BibitemShut {NoStop}%
\bibitem [{\citenamefont {{Chattopadhyay}}\ and\ \citenamefont
  {{Kumar}}(2016)}]{ck16}%
  \BibitemOpen
  \bibfield  {author} {\bibinfo {author} {\bibfnamefont {I.}~\bibnamefont
  {{Chattopadhyay}}}\ and\ \bibinfo {author} {\bibfnamefont {R.}~\bibnamefont
  {{Kumar}}},\ }\href {https://doi.org/10.1093/mnras/stw876} {\bibfield
  {journal} {\bibinfo  {journal} {\mnras}\ }\textbf {\bibinfo {volume} {459}},\
  \bibinfo {pages} {3792} (\bibinfo {year} {2016})},\ \Eprint
  {https://arxiv.org/abs/1605.00752} {arXiv:1605.00752 [astro-ph.HE]}
  \BibitemShut {NoStop}%
\bibitem [{\citenamefont {Le}\ \emph {et~al.}(2016)\citenamefont {Le},
  \citenamefont {Wood}, \citenamefont {Wolff}, \citenamefont {Becker},\ and\
  \citenamefont {Putney}}]{Le2016}%
  \BibitemOpen
  \bibfield  {author} {\bibinfo {author} {\bibfnamefont {T.}~\bibnamefont
  {Le}}, \bibinfo {author} {\bibfnamefont {K.~S.}\ \bibnamefont {Wood}},
  \bibinfo {author} {\bibfnamefont {M.~T.}\ \bibnamefont {Wolff}}, \bibinfo
  {author} {\bibfnamefont {P.~A.}\ \bibnamefont {Becker}},\ and\ \bibinfo
  {author} {\bibfnamefont {J.}~\bibnamefont {Putney}},\ }\href@noop {}
  {\bibfield  {journal} {\bibinfo  {journal} {Astrophys. J.}\ }\textbf
  {\bibinfo {volume} {819}},\ \bibinfo {pages} {112} (\bibinfo {year}
  {2016})}\BibitemShut {NoStop}%
\bibitem [{\citenamefont {Suková}\ \emph {et~al.}(2017)\citenamefont
  {Suková}, \citenamefont {Charzyński},\ and\ \citenamefont
  {Janiuk}}]{Sukova2017}%
  \BibitemOpen
  \bibfield  {author} {\bibinfo {author} {\bibfnamefont {P.}~\bibnamefont
  {Suková}}, \bibinfo {author} {\bibfnamefont {S.}~\bibnamefont
  {Charzyński}},\ and\ \bibinfo {author} {\bibfnamefont {A.}~\bibnamefont
  {Janiuk}},\ }\href@noop {} {\bibfield  {journal} {\bibinfo  {journal} {Mon.
  Not. R. Astron. Soc.}\ }\textbf {\bibinfo {volume} {472}},\ \bibinfo {pages}
  {4327} (\bibinfo {year} {2017})}\BibitemShut {NoStop}%
\bibitem [{\citenamefont {Kumar}\ and\ \citenamefont
  {Chattopadhyay}(2017)}]{kc17}%
  \BibitemOpen
  \bibfield  {author} {\bibinfo {author} {\bibfnamefont {R.}~\bibnamefont
  {Kumar}}\ and\ \bibinfo {author} {\bibfnamefont {I.}~\bibnamefont
  {Chattopadhyay}},\ }\href {https://doi.org/10.1093/mnras/stx1091} {\bibfield
  {journal} {\bibinfo  {journal} {Monthly Notices of the Royal Astronomical
  Society}\ }\textbf {\bibinfo {volume} {469}},\ \bibinfo {pages} {4221}
  (\bibinfo {year} {2017})},\ \Eprint
  {https://arxiv.org/abs/https://academic.oup.com/mnras/article-pdf/469/4/4221/17715028/stx1091.pdf}
  {https://academic.oup.com/mnras/article-pdf/469/4/4221/17715028/stx1091.pdf}
  \BibitemShut {NoStop}%
\bibitem [{\citenamefont {Palit}\ \emph {et~al.}(2019)\citenamefont {Palit},
  \citenamefont {Janiuk},\ and\ \citenamefont {Sukova}}]{Palit2019}%
  \BibitemOpen
  \bibfield  {author} {\bibinfo {author} {\bibfnamefont {I.}~\bibnamefont
  {Palit}}, \bibinfo {author} {\bibfnamefont {A.}~\bibnamefont {Janiuk}},\ and\
  \bibinfo {author} {\bibfnamefont {P.}~\bibnamefont {Sukova}},\ }\href@noop {}
  {\bibfield  {journal} {\bibinfo  {journal} {Mon. Not. R. Astron. Soc.}\
  }\textbf {\bibinfo {volume} {487}},\ \bibinfo {pages} {755} (\bibinfo {year}
  {2019})}\BibitemShut {NoStop}%
\bibitem [{\citenamefont {Palit}\ \emph {et~al.}(2020)\citenamefont {Palit},
  \citenamefont {Janiuk},\ and\ \citenamefont {Czerny}}]{Palit2020b}%
  \BibitemOpen
  \bibfield  {author} {\bibinfo {author} {\bibfnamefont {I.}~\bibnamefont
  {Palit}}, \bibinfo {author} {\bibfnamefont {A.}~\bibnamefont {Janiuk}},\ and\
  \bibinfo {author} {\bibfnamefont {B.}~\bibnamefont {Czerny}},\ }\href@noop {}
  {\bibfield  {journal} {\bibinfo  {journal} {Astrophys. J.}\ }\textbf
  {\bibinfo {volume} {904}},\ \bibinfo {pages} {21} (\bibinfo {year}
  {2020})}\BibitemShut {NoStop}%
\bibitem [{\citenamefont {{Sarkar}}\ \emph {et~al.}(2020)\citenamefont
  {{Sarkar}}, \citenamefont {{Chattopadhyay}},\ and\ \citenamefont
  {{Laurent}}}]{scp20}%
  \BibitemOpen
  \bibfield  {author} {\bibinfo {author} {\bibfnamefont {S.}~\bibnamefont
  {{Sarkar}}}, \bibinfo {author} {\bibfnamefont {I.}~\bibnamefont
  {{Chattopadhyay}}},\ and\ \bibinfo {author} {\bibfnamefont {P.}~\bibnamefont
  {{Laurent}}},\ }\href {https://doi.org/10.1051/0004-6361/202037520}
  {\bibfield  {journal} {\bibinfo  {journal} {\aap}\ }\textbf {\bibinfo
  {volume} {642}},\ \bibinfo {eid} {A209} (\bibinfo {year} {2020})},\ \Eprint
  {https://arxiv.org/abs/2007.00919} {arXiv:2007.00919 [astro-ph.HE]}
  \BibitemShut {NoStop}%
\bibitem [{\citenamefont {Tarafdar}\ \emph {et~al.}(2021)\citenamefont
  {Tarafdar}, \citenamefont {Maity},\ and\ \citenamefont {Das}}]{Tarafdar2021}%
  \BibitemOpen
  \bibfield  {author} {\bibinfo {author} {\bibfnamefont {P.}~\bibnamefont
  {Tarafdar}}, \bibinfo {author} {\bibfnamefont {S.}~\bibnamefont {Maity}},\
  and\ \bibinfo {author} {\bibfnamefont {T.~K.}\ \bibnamefont {Das}},\
  }\href@noop {} {\bibfield  {journal} {\bibinfo  {journal} {Phys. Rev. D}\
  }\textbf {\bibinfo {volume} {103}},\ \bibinfo {pages} {023023} (\bibinfo
  {year} {2021})}\BibitemShut {NoStop}%
\bibitem [{\citenamefont {{Sarkar}}\ and\ \citenamefont
  {{Chattopadhyay}}(2019)}]{sc19a}%
  \BibitemOpen
  \bibfield  {author} {\bibinfo {author} {\bibfnamefont {S.}~\bibnamefont
  {{Sarkar}}}\ and\ \bibinfo {author} {\bibfnamefont {I.}~\bibnamefont
  {{Chattopadhyay}}},\ }\href {https://doi.org/10.1142/S0218271819500378}
  {\bibfield  {journal} {\bibinfo  {journal} {International Journal of Modern
  Physics D}\ }\textbf {\bibinfo {volume} {28}},\ \bibinfo {pages} {1950037}
  (\bibinfo {year} {2019})},\ \Eprint {https://arxiv.org/abs/1811.05947}
  {arXiv:1811.05947 [astro-ph.HE]} \BibitemShut {NoStop}%
\bibitem [{\citenamefont {{Ryu}}\ \emph {et~al.}(2006)\citenamefont {{Ryu}},
  \citenamefont {{Chattopadhyay}},\ and\ \citenamefont
  {{Choi}}}]{2006ApJS..166..410R}%
  \BibitemOpen
  \bibfield  {author} {\bibinfo {author} {\bibfnamefont {D.}~\bibnamefont
  {{Ryu}}}, \bibinfo {author} {\bibfnamefont {I.}~\bibnamefont
  {{Chattopadhyay}}},\ and\ \bibinfo {author} {\bibfnamefont {E.}~\bibnamefont
  {{Choi}}},\ }\href {https://doi.org/10.1086/505937} {\bibfield  {journal}
  {\bibinfo  {journal} {\apjs}\ }\textbf {\bibinfo {volume} {166}},\ \bibinfo
  {pages} {410} (\bibinfo {year} {2006})},\ \Eprint
  {https://arxiv.org/abs/astro-ph/0605550} {arXiv:astro-ph/0605550 [astro-ph]}
  \BibitemShut {NoStop}%
\bibitem [{\citenamefont {{Chattopadhyay}}\ and\ \citenamefont
  {{Ryu}}(2009)}]{2009ApJ...694..492C}%
  \BibitemOpen
  \bibfield  {author} {\bibinfo {author} {\bibfnamefont {I.}~\bibnamefont
  {{Chattopadhyay}}}\ and\ \bibinfo {author} {\bibfnamefont {D.}~\bibnamefont
  {{Ryu}}},\ }\href {https://doi.org/10.1088/0004-637X/694/1/492} {\bibfield
  {journal} {\bibinfo  {journal} {\apj}\ }\textbf {\bibinfo {volume} {694}},\
  \bibinfo {pages} {492} (\bibinfo {year} {2009})},\ \Eprint
  {https://arxiv.org/abs/0812.2607} {arXiv:0812.2607 [astro-ph]} \BibitemShut
  {NoStop}%
\bibitem [{\citenamefont {Barcel{\'o}}\ \emph {et~al.}(2011)\citenamefont
  {Barcel{\'o}}, \citenamefont {Liberati},\ and\ \citenamefont
  {Visser}}]{Barcelo2011}%
  \BibitemOpen
  \bibfield  {author} {\bibinfo {author} {\bibfnamefont {C.}~\bibnamefont
  {Barcel{\'o}}}, \bibinfo {author} {\bibfnamefont {S.}~\bibnamefont
  {Liberati}},\ and\ \bibinfo {author} {\bibfnamefont {M.}~\bibnamefont
  {Visser}},\ }\href {https://doi.org/10.12942/lrr-2011-3} {\bibfield
  {journal} {\bibinfo  {journal} {Living Reviews in Relativity}\ }\textbf
  {\bibinfo {volume} {14}},\ \bibinfo {pages} {3} (\bibinfo {year}
  {2011})}\BibitemShut {NoStop}%
\bibitem [{\citenamefont {Shaikh}\ \emph {et~al.}(2017)\citenamefont {Shaikh},
  \citenamefont {Firdousi},\ and\ \citenamefont {Das}}]{Shaikh_2017}%
  \BibitemOpen
  \bibfield  {author} {\bibinfo {author} {\bibfnamefont {M.~A.}\ \bibnamefont
  {Shaikh}}, \bibinfo {author} {\bibfnamefont {I.}~\bibnamefont {Firdousi}},\
  and\ \bibinfo {author} {\bibfnamefont {T.~K.}\ \bibnamefont {Das}},\ }\href
  {https://doi.org/10.1088/1361-6382/aa7b19} {\bibfield  {journal} {\bibinfo
  {journal} {Classical and Quantum Gravity}\ }\textbf {\bibinfo {volume}
  {34}},\ \bibinfo {pages} {155008} (\bibinfo {year} {2017})}\BibitemShut
  {NoStop}%
\bibitem [{\citenamefont {Tarafdar}\ \emph {et~al.}(2019)\citenamefont
  {Tarafdar}, \citenamefont {Bollimpalli}, \citenamefont {Nag},\ and\
  \citenamefont {Das}}]{PhysRevD.100.043024}%
  \BibitemOpen
  \bibfield  {author} {\bibinfo {author} {\bibfnamefont {P.}~\bibnamefont
  {Tarafdar}}, \bibinfo {author} {\bibfnamefont {D.~A.}\ \bibnamefont
  {Bollimpalli}}, \bibinfo {author} {\bibfnamefont {S.}~\bibnamefont {Nag}},\
  and\ \bibinfo {author} {\bibfnamefont {T.~K.}\ \bibnamefont {Das}},\ }\href
  {https://doi.org/10.1103/PhysRevD.100.043024} {\bibfield  {journal} {\bibinfo
   {journal} {Phys. Rev. D}\ }\textbf {\bibinfo {volume} {100}},\ \bibinfo
  {pages} {043024} (\bibinfo {year} {2019})}\BibitemShut {NoStop}%
\bibitem [{\citenamefont {Maity}\ \emph {et~al.}(2022)\citenamefont {Maity},
  \citenamefont {Shaikh}, \citenamefont {Tarafdar},\ and\ \citenamefont
  {Das}}]{PhysRevD.106.044062}%
  \BibitemOpen
  \bibfield  {author} {\bibinfo {author} {\bibfnamefont {S.}~\bibnamefont
  {Maity}}, \bibinfo {author} {\bibfnamefont {M.~A.}\ \bibnamefont {Shaikh}},
  \bibinfo {author} {\bibfnamefont {P.}~\bibnamefont {Tarafdar}},\ and\
  \bibinfo {author} {\bibfnamefont {T.~K.}\ \bibnamefont {Das}},\ }\href
  {https://doi.org/10.1103/PhysRevD.106.044062} {\bibfield  {journal} {\bibinfo
   {journal} {Phys. Rev. D}\ }\textbf {\bibinfo {volume} {106}},\ \bibinfo
  {pages} {044062} (\bibinfo {year} {2022})}\BibitemShut {NoStop}%
\bibitem [{\citenamefont {{Artemova}}\ \emph {et~al.}(1996)\citenamefont
  {{Artemova}}, \citenamefont {{Bjoernsson}},\ and\ \citenamefont
  {{Novikov}}}]{1996ApJ...461..565A}%
  \BibitemOpen
  \bibfield  {author} {\bibinfo {author} {\bibfnamefont {I.~V.}\ \bibnamefont
  {{Artemova}}}, \bibinfo {author} {\bibfnamefont {G.}~\bibnamefont
  {{Bjoernsson}}},\ and\ \bibinfo {author} {\bibfnamefont {I.~D.}\ \bibnamefont
  {{Novikov}}},\ }\href {https://doi.org/10.1086/177084} {\bibfield  {journal}
  {\bibinfo  {journal} {\apj}\ }\textbf {\bibinfo {volume} {461}},\ \bibinfo
  {pages} {565} (\bibinfo {year} {1996})}\BibitemShut {NoStop}%
\bibitem [{\citenamefont {Dhruv}\ \emph {et~al.}(2025)\citenamefont {Dhruv},
  \citenamefont {Prather}, \citenamefont {Wong},\ and\ \citenamefont
  {Gammie}}]{Dhruv_2025}%
  \BibitemOpen
  \bibfield  {author} {\bibinfo {author} {\bibfnamefont {V.}~\bibnamefont
  {Dhruv}}, \bibinfo {author} {\bibfnamefont {B.}~\bibnamefont {Prather}},
  \bibinfo {author} {\bibfnamefont {G.~N.}\ \bibnamefont {Wong}},\ and\
  \bibinfo {author} {\bibfnamefont {C.~F.}\ \bibnamefont {Gammie}},\ }\href
  {https://doi.org/10.3847/1538-4365/adaea6} {\bibfield  {journal} {\bibinfo
  {journal} {The Astrophysical Journal Supplement Series}\ }\textbf {\bibinfo
  {volume} {277}},\ \bibinfo {pages} {16} (\bibinfo {year} {2025})}\BibitemShut
  {NoStop}%
\bibitem [{\citenamefont {Porth}\ \emph {et~al.}(2019)\citenamefont {Porth}
  \emph {et~al.}}]{Porth2019}%
  \BibitemOpen
  \bibfield  {author} {\bibinfo {author} {\bibfnamefont {O.}~\bibnamefont
  {Porth}} \emph {et~al.},\ }\href {https://doi.org/10.3847/1538-4365/ab29fd}
  {\bibfield  {journal} {\bibinfo  {journal} {The Astrophysical Journal
  Supplement Series}\ }\textbf {\bibinfo {volume} {243}},\ \bibinfo {pages}
  {26} (\bibinfo {year} {2019})}\BibitemShut {NoStop}%
\bibitem [{\citenamefont {Mitra}\ and\ \citenamefont {Das}(2024)}]{Mitra_2024}%
  \BibitemOpen
  \bibfield  {author} {\bibinfo {author} {\bibfnamefont {S.}~\bibnamefont
  {Mitra}}\ and\ \bibinfo {author} {\bibfnamefont {S.}~\bibnamefont {Das}},\
  }\href {https://doi.org/10.3847/1538-4357/ad55cb} {\bibfield  {journal}
  {\bibinfo  {journal} {The Astrophysical Journal}\ }\textbf {\bibinfo {volume}
  {971}},\ \bibinfo {pages} {28} (\bibinfo {year} {2024})}\BibitemShut
  {NoStop}%
\bibitem [{\citenamefont {Sharma}\ \emph {et~al.}(2025)\citenamefont {Sharma},
  \citenamefont {Medeiros}, \citenamefont {Wong}, \citenamefont {Chan},
  \citenamefont {Halevi}, \citenamefont {Mullen},\ and\ \citenamefont
  {Stone}}]{Sharma_2025}%
  \BibitemOpen
  \bibfield  {author} {\bibinfo {author} {\bibfnamefont {A.}~\bibnamefont
  {Sharma}}, \bibinfo {author} {\bibfnamefont {L.}~\bibnamefont {Medeiros}},
  \bibinfo {author} {\bibfnamefont {G.~N.}\ \bibnamefont {Wong}}, \bibinfo
  {author} {\bibfnamefont {C.-k.}\ \bibnamefont {Chan}}, \bibinfo {author}
  {\bibfnamefont {G.}~\bibnamefont {Halevi}}, \bibinfo {author} {\bibfnamefont
  {P.~D.}\ \bibnamefont {Mullen}},\ and\ \bibinfo {author} {\bibfnamefont
  {J.~M.}\ \bibnamefont {Stone}},\ }\href
  {https://doi.org/10.3847/1538-4357/adc104} {\bibfield  {journal} {\bibinfo
  {journal} {The Astrophysical Journal}\ }\textbf {\bibinfo {volume} {985}},\
  \bibinfo {pages} {40} (\bibinfo {year} {2025})}\BibitemShut {NoStop}%
\bibitem [{\citenamefont {Dexter}(2016)}]{10.1093/mnras/stw1526}%
  \BibitemOpen
  \bibfield  {author} {\bibinfo {author} {\bibfnamefont {J.}~\bibnamefont
  {Dexter}},\ }\href {https://doi.org/10.1093/mnras/stw1526} {\bibfield
  {journal} {\bibinfo  {journal} {Monthly Notices of the Royal Astronomical
  Society}\ }\textbf {\bibinfo {volume} {462}},\ \bibinfo {pages} {115}
  (\bibinfo {year} {2016})},\ \Eprint
  {https://arxiv.org/abs/https://academic.oup.com/mnras/article-pdf/462/1/115/18469081/stw1526.pdf}
  {https://academic.oup.com/mnras/article-pdf/462/1/115/18469081/stw1526.pdf}
  \BibitemShut {NoStop}%
\bibitem [{\citenamefont {Schnittman}\ and\ \citenamefont
  {Krolik}(2013)}]{Schnittman_2013}%
  \BibitemOpen
  \bibfield  {author} {\bibinfo {author} {\bibfnamefont {J.~D.}\ \bibnamefont
  {Schnittman}}\ and\ \bibinfo {author} {\bibfnamefont {J.~H.}\ \bibnamefont
  {Krolik}},\ }\href {https://doi.org/10.1088/0004-637X/777/1/11} {\bibfield
  {journal} {\bibinfo  {journal} {The Astrophysical Journal}\ }\textbf
  {\bibinfo {volume} {777}},\ \bibinfo {pages} {11} (\bibinfo {year}
  {2013})}\BibitemShut {NoStop}%
\bibitem [{\citenamefont {Hu}\ \emph {et~al.}(2022)\citenamefont {Hu},
  \citenamefont {Hou}, \citenamefont {Yan}, \citenamefont {Guo},\ and\
  \citenamefont {Chen}}]{Hu2022}%
  \BibitemOpen
  \bibfield  {author} {\bibinfo {author} {\bibfnamefont {Z.}~\bibnamefont
  {Hu}}, \bibinfo {author} {\bibfnamefont {Y.}~\bibnamefont {Hou}}, \bibinfo
  {author} {\bibfnamefont {H.}~\bibnamefont {Yan}}, \bibinfo {author}
  {\bibfnamefont {M.}~\bibnamefont {Guo}},\ and\ \bibinfo {author}
  {\bibfnamefont {B.}~\bibnamefont {Chen}},\ }\href
  {https://doi.org/10.1140/epjc/s10052-022-11144-9} {\bibfield  {journal}
  {\bibinfo  {journal} {The European Physical Journal C}\ }\textbf {\bibinfo
  {volume} {82}},\ \bibinfo {pages} {1166} (\bibinfo {year}
  {2022})}\BibitemShut {NoStop}%
\bibitem [{\citenamefont {Zhu}\ \emph {et~al.}(2015)\citenamefont {Zhu},
  \citenamefont {Narayan}, \citenamefont {Sadowski},\ and\ \citenamefont
  {Psaltis}}]{10.1093/mnras/stv1046}%
  \BibitemOpen
  \bibfield  {author} {\bibinfo {author} {\bibfnamefont {Y.}~\bibnamefont
  {Zhu}}, \bibinfo {author} {\bibfnamefont {R.}~\bibnamefont {Narayan}},
  \bibinfo {author} {\bibfnamefont {A.}~\bibnamefont {Sadowski}},\ and\
  \bibinfo {author} {\bibfnamefont {D.}~\bibnamefont {Psaltis}},\ }\href
  {https://doi.org/10.1093/mnras/stv1046} {\bibfield  {journal} {\bibinfo
  {journal} {Monthly Notices of the Royal Astronomical Society}\ }\textbf
  {\bibinfo {volume} {451}},\ \bibinfo {pages} {1661} (\bibinfo {year}
  {2015})},\ \Eprint
  {https://arxiv.org/abs/https://academic.oup.com/mnras/article-pdf/451/2/1661/5705021/stv1046.pdf}
  {https://academic.oup.com/mnras/article-pdf/451/2/1661/5705021/stv1046.pdf}
  \BibitemShut {NoStop}%
\bibitem [{\citenamefont {{Pihajoki}}\ \emph {et~al.}(2017)\citenamefont
  {{Pihajoki}}, \citenamefont {{Rantala}},\ and\ \citenamefont
  {{Johansson}}}]{2017IAUS..324..347P}%
  \BibitemOpen
  \bibfield  {author} {\bibinfo {author} {\bibfnamefont {P.}~\bibnamefont
  {{Pihajoki}}}, \bibinfo {author} {\bibfnamefont {A.}~\bibnamefont
  {{Rantala}}},\ and\ \bibinfo {author} {\bibfnamefont {P.~H.}\ \bibnamefont
  {{Johansson}}},\ }in\ \href {https://doi.org/10.1017/S1743921316012990}
  {\emph {\bibinfo {booktitle} {New Frontiers in Black Hole Astrophysics}}},\
  \bibinfo {series} {IAU Symposium}, Vol.\ \bibinfo {volume} {324},\ \bibinfo
  {editor} {edited by\ \bibinfo {editor} {\bibfnamefont {A.}~\bibnamefont
  {{Gomboc}}}}\ (\bibinfo {year} {2017})\ pp.\ \bibinfo {pages} {347--350},\
  \Eprint {https://arxiv.org/abs/1612.02828} {arXiv:1612.02828 [astro-ph.IM]}
  \BibitemShut {NoStop}%
\bibitem [{\citenamefont {Shcherbakov}\ and\ \citenamefont
  {Huang}(2011)}]{10.1111/j.1365-2966.2010.17502.x}%
  \BibitemOpen
  \bibfield  {author} {\bibinfo {author} {\bibfnamefont {R.~V.}\ \bibnamefont
  {Shcherbakov}}\ and\ \bibinfo {author} {\bibfnamefont {L.}~\bibnamefont
  {Huang}},\ }\href {https://doi.org/10.1111/j.1365-2966.2010.17502.x}
  {\bibfield  {journal} {\bibinfo  {journal} {Monthly Notices of the Royal
  Astronomical Society}\ }\textbf {\bibinfo {volume} {410}},\ \bibinfo {pages}
  {1052} (\bibinfo {year} {2011})},\ \Eprint
  {https://arxiv.org/abs/https://academic.oup.com/mnras/article-pdf/410/2/1052/3438342/mnras0410-1052.pdf}
  {https://academic.oup.com/mnras/article-pdf/410/2/1052/3438342/mnras0410-1052.pdf}
  \BibitemShut {NoStop}%
\bibitem [{\citenamefont {{Chakrabarti}}\ and\ \citenamefont
  {{Khanna}}(1992)}]{1992MNRAS.256..300C}%
  \BibitemOpen
  \bibfield  {author} {\bibinfo {author} {\bibfnamefont {S.~K.}\ \bibnamefont
  {{Chakrabarti}}}\ and\ \bibinfo {author} {\bibfnamefont {R.}~\bibnamefont
  {{Khanna}}},\ }\href {https://doi.org/10.1093/mnras/256.2.300} {\bibfield
  {journal} {\bibinfo  {journal} {\mnras}\ }\textbf {\bibinfo {volume} {256}},\
  \bibinfo {pages} {300} (\bibinfo {year} {1992})}\BibitemShut {NoStop}%
\bibitem [{\citenamefont {L{\o}v{\aa}s}(1998)}]{lovaas1998modified}%
  \BibitemOpen
  \bibfield  {author} {\bibinfo {author} {\bibfnamefont {T.}~\bibnamefont
  {L{\o}v{\aa}s}},\ }\href@noop {} {\bibfield  {journal} {\bibinfo  {journal}
  {International Journal of Modern Physics D}\ }\textbf {\bibinfo {volume}
  {7}},\ \bibinfo {pages} {471} (\bibinfo {year} {1998})}\BibitemShut {NoStop}%
\bibitem [{\citenamefont {{Semer{\'a}k}}\ and\ \citenamefont
  {{Karas}}(1999)}]{1999A&A...343..325S}%
  \BibitemOpen
  \bibfield  {author} {\bibinfo {author} {\bibfnamefont {O.}~\bibnamefont
  {{Semer{\'a}k}}}\ and\ \bibinfo {author} {\bibfnamefont {V.}~\bibnamefont
  {{Karas}}},\ }\href {https://doi.org/10.48550/arXiv.astro-ph/9901289}
  {\bibfield  {journal} {\bibinfo  {journal} {\aap}\ }\textbf {\bibinfo
  {volume} {343}},\ \bibinfo {pages} {325} (\bibinfo {year} {1999})},\ \Eprint
  {https://arxiv.org/abs/astro-ph/9901289} {arXiv:astro-ph/9901289 [astro-ph]}
  \BibitemShut {NoStop}%
\bibitem [{\citenamefont {Mukhopadhyay}(2002)}]{Mukhopadhyay_2002}%
  \BibitemOpen
  \bibfield  {author} {\bibinfo {author} {\bibfnamefont {B.}~\bibnamefont
  {Mukhopadhyay}},\ }\href {https://doi.org/10.1086/344227} {\bibfield
  {journal} {\bibinfo  {journal} {The Astrophysical Journal}\ }\textbf
  {\bibinfo {volume} {581}},\ \bibinfo {pages} {427} (\bibinfo {year}
  {2002})}\BibitemShut {NoStop}%
\bibitem [{\citenamefont {Chakrabarti}\ and\ \citenamefont
  {Mondal}(2006)}]{10.1111/j.1365-2966.2006.10350.x}%
  \BibitemOpen
  \bibfield  {author} {\bibinfo {author} {\bibfnamefont {S.~K.}\ \bibnamefont
  {Chakrabarti}}\ and\ \bibinfo {author} {\bibfnamefont {S.}~\bibnamefont
  {Mondal}},\ }\href {https://doi.org/10.1111/j.1365-2966.2006.10350.x}
  {\bibfield  {journal} {\bibinfo  {journal} {Monthly Notices of the Royal
  Astronomical Society}\ }\textbf {\bibinfo {volume} {369}},\ \bibinfo {pages}
  {976} (\bibinfo {year} {2006})},\ \Eprint
  {https://arxiv.org/abs/https://academic.oup.com/mnras/article-pdf/369/2/976/11182178/mnras0369-0976.pdf}
  {https://academic.oup.com/mnras/article-pdf/369/2/976/11182178/mnras0369-0976.pdf}
  \BibitemShut {NoStop}%
\bibitem [{\citenamefont {Ghosh}\ and\ \citenamefont
  {Mukhopadhyay}(2007)}]{ghosh2007generalized}%
  \BibitemOpen
  \bibfield  {author} {\bibinfo {author} {\bibfnamefont {S.}~\bibnamefont
  {Ghosh}}\ and\ \bibinfo {author} {\bibfnamefont {B.}~\bibnamefont
  {Mukhopadhyay}},\ }\href@noop {} {\bibfield  {journal} {\bibinfo  {journal}
  {The Astrophysical Journal}\ }\textbf {\bibinfo {volume} {667}},\ \bibinfo
  {pages} {367} (\bibinfo {year} {2007})}\BibitemShut {NoStop}%
\bibitem [{\citenamefont {Ghosh}\ \emph {et~al.}(2014)\citenamefont {Ghosh},
  \citenamefont {Sarkar},\ and\ \citenamefont {Bhadra}}]{Ghosh_2014}%
  \BibitemOpen
  \bibfield  {author} {\bibinfo {author} {\bibfnamefont {S.}~\bibnamefont
  {Ghosh}}, \bibinfo {author} {\bibfnamefont {T.}~\bibnamefont {Sarkar}},\ and\
  \bibinfo {author} {\bibfnamefont {A.}~\bibnamefont {Bhadra}},\ }\href
  {https://doi.org/10.1093/mnras/stu2046} {\bibfield  {journal} {\bibinfo
  {journal} {Monthly Notices of the Royal Astronomical Society}\ }\textbf
  {\bibinfo {volume} {445}},\ \bibinfo {pages} {4460–4476} (\bibinfo {year}
  {2014})}\BibitemShut {NoStop}%
\bibitem [{\citenamefont {{Karas}}\ and\ \citenamefont
  {{Abramowicz}}(2014)}]{2014bhns.work..121K}%
  \BibitemOpen
  \bibfield  {author} {\bibinfo {author} {\bibfnamefont {V.}~\bibnamefont
  {{Karas}}}\ and\ \bibinfo {author} {\bibfnamefont {M.~A.}\ \bibnamefont
  {{Abramowicz}}},\ }in\ \href {https://doi.org/10.48550/arXiv.1412.6832}
  {\emph {\bibinfo {booktitle} {Proceedings of RAGtime 10-13: Workshops on
  black holes and neutron stars}}}\ (\bibinfo {year} {2014})\ pp.\ \bibinfo
  {pages} {121--128},\ \Eprint {https://arxiv.org/abs/1412.6832}
  {arXiv:1412.6832 [astro-ph.HE]} \BibitemShut {NoStop}%
\bibitem [{\citenamefont {Chakrabarti}\ and\ \citenamefont
  {Das}(2001)}]{https://doi.org/10.1046/j.1365-8711.2001.04758.x}%
  \BibitemOpen
  \bibfield  {author} {\bibinfo {author} {\bibfnamefont {S.~K.}\ \bibnamefont
  {Chakrabarti}}\ and\ \bibinfo {author} {\bibfnamefont {S.}~\bibnamefont
  {Das}},\ }\href
  {https://doi.org/https://doi.org/10.1046/j.1365-8711.2001.04758.x} {\bibfield
   {journal} {\bibinfo  {journal} {Monthly Notices of the Royal Astronomical
  Society}\ }\textbf {\bibinfo {volume} {327}},\ \bibinfo {pages} {808}
  (\bibinfo {year} {2001})},\ \Eprint
  {https://arxiv.org/abs/https://onlinelibrary.wiley.com/doi/pdf/10.1046/j.1365-8711.2001.04758.x}
  {https://onlinelibrary.wiley.com/doi/pdf/10.1046/j.1365-8711.2001.04758.x}
  \BibitemShut {NoStop}%
\bibitem [{\citenamefont {Nag}\ \emph {et~al.}(2012)\citenamefont {Nag},
  \citenamefont {Acharya}, \citenamefont {Ray},\ and\ \citenamefont
  {Das}}]{NAG2012285}%
  \BibitemOpen
  \bibfield  {author} {\bibinfo {author} {\bibfnamefont {S.}~\bibnamefont
  {Nag}}, \bibinfo {author} {\bibfnamefont {S.}~\bibnamefont {Acharya}},
  \bibinfo {author} {\bibfnamefont {A.~K.}\ \bibnamefont {Ray}},\ and\ \bibinfo
  {author} {\bibfnamefont {T.~K.}\ \bibnamefont {Das}},\ }\href
  {https://doi.org/https://doi.org/10.1016/j.newast.2011.06.013} {\bibfield
  {journal} {\bibinfo  {journal} {New Astronomy}\ }\textbf {\bibinfo {volume}
  {17}},\ \bibinfo {pages} {285} (\bibinfo {year} {2012})}\BibitemShut
  {NoStop}%
\bibitem [{\citenamefont {Bilić}\ \emph {et~al.}(2013)\citenamefont {Bilić},
  \citenamefont {Choudhary}, \citenamefont {Das},\ and\ \citenamefont
  {Nag}}]{Bilić_2014}%
  \BibitemOpen
  \bibfield  {author} {\bibinfo {author} {\bibfnamefont {N.}~\bibnamefont
  {Bilić}}, \bibinfo {author} {\bibfnamefont {A.}~\bibnamefont {Choudhary}},
  \bibinfo {author} {\bibfnamefont {T.~K.}\ \bibnamefont {Das}},\ and\ \bibinfo
  {author} {\bibfnamefont {S.}~\bibnamefont {Nag}},\ }\href
  {https://doi.org/10.1088/0264-9381/31/3/035002} {\bibfield  {journal}
  {\bibinfo  {journal} {Classical and Quantum Gravity}\ }\textbf {\bibinfo
  {volume} {31}},\ \bibinfo {pages} {035002} (\bibinfo {year}
  {2013})}\BibitemShut {NoStop}%
\bibitem [{\citenamefont {{Hubeny}}\ and\ \citenamefont
  {{Hubeny}}(1998)}]{1998ApJ...505..558H}%
  \BibitemOpen
  \bibfield  {author} {\bibinfo {author} {\bibfnamefont {I.}~\bibnamefont
  {{Hubeny}}}\ and\ \bibinfo {author} {\bibfnamefont {V.}~\bibnamefont
  {{Hubeny}}},\ }\href {https://doi.org/10.1086/306207} {\bibfield  {journal}
  {\bibinfo  {journal} {\apj}\ }\textbf {\bibinfo {volume} {505}},\ \bibinfo
  {pages} {558} (\bibinfo {year} {1998})},\ \Eprint
  {https://arxiv.org/abs/astro-ph/9804288} {arXiv:astro-ph/9804288 [astro-ph]}
  \BibitemShut {NoStop}%
\bibitem [{\citenamefont {{Davis}}\ and\ \citenamefont
  {{Hubeny}}(2006)}]{2006ApJS..164..530D}%
  \BibitemOpen
  \bibfield  {author} {\bibinfo {author} {\bibfnamefont {S.~W.}\ \bibnamefont
  {{Davis}}}\ and\ \bibinfo {author} {\bibfnamefont {I.}~\bibnamefont
  {{Hubeny}}},\ }\href {https://doi.org/10.1086/503549} {\bibfield  {journal}
  {\bibinfo  {journal} {\apjs}\ }\textbf {\bibinfo {volume} {164}},\ \bibinfo
  {pages} {530} (\bibinfo {year} {2006})},\ \Eprint
  {https://arxiv.org/abs/astro-ph/0602499} {arXiv:astro-ph/0602499 [astro-ph]}
  \BibitemShut {NoStop}%
\bibitem [{\citenamefont {{Beskin}}(1997)}]{1997PhyU...40..659B}%
  \BibitemOpen
  \bibfield  {author} {\bibinfo {author} {\bibfnamefont {V.~S.}\ \bibnamefont
  {{Beskin}}},\ }\href {https://doi.org/10.1070/PU1997v040n07ABEH000250}
  {\bibfield  {journal} {\bibinfo  {journal} {Physics Uspekhi}\ }\textbf
  {\bibinfo {volume} {40}},\ \bibinfo {pages} {659} (\bibinfo {year}
  {1997})}\BibitemShut {NoStop}%
\bibitem [{\citenamefont {Beskin}\ and\ \citenamefont
  {Tchekhovskoy}(2005)}]{Beskin_2005}%
  \BibitemOpen
  \bibfield  {author} {\bibinfo {author} {\bibfnamefont {V.}~\bibnamefont
  {Beskin}}\ and\ \bibinfo {author} {\bibfnamefont {A.}~\bibnamefont
  {Tchekhovskoy}},\ }\href {https://doi.org/10.1051/0004-6361:20041592}
  {\bibfield  {journal} {\bibinfo  {journal} {Astronomy \& Astrophysics}\
  }\textbf {\bibinfo {volume} {433}},\ \bibinfo {pages} {619–628} (\bibinfo
  {year} {2005})}\BibitemShut {NoStop}%
\bibitem [{\citenamefont {{Beskin}}(2009)}]{2009mfca.book.....B}%
  \BibitemOpen
  \bibfield  {author} {\bibinfo {author} {\bibfnamefont {V.~S.}\ \bibnamefont
  {{Beskin}}},\ }\href@noop {} {\emph {\bibinfo {title} {{MHD Flows in Compact
  Astrophysical Objects: Accretion, Winds and Jets}}}}\ (\bibinfo {year}
  {2009})\BibitemShut {NoStop}%
\bibitem [{\citenamefont {Balbus}\ and\ \citenamefont
  {Hawley}(1998)}]{RevModPhys.70.1}%
  \BibitemOpen
  \bibfield  {author} {\bibinfo {author} {\bibfnamefont {S.~A.}\ \bibnamefont
  {Balbus}}\ and\ \bibinfo {author} {\bibfnamefont {J.~F.}\ \bibnamefont
  {Hawley}},\ }\href {https://doi.org/10.1103/RevModPhys.70.1} {\bibfield
  {journal} {\bibinfo  {journal} {Rev. Mod. Phys.}\ }\textbf {\bibinfo {volume}
  {70}},\ \bibinfo {pages} {1} (\bibinfo {year} {1998})}\BibitemShut {NoStop}%
\bibitem [{\citenamefont {Paul}\ \emph {et~al.}(2025)\citenamefont {Paul},
  \citenamefont {Chakraborty}, \citenamefont {Ghose},\ and\ \citenamefont
  {Das}}]{paul2025gravity}%
  \BibitemOpen
  \bibfield  {author} {\bibinfo {author} {\bibfnamefont {T.}~\bibnamefont
  {Paul}}, \bibinfo {author} {\bibfnamefont {A.}~\bibnamefont {Chakraborty}},
  \bibinfo {author} {\bibfnamefont {S.}~\bibnamefont {Ghose}},\ and\ \bibinfo
  {author} {\bibfnamefont {T.~K.}\ \bibnamefont {Das}},\ }\href@noop {}
  {\bibfield  {journal} {\bibinfo  {journal} {arXiv e-prints}\ ,\ \bibinfo
  {pages} {arXiv}} (\bibinfo {year} {2025})}\BibitemShut {NoStop}%
\bibitem [{\citenamefont {{Bondi}}(1952)}]{1952MNRAS.112..195B}%
  \BibitemOpen
  \bibfield  {author} {\bibinfo {author} {\bibfnamefont {H.}~\bibnamefont
  {{Bondi}}},\ }\href {https://doi.org/10.1093/mnras/112.2.195} {\bibfield
  {journal} {\bibinfo  {journal} {\mnras}\ }\textbf {\bibinfo {volume} {112}},\
  \bibinfo {pages} {195} (\bibinfo {year} {1952})}\BibitemShut {NoStop}%
\bibitem [{\citenamefont {Pal}\ \emph {et~al.}(2025)\citenamefont {Pal},
  \citenamefont {Ghose}, \citenamefont {Sarkar},\ and\ \citenamefont
  {Das}}]{pal2025effectspindynamicsmulticomponent}%
  \BibitemOpen
  \bibfield  {author} {\bibinfo {author} {\bibfnamefont {K.}~\bibnamefont
  {Pal}}, \bibinfo {author} {\bibfnamefont {S.}~\bibnamefont {Ghose}}, \bibinfo
  {author} {\bibfnamefont {S.}~\bibnamefont {Sarkar}},\ and\ \bibinfo {author}
  {\bibfnamefont {T.~K.}\ \bibnamefont {Das}},\ }\href
  {https://arxiv.org/abs/2503.04321} {\bibinfo {title} {Effect of spin on the
  dynamics of multi-component trans-relativistic accretion flows around kerr
  black holes}} (\bibinfo {year} {2025}),\ \Eprint
  {https://arxiv.org/abs/2503.04321} {arXiv:2503.04321 [astro-ph.HE]}
  \BibitemShut {NoStop}%
\bibitem [{\citenamefont {Unruh}(1981)}]{Unruh1981}%
  \BibitemOpen
  \bibfield  {author} {\bibinfo {author} {\bibfnamefont {W.~G.}\ \bibnamefont
  {Unruh}},\ }\href {https://doi.org/10.1103/PhysRevLett.46.1351} {\bibfield
  {journal} {\bibinfo  {journal} {Phys. Rev. Lett.}\ }\textbf {\bibinfo
  {volume} {46}},\ \bibinfo {pages} {1351} (\bibinfo {year}
  {1981})}\BibitemShut {NoStop}%
\bibitem [{\citenamefont {Visser}(1998)}]{Visser1998}%
  \BibitemOpen
  \bibfield  {author} {\bibinfo {author} {\bibfnamefont {M.}~\bibnamefont
  {Visser}},\ }\href {https://doi.org/10.1088/0264-9381/15/6/024} {\bibfield
  {journal} {\bibinfo  {journal} {Classical and Quantum Gravity}\ }\textbf
  {\bibinfo {volume} {15}},\ \bibinfo {pages} {1767} (\bibinfo {year}
  {1998})}\BibitemShut {NoStop}%
\bibitem [{\citenamefont {Das}(2004)}]{Das2004}%
  \BibitemOpen
  \bibfield  {author} {\bibinfo {author} {\bibfnamefont {T.~K.}\ \bibnamefont
  {Das}},\ }\href {https://doi.org/10.1088/0264-9381/21/22/016} {\bibfield
  {journal} {\bibinfo  {journal} {Classical and Quantum Gravity}\ }\textbf
  {\bibinfo {volume} {21}},\ \bibinfo {pages} {5253} (\bibinfo {year}
  {2004})}\BibitemShut {NoStop}%
\end{thebibliography}
\end{document}